\documentclass[%
 reprint,
 amsmath,amssymb,
 aps,
]{revtex4-1}

\usepackage{graphicx}
\usepackage{dcolumn}
\usepackage{bm}

\usepackage[usenames,dvipsnames]{xcolor}
\usepackage[compact]{titlesec}
\usepackage{hyperref}
\usepackage{cleveref}
\usepackage{bigints}
\usepackage{balance}
\usepackage{indentfirst}
\usepackage{subfigure, float}

\usepackage{longtable} 
\usepackage{array}
\usepackage{multirow}

\usepackage{xr}

\newcommand{\mcP}{\mathcal{P}}

\newcommand{\bbZ}{\mathbb{Z}}
\newcommand{\bbR}{\mathbb{R}}

\newcommand{\bh}{\bm{h}}

\newcommand{\bk}{\bm{k}}
\newcommand{\br}{\bm{r}}
\newcommand{\hphi}{\hat{\phi}}

\newcommand{\mcF}{\mathcal{F}}

\newcommand\tbbint{{-\mkern -16mu\int}}

\newcommand\dbbint{{-\mkern -19mu\int}}

\newcommand\bbint{
	{\mathchoice{\dbbint}{\tbbint}{\tbbint}{\tbbint}}
}

\crefname{figure}{fig.}{fig.}
\crefname{equation}{eq.}{eq.}

\newcommand{\Rmnum}[1]{\uppercase\expandafter{\romannumeral #1}} 

\definecolor{RoyalBlue2}{RGB}{67,110,238}
\definecolor{Orange1}{RGB}{255, 165, 0}
\definecolor{Purple2}{RGB}{145,44,238}
\definecolor{DarkGreen}{RGB}{0,100,0}
\definecolor{DarkTurquoise}{RGB}{0,206,209}

\footnotetext{\ddag~ These authors contributed equally to this work.}
\footnotetext{* Corresponding authors. E-mail addresses: pzhang@pku.edu.cn (P. Zhang), 
	shi@mcmaster.ca (A.-C. Shi), kaijiang@xtu.edu.cn (K. Jiang).}

\begin{document}

\graphicspath{{swfigEnd/}}

\title{Designing a minimal Landau theory to stabilize desired quasicrystals}

\author{
	Wei Si,\textit{$^{a,\,b\,\ddag}$}
	Shifeng Li,\textit{$^{a,\,c\,\ddag}$} 
	Pingwen Zhang,\textit{$^{b,\,c}$\,$^{\ast}$}
	An-Chang Shi,\textit{$^{d}$\,$^{\ast}$}
	and Kai Jiang \textit{$^{a}$\,$^{\ast}$ }
}
\affiliation{\textit{
	$^{a}$
	School of Mathematics and Computational Science, Hunan Key Laboratory for
	Computation and Simulation in Science and Engineering, Xiangtan University,
	Hunan, 411105, China. 
	\\
	$^{b}$
	LMAM, CAPT and School of Mathematical Sciences, Peking University, Beijing 100871, China.
	\\
	$^{c}$
	School of Mathematics and Statistics, Wuhan University, Wuhan, 430072, China.
	\\
	$^{d}$
	Department of Physics and Astronomy, McMaster University, Hamilton L8S 4M1, Canada.
}}


\begin{abstract}
	Interparticle interactions with multiple length scales play a pivotal role
	in the formation and stability of quasicrystals. Choosing a minimal set of
	length scales to stabilize a given quasicrystal is a challenging problem.
	To address this challenge, we propose an intelligent screening
	method (ISM) to design a Landau theory with a minimal number of
	length scales -- referred to as the minimal Landau theory -- that includes
	only the essential length scales necessary to stabilize quasicrystals. 
	Based on a generalized multiple-length-scale Landau theory, ISM first
	evaluates various spectral configurations of candidate structures under a
	hard constraint. It then identifies the configuration with the lowest free
	energy. Using this optimal configuration, ISM calculates phase diagrams to
	explore the thermodynamic stability of desired quasicrystals.  ISM can
	design a minimal Landau theory capable of stabilizing the desired
	quasicrystals by incrementally increasing the number of length scales.  Our
	application of ISM has not only confirmed known behaviors in 10- and 12-fold
	quasicrystals but also led to a significant prediction that quasicrystals
	with 8-, 14-, 16-, and 18-fold symmetry could be stable within
	three-length-scale Landau models. 
	
\end{abstract}


\maketitle


\section{Introduction}
\label{sec:intrd}

Quasicrystals (QCs) are ordered structures that exhibit
rotational symmetry but lack translational symmetry. Since the first discovery
of QCs in Al-Mn alloys\,\cite{shechtman1984metallic}, QCs have attracted
tremendous attention in material science and condensed matter
physics\,\cite{steurer2004twenty, zeng2004supramolecular, duan2018stability,
archer2022rectangle, liu2022expanding, suzuki2022largest, zeng2023columnar,
fayen2023self}. In recent years, QCs have been discovered in a variety of soft
condensed matter, including micelle-forming liquid
crystals\,\cite{loewen1994melting, zeng2004supramolecular, Percec2009,
zeng2023columnar, cao2023understanding}, block copolymers\,\cite{Hayashida2007,
Mai2012, Zhang2012, duan2018stability, xie2020regulate, suzuki2022largest},
colloidal suspensions\,\cite{Fischer2011}, and binary mixtures of
nanoparticles\,\cite{Sztrum2006, Talapin2009}. To date, numerous QCs with $8$-,
$10$-, $12$-, and $18$-fold rotational symmetries have been frequently reported
in both metallic alloys\,\cite{steurer2004twenty} and
soft matters\,\cite{zeng2004supramolecular, Dotera2006, Hayashida2007,
Talapin2009, Fischer2011, Zhang2012, miyamori2020periodic, lindsay2020a15,
suzuki2022largest, liu2022expanding, zeng2023columnar, fayen2023self}. Much
effort has been devoted to studying the properties of QCs, predicting their
stability, and developing methods to control their
formation\,\cite{tsai2008icosahedral, linden2012formation}.

Landau theories have been extensively employed to study the formation, stability
and phase transition of ordered phases, including periodic crystals and
QCs\,\cite{Dotera2007, Lifshitz2007, Barkan2011, yin2021transition,
cui2023efficient}. Generally, a Landau free-energy functional consists of a
polynomial-type bulk energy and a nonlocal pairwise interaction,
\begin{equation}
	\begin{aligned}
		\mcF[\phi(\br)] &= \bbint \left[ d_{2} \phi(\br)^{2} 
		+ d_{3} \phi(\br)^{3} + d_{4} \phi(\br)^{4} 
		+ \ldots \right] \,d\br \\
		&~~~~+ \frac{1}{2} \bbint \bbint \phi(\br) 
			C(|\br-\br'|) \phi(\br') \,d\br d\br',
	\end{aligned}
	\label{eq:general.func}
\end{equation}
where $\phi(\br)$ is an order parameter describing the particle distribution,
$C(r)$ is the correlation potential that is finite for the distance $r$ between
particles\,\cite{likos2001effective}. $\bbint =
\lim\limits_{\Omega\to\mathbb{R}^3} \frac{1}{V(\Omega)} \int_{\Omega}$ and
$V(\Omega)$ is the volume of the region $\Omega$. In the case of a periodic
phase, the integral is equivalent to an integral over its unit cell. The power
series in the first term of \Cref{eq:general.func} is typically truncated to
the fourth order\,\cite{Brazovskii1975, Swift1977, Lifshitz1997}. The quadratic
term contributes to the growth of instability, while the quartic term
establishes a lower bound for the free energy. The cubic term breaks the
$\phi\to-\phi$ symmetry.

An understanding of how to stabilize an ordered structure comes from
representing the second term of \eqref{eq:general.func} in reciprocal space,
\begin{equation*}
	\frac{1}{2} \bbint \hat{C}(k) \left|\hat{\phi}(\bk)\right|^{2} \,d\bk,
	~~ k = |\bk|,
\end{equation*}
where $\bk$ is the reciprocal lattice vector (RLV), $\hat{\phi}(\bk) = \bbint
\exp(-i\bk\cdot\br) \phi(\br) \,d\br$ is the Fourier transform of $\phi(\br)$,
and $\hat{C}(k)$ is the Fourier transform of $C(r)$.
The Fourier coefficients $\hat{\phi}(\bk)$ with wave numbers at the minima of
$\hat{C}(k)$ are energetically favored.
Given the $N$-fold rotational symmetry, the correlation potential can be
approximated by a polynomial with roots $d_1, d_2, \ldots$,
\begin{equation}
	\hat{C}(k) \approx
	c (k^2-d_1^2) (k^2-d_2^2) \cdots,~~c>0.
	\label{eq:C.poly}
\end{equation}
When $\hat{C}(k)$ is truncated to second order, i.e., $c(k^2-d_1^2)$, which
has the minima at $k=0$, the model can be used to simulate the
solidification process of binary mixtures, such as phase-field
models\,\cite{echebarria2004quantitative}.

More complex phase behaviors related to multiple length scales can be investigated by the
correlation potential with multiple roots\,\cite{Brazovskii1975, Swift1977,
Lifshitz1997, jiang2017stability, Savitz2018, jiang2020stability}. For a single
length scale, the potential \eqref{eq:C.poly} should be truncated to fourth
order and rewritten as $c(k^2-1)^2$ by scaling $k$ in units of
$\sqrt{(d_1^2+d_2^2)/2}$ and omitting constant terms. This potential discourages
RLVs with wave numbers deviating from the length scale $1$. The
single-length-scale potential has been
extensively utilized to explain phase behaviors in periodic systems, such as
Landau-Brazovskii model\,\cite{Brazovskii1975} and Swift-Hohenberg
model\,\cite{Swift1977}. To achieve two length scales, $\hat{C}(k)$ must be
truncated to eighth order and rewritten as $c[(k^2-q_1^2)(k^2-q_2^2)]^2$, which
features two equal-depth minima at $k=q_1$ and $k=q_2$.  This two-length-scale
potential was firstly proposed by Lifshitz and Petrich (LP) to describe
quasiperiodic pattern-forming dynamics\,\cite{Lifshitz1997} and to find stable
$12$-fold QCs by setting $q_2/q_1 = 2\cos(\pi/12)$. Over the years, this
two-length-scale type
potential has been widely used to study the thermodynamic stability of QCs and
found stable $10$-, $12$-fold QCs and three-dimensional icosahedral
QCs\,\cite{archer2013quasicrystalline, Dotera2014, Barkan2014, Jiang2015,
subramanian2016three, jiang2017stability, Savitz2018, jiang2020stability,
liang2022molecular, archer2022rectangle}.

However, many QCs are metastable or unstable in two-length-scale models, such as
$8$- and $18$-fold QCs. This raises the question: could these QCs be stabilized
by introducing more length scales? Lifshitz and Petrich speculated that
correlation potentials with three or four length scales could stabilize
higher-order symmetric QCs, such as $18$- or $24$-fold QCs\,\cite{Lifshitz1997}.
Savitz \textit{et al.} have demonstrated that this conjecture might be correct
and found stable $8$- and $18$-fold QCs in four-length-scale Landau
models\,\cite{Savitz2018}. However, increasing the number of length scales
introduces greater complexity in interparticle interactions within physical
systems and poses a significant challenge for theoretical
analysis\,\cite{reich1998dodecagonal, zeng2004supramolecular, Hayashida2007,
Zhang2012, mkhonta2013exploring}. Therefore, it is crucial to design a Landau
theory with the minimal number of length scales, i.e., the minimal Landau model
to stabilize desired QCs.

Designing a minimal Landau theory requires a general Landau model with multiple
length scales. To incorporate $m$ length scales, we could truncate
\Cref{eq:C.poly} to the $4m$-th order and adjust parameters to achieve
equal-depth minima at $k=q_1,\ldots,q_m$, thereby rewriting the correlation
potential in the form,
\begin{equation}
	\hat{C}_{m}(k) = c \Big[\prod_{j=1}^{m}(k^2-q_{j}^{2}) \Big]^{2},
	~~ c>0.
	\label{eq:m.model.k}
\end{equation}
This pair potential favors the RLVs with wave numbers close to the length scales
$q_1,\ldots,q_m$ but suppresses the other RLVs. Substituting
\Cref{eq:m.model.k} into \Cref{eq:general.func} and truncating the polynomial to
fourth order lead to a generalized $m$-length-scale free-energy
functional\,\cite{mkhonta2013exploring, Savitz2018}
\begin{equation}
	\begin{aligned}
		\mcF_{m}[\phi(\br)]
		&= \bbint\Big( -\frac{\epsilon}{2}\phi^{2} 
			- \frac{\alpha}{3} \phi^{3} + \frac{1}{4}\phi^{4} \Big)\,d\br 
		\\ & ~~~~
		+ \frac{c}{2} \bbint \Big[ \prod_{j=1}^{m}(\nabla^{2}+q_{j}^{2})
			\phi(\br) \Big]^{2}\,d\br,
	\end{aligned}
	\label{eq:m.model}
\end{equation}
where the parameter $\epsilon$ is temperature-related, $\alpha$ measures the
intensity of three-body interaction, $c>0$ is a penalty factor, and $q_{1},
\ldots,q_{m}$ are length scales. The function $\phi(\br)$ satisfies the
mean-zero constraint, $\bbint\phi(\br)\,d\br = 0$, corresponding to a
mass-conserved system.

We organize the rest of this paper as follows. In \Cref{sec:ISM}, we propose an
intelligent screening method to design a minimal Landau theory for desired QCs.
In \Cref{sec:results}, we apply this method to design minimal Landau theories
for $2n$-fold QCs ($n=4,5,\ldots,9$). We find that three-length-scale Landau
models can stabilize $8$-, $14$-, $16$-, and $18$-fold QCs. Finally, we
summarize this work in \Cref{sec:conclu}.

\section{Intelligent screening method (ISM)}
\label{sec:ISM}

Designing a minimal Landau theory to stabilize desired QCs requires
considering various candidate structures. For a candidate phase, there are
numerous possible configurations of RLVs describing its spectral distribution.
To identify the optimal configuration with the lowest free energy, it is crucial
to analyze the contributions of RLVs.

RLVs could be categorized into primary and non-primary RLVs. The primary RLVs
exhibit strong intensities and have wave numbers equal to length scales.
The remaining RLVs are the non-primary RLVs. Numerous studies indicate
that the primary RLVs determine the main characteristics of ordered structures and
the non-primary RLVs influence local details\,\cite{Jiang2015,
jiang2017stability, ratliff2019wave, Savitz2018}. The contributions of primary
and non-primary RLVs can be studied from two perspectives: hard constraint (HC)
with $c\to\infty$, and soft constraint (SC) with a finite $c$. Under the HC,
$\hat{C}_{m}(k)$ is zero if the wave number $k$ belongs to the set
$\{q_1,\ldots,q_m\}$, and otherwise it is infinite. This implies that all
non-primary RLVs are forbidden under this constraint. The SC relaxes the
restriction on wave numbers, permitting the emergence of non-primary RLVs, which
may be favored by a realistic system.

\begin{figure*}[htbp]
	\centering
	\includegraphics[scale=0.45]{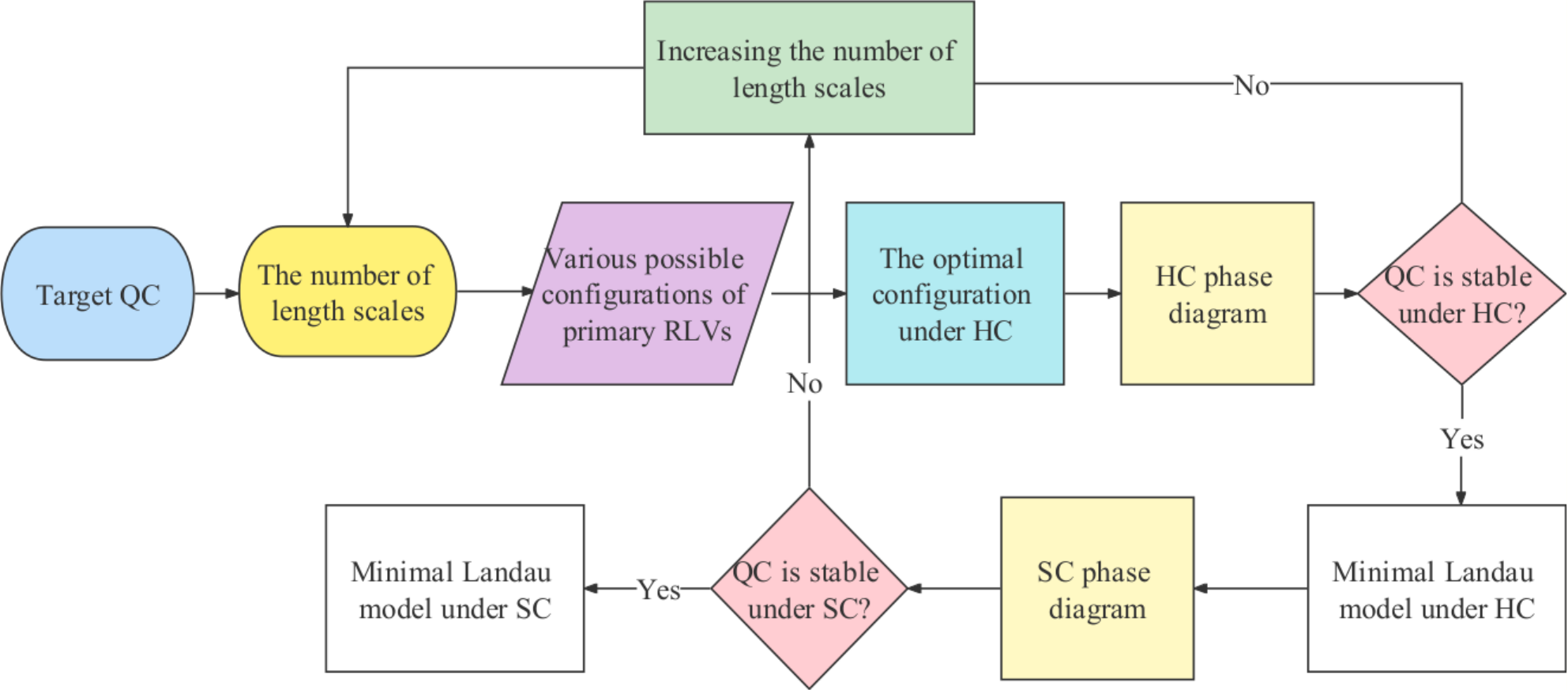}
	\caption{
	Flowchart of ISM. The number of length scales increases gradually from 1.}
	\label{fig:flowChart}
\end{figure*}

Based on the two constraints, we propose ISM to design a minimal Landau
theory to stabilize the desired QCs, as shown in \Cref{fig:flowChart}.
Given a target QC and the number of length scales, there might be
many configurations of primary RLVs. Let us consider HC first.
For a specific configuration, HC only allows a finite number of primary RLVs,
thus the free energy functional can be written as a polynomial function.
The polynomial function can be easily minimized by computer-assisted symbolic
calculation to obtain the free energy of the configuration. Among all configurations, 
ISM selects the optimal one with the lowest free energy by directly comparing
the free energies. Using the optimal configuration, we can study the
thermodynamic stability of the target QC by constructing phase diagrams under
HC. If the target QC is metastable or unstable, we increase the number of length
scales to obtain more configurations. The optimal configuration among these may
lead to the formation of stable QC. The free energy functional $\mcF_{m}$
with these length scales is called a HC minimal Landau model. Based on the
optimal configuration and the HC minimal Landau model, we further design a
minimal Landau model under SC. Since SC permits the emergence of non-primary
RLVs, we use numerical methods to calculate SC phase diagrams to study the
stability of target QC. We obtain the SC minimal Landau model if the target QC
is stable, otherwise, we repeat the above process with more length scales. We
can always obtain a minimal Landau theory to stabilize target QC since more
length scales have more primary RLVs that may form more triplets to lower the
free energy.

\begin{figure}[htbp]
	\centering
	\includegraphics[scale=0.12]{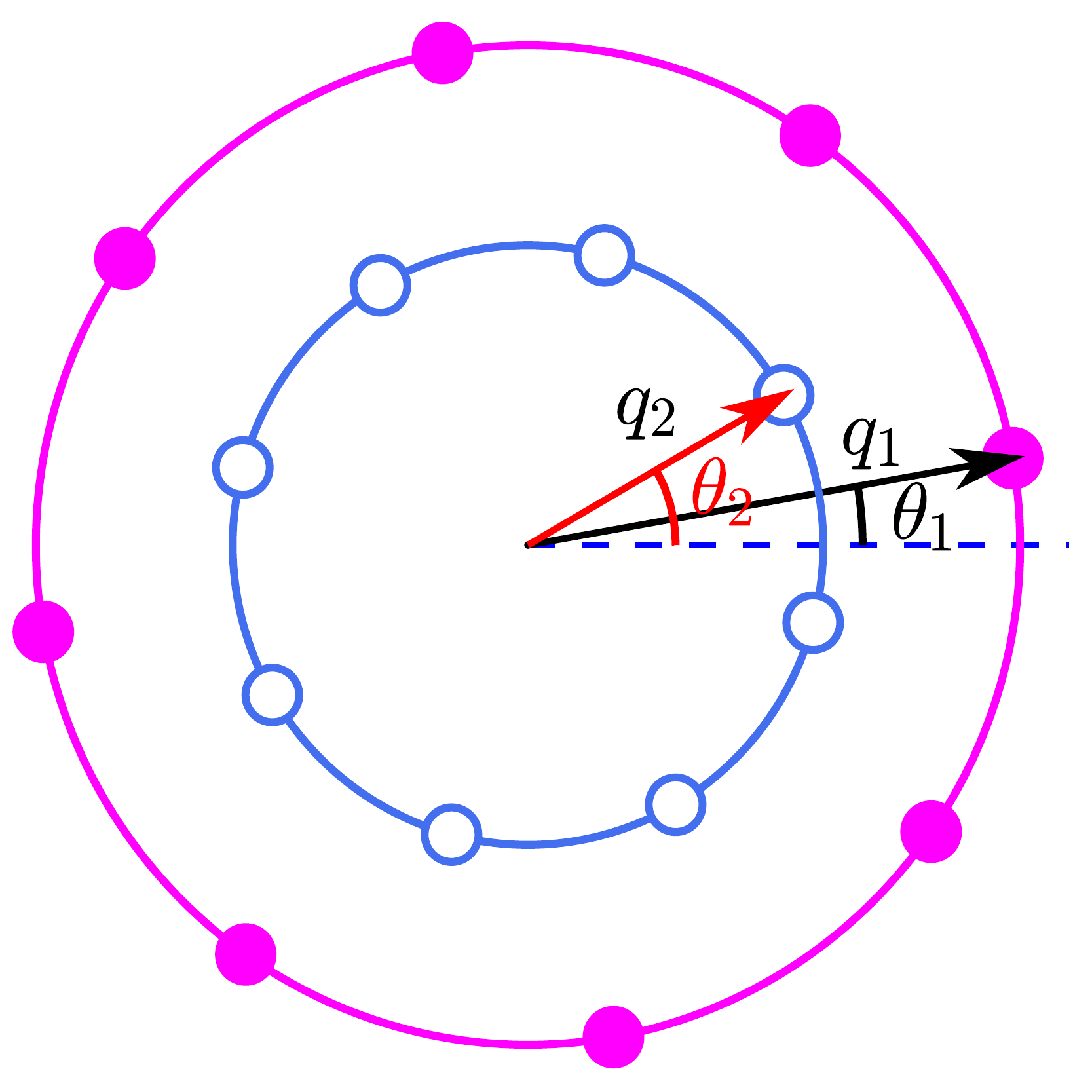}
    \caption{
		Primary RLVs of the $8$-fold QC with two length scales $q_1$ and
		$q_2$. Magenta (royal blue) dots and the origin form the primary RLVs.
		$\theta_1$ and $\theta_2$ are offset angles.
		}
	\label{fig:illustrate}
\end{figure}

In order to obtain the possible configurations of primary RLVs, we first analyze
the elements of primary RLVs. For a candidate phase, its primary RLVs are
determined by the length scales and the relative positions. As an example, we
consider two-length-scale $8$-fold QCs, as demonstrated in \Cref{fig:illustrate}. 
The length scales are consistent with the radii of the circles $q_1$ and $q_2$.
The relative positions depend on the offset angles $\theta_1$ and $\theta_2$.
The offset angle of each circle is defined as the minimal angle at which the
primary RLV rotates clockwise in the horizontal direction. Note that the primary
RLVs on each circle are equivalent due to rotational symmetry. Without loss of
generality, we set $\theta_1=0$ and the rest is $\theta_2-\theta_1$. For an
$m$-length-scale $N$-fold candidate phase, we have
\begin{align}
	& q_{j} = \cos(w_{j}\pi/N), ~~~~~~ j=1,\ldots,m, \label{eq:q}
	\\
	& \theta_{1} = 0, ~~ \theta_{j'} = s_{j'}\pi/N, ~~ j'=2,\ldots,m. \label{eq:theta}
\end{align}
The primary RLVs given by \eqref{eq:q} and \eqref{eq:theta} can form more
triplets, which could lower the free energy. $w_{j} \in [0, N/2)$ due to the
periodicity of the cosine function, and $s_{j'} \in [0,2)$ because of rotational
symmetry. We obtain various possible configurations of primary RLVs by
discretizing $w_{j}$ and $s_{j'}$ in the valid ranges.

Under HC, the free energy functional $\mcF_{m}$ can be written as a polynomial
function. Since $\hat{C}_{m}(k)$ should be zero to ensure finite free energy,
$\mcF_{m}$ preserves bulk energy part
\begin{equation}
	\mcF_{m} = \bbint \left( -\frac{\epsilon^{*}}{2}\phi^{2} -
	\frac{1}{3}\phi^{3} + \frac{1}{4}\phi^{4} \right)\,d\br,
	\label{eq:m.model.hard.rescale}
\end{equation}
where $\epsilon^{*} = \epsilon/\alpha^{2}$. Note that
\Cref{eq:m.model.hard.rescale} has scaled the cubic coefficient $\alpha$ to
unity by measuring the field $\phi$ in units of $\alpha$ and the energy in units
of $\alpha^{4}$. Using the Fourier transformation of $\phi(\br)$,
\Cref{eq:m.model.hard.rescale} becomes
\begin{equation}
	\begin{aligned}
		& ~~~~ \mcF_{m}(\hphi_{q_1},\ldots,\hphi_{q_m})
		\\
		& = -\frac{\epsilon^{*}}{2}
		\sum_{\bk_1+\bk_2=\bm{0}} \hphi_{k_1} \hphi_{k_2}
		\\ & ~~~~ 
		- \frac{1}{3} \sum_{\bk_{1}+\bk_{2}+\bk_3=\bm{0}}
		\hphi_{k_1} \hphi_{k_2} \hphi_{k_3}
		\\ & ~~~~ 
		+ \frac{1}{4} \sum_{\bk_1+\bk_2+\bk_3+\bk_4=\bm{0}}
		\hphi_{k_1} \hphi_{k_2} \hphi_{k_3} \hphi_{k_4},
	\end{aligned}
	\label{eq:m.model.hard.rescale.k}
\end{equation}
where all $k_i = |\bk_i|$ belong to the set $\{q_1,\ldots,q_m\}$, $i=1,2,3,4$,
and $\hphi_{q_j}$ denotes the Fourier coefficient with wave number $q_j$.
In \Cref{eq:m.model.hard.rescale.k}, the three-RLV interaction in the third row
is beneficial to lower the free energy, but the four-RLV interaction in the
fourth row increases the free energy. Since the primary RLVs are finite for
a specific configuration, we can calculate the summations in
\Cref{eq:m.model.hard.rescale.k} by symbolic computation. We then minimize
the polynomial function with respective to $\hphi_{q_1},\ldots,\hphi_{q_m}$ to
obtain the free energy of this configuration. Among all possible configurations,
we select the optimal configuration with the lowest free energy. Using the
optimal configuration, we study the thermodynamic stability of desired QCs by
constructing a phase diagram under HC.

Under SC, ISM can also examine the thermodynamic stability of desired QCs by
combining with numerical methods. An accurate and efficient numerical approach to
study QCs is the projection method\,\cite{Jiang2014, Jiang2023numerical}.
The projection method embeds the QC into a high-dimensional periodic system
which can be efficiently calculated by fast Fourier transformation, and then
obtains the QC by projecting it back to the original space. The specific formula
of the projection method is
\begin{equation}
	\phi(\br) = \sum_{\bh\in\bbZ^{d_1}} \hat{\phi}(\bh) 
	e^{i(\mcP\cdot\bh)^{T} \cdot \br},
	~~~~ \br \in \bbR^{d_0}, ~~ d_0 \leq d_1,
\end{equation}
where $\mcP$ is a $d_0\times d_1$-order projection matrix. $d_0$ is the
dimension of the original space, and $d_1$ is the dimension of the
high-dimensional space dependent on the symmetry of the QC. A special case of
$d_0 = d_1$ implies that the projection method is the common Fourier
pseudo-spectral approach for periodic crystals.
Moreover, the $m$-length-scale model under SC can be rescaled to reduce the
number of model parameters.
Let $q_*$ be any element of the set $\{q_j\}_{j=1}^{m}$. $c$ is rescaled to
unit by measuring the field $\phi$ in units of $\sqrt{c} q_*^{2m}$, and
consequently the energy is measured in units of $c^2 q_*^{8m}$, thus
\Cref{eq:m.model} becomes
\begin{equation}
	\begin{aligned}
		\mcF_{m}[\phi(\br)]
		&= \bbint\left( -\frac{\epsilon}{2}\phi^{2} 
			- \frac{\alpha}{3} \phi^{3} + \frac{1}{4}\phi^{4} \right)\,d\br \\
		&+ \frac{1}{2} \bbint \Big[ \prod_{j=1}^{m}(\nabla^{2}+q_{j}^{2}/q_*^2) 
			\phi(\br) \Big]^{2}\,d\br,
	\end{aligned}
	\label{eq:m.model.soft.rescale}
\end{equation}
where $\epsilon$ and $\alpha$ are measured in units of $cq_*^{4m}$ and
$\sqrt{c}q_*^{2m}$, respectively. In this work, $q_*$ takes the minimal value
of $\{q_j\}_{j=1}^{m}$. Its stationary solutions can be quickly and robustly
obtained by recently developed optimization methods\,\cite{jiang2020efficient,
bao2022adaptive, Bao2024ConvergenceAF}. And its phase diagram can be
automatically and efficiently generated by our developed open-source
software\,\cite{agpd}.

\section{Results and discussions}
\label{sec:results}

\begin{figure}[htbp]
	\centering
	\subfigure[HEX$^1$]{
		\includegraphics[scale=0.08]{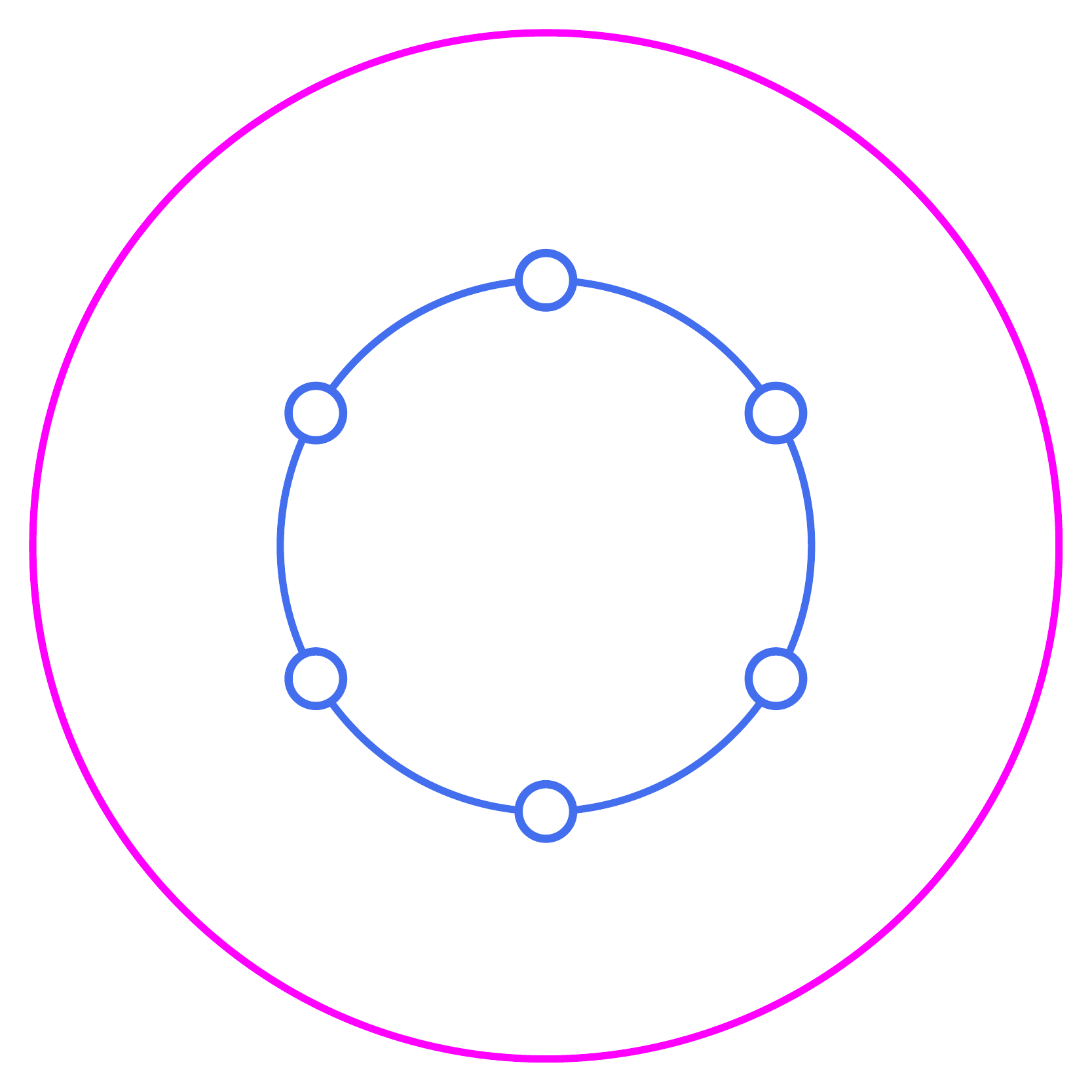}}
	\subfigure[SQU$^1$]{
		\includegraphics[scale=0.08]{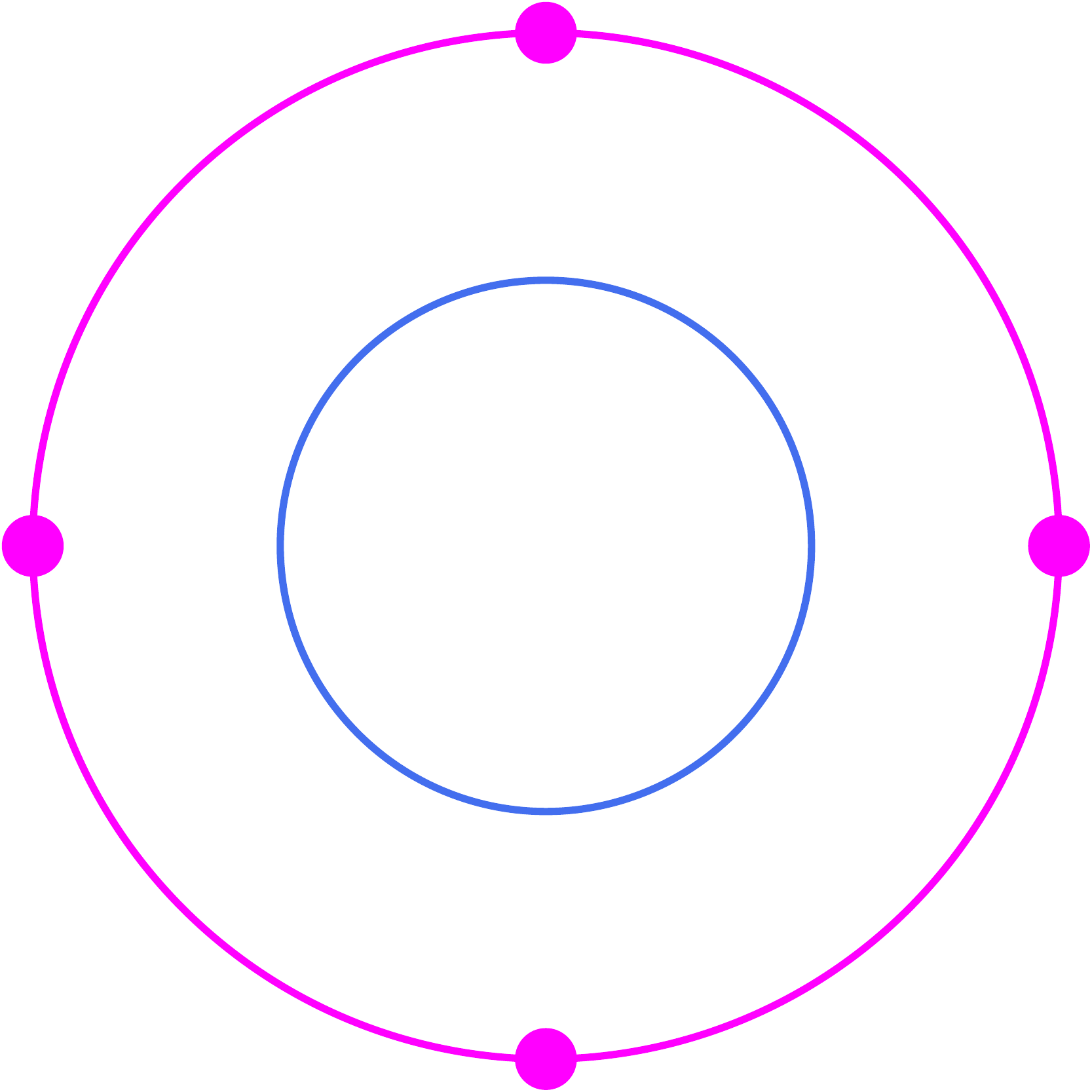}}
	\subfigure[LAM$^1$]{
		\includegraphics[scale=0.08]{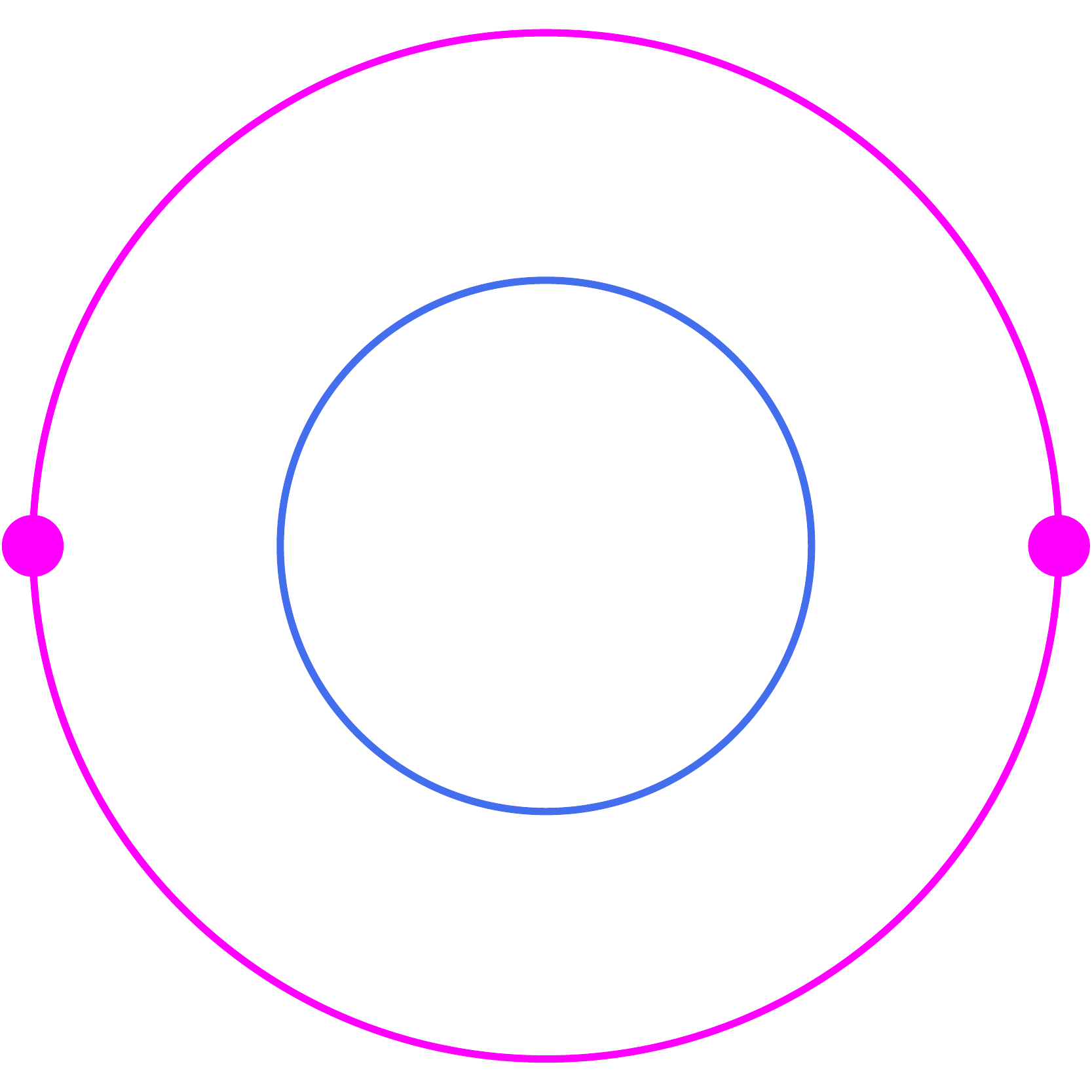}}
	\vfill
	\subfigure[HEX$^2$]{
		\includegraphics[scale=0.08]{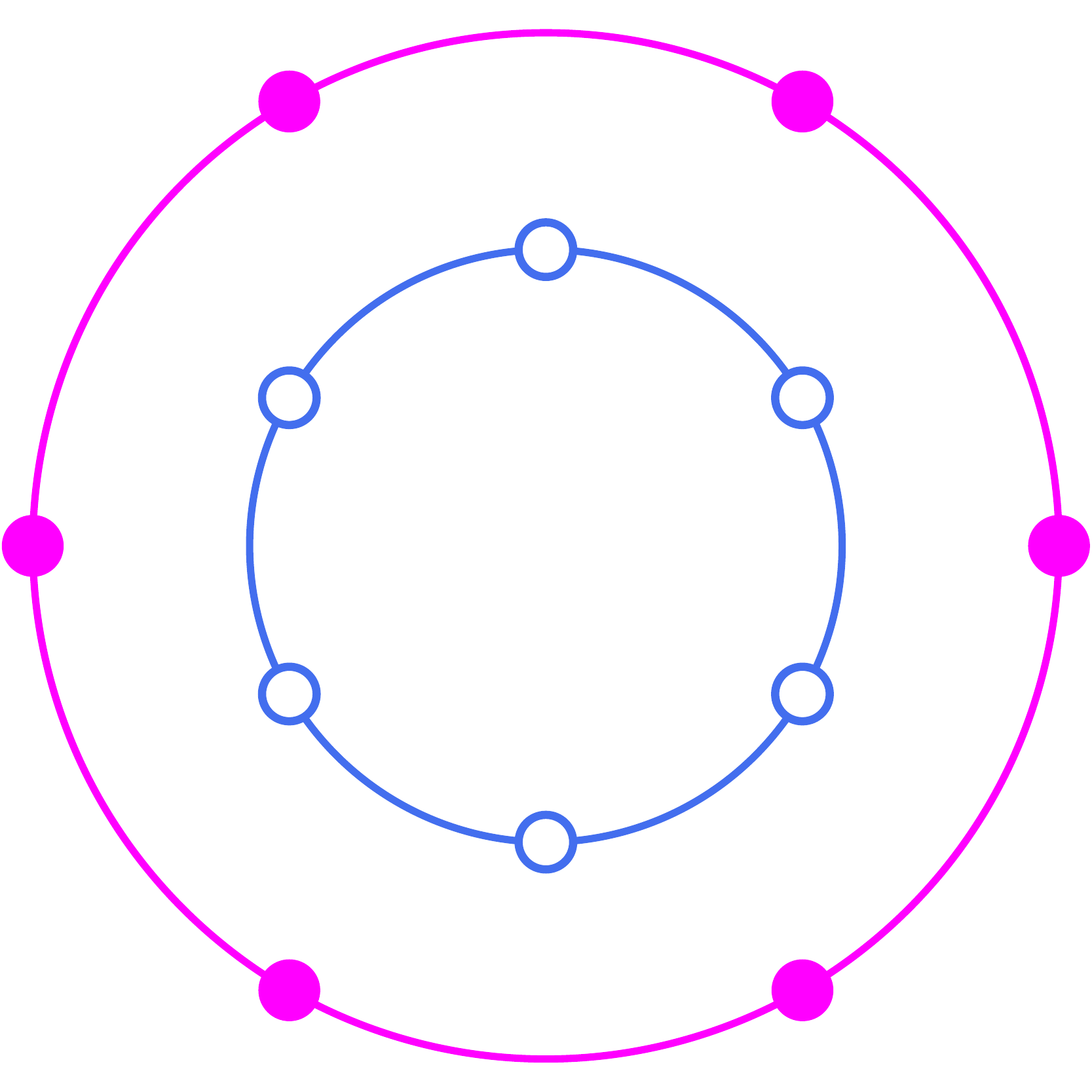}}
	\subfigure[SQU$^2$]{
		\includegraphics[scale=0.08]{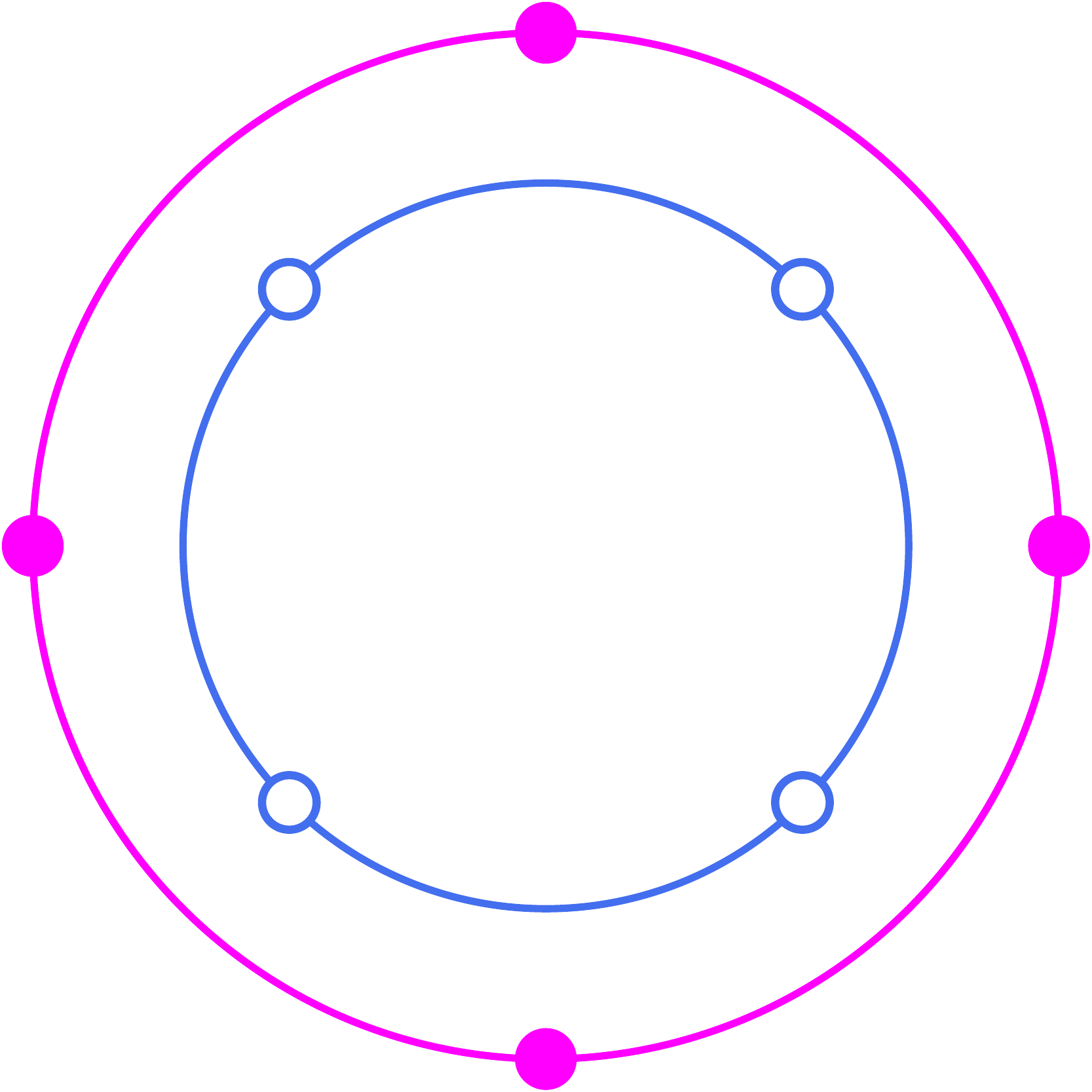}}
	\caption{\label{fig:single.crystals}
		Optimal primary RLVs of (a)-(c) single-length-scale and (d)-(e)
		two-length-scale competing crystals. Superscripts denote the number of
		valid length scales. The radius of inner (royal blue) circle is $q_1$
		and the radius of outer (magenta) circle is $q_2$.  $q_2/q_1$ is equal
		to (d) $2\cos(\pi/6)$; (e) $2\cos(\pi/4)$.
	}
\end{figure}

Applying the ISM, we design minimal Landau theories for two-dimensional
$2n$-fold QCs ($n=4,5,\ldots,9$). These QCs are named as octagonal (O),
decagonal (D), dodecagonal (DD), tetradecagonal (TD), hexadecagonal (HD) and
octadecagonal (OD) QCs, respectively. Note that the primary RLVs with a single
length scale could not generate sufficient three-RLV interactions to stabilize
QCs in the Landau free-energy functionals\,\cite{Lifshitz1997, ratliff2019wave}.
We present the numerical results of two, three and four length scales in
Supplementary Material (SM). The results include optimal configurations
of primary RLVs, HC free energy, HC phase diagrams, and the SC phase diagrams of
minimal Landau models. The results demonstrate that $10$- and $12$-fold QCs can
be stabilized in the Landau model with at least two length scales, which is
consistent with previous findings\,\cite{Lifshitz1997, Jiang2015, Barkan2014,
jiang2017stability, Dotera2014}, implying the effectiveness of ISM. The results
of $8$-, $14$-, $16$- and $18$-fold QCs give some exciting predictions, which will
be introduced in this section.

In what follows, we consider three competing crystals, including lamellar (LAM),
square (SQU) and hexagonal (HEX) crystals to study the thermodynamic stability
of an $m$-length-scale QC. The competing crystals have the length scales
consistent with the length scales of the QC. The offset angles follow
\Cref{eq:theta}, where $N=2$ for LAM, $N=4$ for SQU, and $N=6$ for HEX.
Under HC, the free energy functional can be written as a polynomial function, thus
we can easily obtain the optimal primary RLVs of these crystals. It should be
noted that if the Fourier coefficients at some primary RLVs are very weak or
even vanished, i.e., these primary RLVs become non-primary RLVs, the number of
valid length scales of the crystals denoted by superscripts is less than $m$.
Numerical simulations demonstrate the optimal configuration of primary RLVs has
one or two valid length scales for HEX and SQU but only one valid length scale
for LAM, as shown in \Cref{fig:single.crystals}. 
For the case of one valid length scale, the HC free energies are
\begin{equation}
	\mcF_{\text{HEX}^1}(\hphi_j,\epsilon^{*}) = -3\epsilon^{*}\hphi_j^{2} 
		- 4\hphi_j^{3} + \frac{45}{2}\hphi_j^{4},
	\label{eq:energy.single.6}
\end{equation}
\begin{equation}
	\begin{aligned}
		\mcF_{\text{SQU}^1}(\hphi_{j},\epsilon^{*}) 
		= -2\epsilon^{*} \hphi_{j}^{2} + 9\hphi_{j}^{4},
	\end{aligned}
	\label{eq:energy.single.4}
\end{equation}
\begin{equation}
	\mcF_{\text{LAM}^1}(\hphi_{j},\epsilon^{*}) = -\epsilon^{*}\hphi_{j}^{2} 
		+ \frac{3}{2}\hphi_{j}^{4}.
	\label{eq:energy.single.2}
\end{equation}
For the case of two valid length scales, the length scales satisfy a special
ratio to form more three-RLV interactions: $q_2/q_1=2\cos(\pi/6)$ for HEX$^2$
and $q_2/q_1=2\cos(\pi/4)$ for SQU$^2$. Their HC free energies are given by
\begin{equation}
	\begin{aligned}
		& \mcF_{\text{HEX}^2}(\{\hphi_{j}\}_{j=1}^{2},\epsilon^{*}) 
		= -3\epsilon^{*} (\hphi_{1}^{2} + \hphi_{2}^{2})
		- 12\hphi_{1}^{2}\hphi_{2}
		\\
		& 
		- 4\hphi_{1}^{3} - 4\hphi_{2}^{3}
		+ \frac{45}{2} (\hphi_{1}^{4} + \hphi_{2}^{4})
		+ 36\hphi_{1}^{3}\hphi_{2} + 90\hphi_{1}^{2}\hphi_{2}^{2},
	\end{aligned}
	\label{eq:energy.double.6}
\end{equation}
\begin{equation}
	\begin{aligned}
		\mcF_{\text{SQU}^2}(\{\hphi_{j}\}_{j=1}^{2},\epsilon^{*}) 
		& = -2\epsilon^{*} (\hphi_{1}^{2} + \hphi_{2}^{2})
		- 8\hphi_{1}^{2}\hphi_{2}
		\\
		& ~~~~
		+ 9(\hphi_{1}^{4} + 4\hphi_{1}^{2}\hphi_{2}^{2} + \hphi_{2}^{4}).
	\end{aligned}
	\label{eq:energy.double.4}
\end{equation}

\subsection{Octagonal (O) QCs}
\label{subsec:8fold}

Two-dimensional OQCs have been frequently observed in materials since the first
discovery in V-Ni-Si and Cr-Ni-Si alloys\,\cite{wang1987two}. Their
electron-diffraction patterns reveal that OQCs have multiple length
scales\,\cite{wang1987two}. Theoretical studies have pointed out that the
formation of OQCs may require correlation potentials with multiple length
scales. Concretely, the two-length-scale Landau model, such as the LP model,
could obtain a metastable OQC\,\cite{Jiang2015}, and the four-length-scale model
could stabilize OQC\,\cite{Savitz2018}. In this section, we apply the ISM and design
a minimal Landau theory to stabilize OQC.

\begin{figure}[htbp]
	\centering
	\includegraphics[scale=0.135]{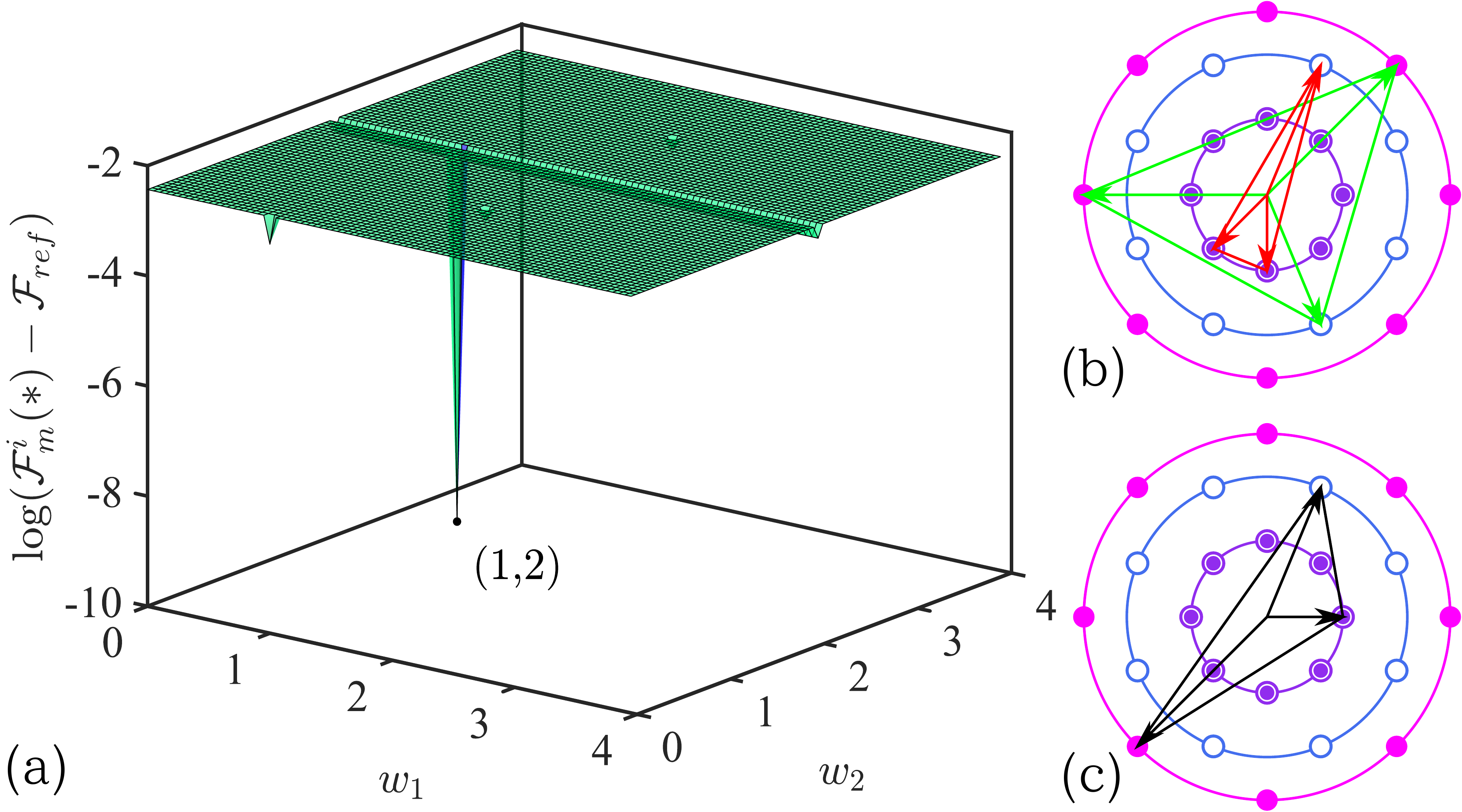}
	\caption{
		(a) Free energies of OQC$^3$ as a function of $w_{1}$ and $w_{2}$.
		$w_{3} = 3$, $s_{2}=1$, $s_{3}=0$.
		$\mcF_m^i(*)$ is the minimal value of the HC model with $\epsilon^{*}=0$.
		$(w_1,w_2)=(1,2)$ is the peak point with free energy $\mcF_{ref} =
		-3.7285\times10^{-3}$.
		At the peak point, the primary RLVs with length scales $\cos(\pi/8)$, $\cos(\pi/4)$,
		$\cos(3\pi/8)$ have two types of three-RLV interactions, as shown in (b)-(c).
		}
	\label{fig:8.ISM.mechanism}
\end{figure}

\begin{figure*}[htbp]
	\centering
	\includegraphics[scale=0.14]{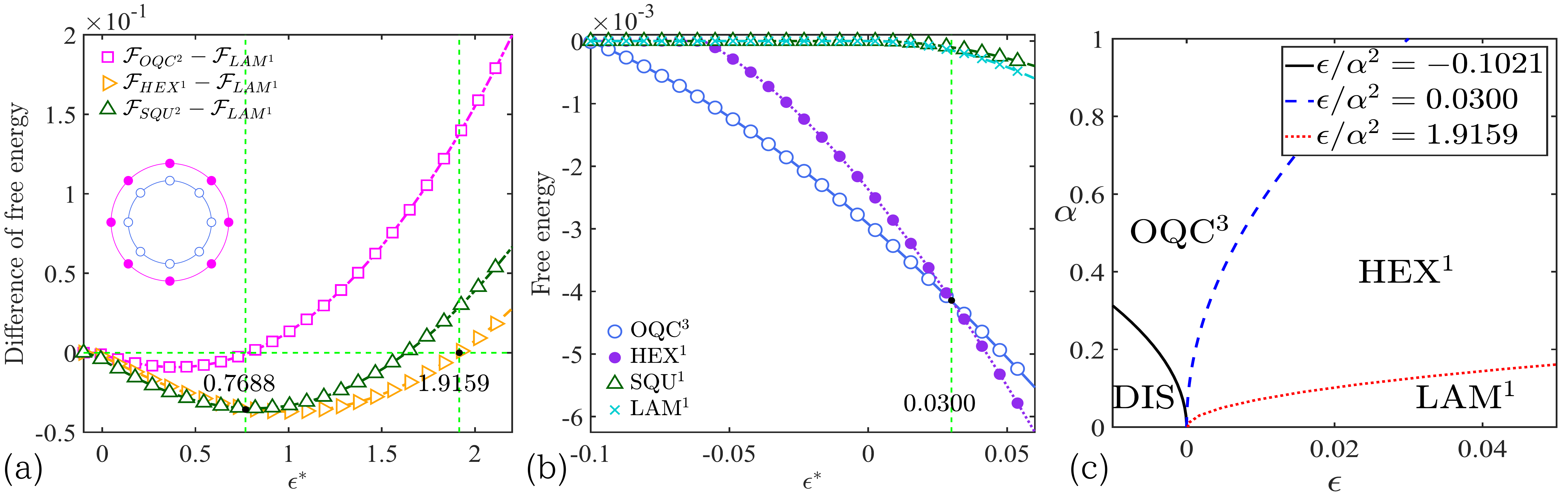}
	\caption{
		(a) HC free energy of two-length-scale candidate phases with the energy
		of LAM$^1$ as the baseline.
		Two length scales are $1$ and $\cos(\pi/4)$.
		The embedded pattern is the optimal primary RLVs of OQC$^2$.
		(b) Free energies of candidate structures in the HC model with
		three length scales $\cos(\pi/8)$, $\cos(\pi/4)$, and $\cos(3\pi/8)$.
		(c) HC phase diagram of the three-length-scale model in
		$\epsilon$-$\alpha$ plane.
		DIS stands for the disordered phase with zero free energy.
		}
	\label{fig:energy.8}
\end{figure*}

ISM firstly finds the optimal configuration of primary RLVs of $m$-length-scale
OQC. The length scales and offset angles are
\begin{equation*}
	\begin{aligned}
		& q_{j} = \cos(w_{j}\pi/8), ~~ w_j \in [0,4), ~~~~~~ j=1,\ldots,m, \\
		& \theta_{1} = 0, ~~ \theta_{j'} = s_{j'}\pi/8, ~~ 
			s_{j'} \in [0,2), ~~ j'=2,\ldots,m.
	\end{aligned}
\end{equation*}
For a specific configuration involving variables
$w_1^i,\ldots,w_m^i,s_2^i,\ldots,s_m^i$, the Landau model is written as a
polynomial with the minimal value $\mcF_{m}^{i}(*)$. The configuration yielding
the lowest $\mcF_{m}^{i}(*)$ is considered optimal. Taking $m=3$ as an example,
\Cref{fig:8.ISM.mechanism}\,(a) plots $\mcF_{m}^{i}(*)$ against $w_1$ and $w_2$
when fixing $w_{3} = 3$, $s_{2}=1$, and $s_{3}=0$. The energy surface is almost flat
except for a few peaks. This implies that only a few configurations can
significantly lower free energies. The lowest peak whose energy is denoted as
$\mcF_{ref}$ occurs at $(w_1,w_2) = (1,2)$. The corresponding
primary RLVs have length scales of $\cos(\pi/8)$, $\cos(\pi/4)$, and
$\cos(3\pi/8)$. We have confirmed that this configuration remains optimal as
$w_3$, $s_2$ and $s_3$ change. In this optimal configuration, there
are two different ways to form the three-RLV interaction. One involves two
RLVs with equal wave numbers and another RLV with a different wave number, as
illustrated in \Cref{fig:8.ISM.mechanism}\,(b). Another way involves three RLVs
with different wave numbers, as shown in \Cref{fig:8.ISM.mechanism}\,(c). The HC
free energy of the optimal configuration has the following expression
\begin{equation}
	\begin{aligned}
		& \mcF_{\text{OQC}^3} (\{\hphi\}_{j=1}^{3},\epsilon^{*})
		= -4\epsilon^{*} \sum_{j=1}^{3}\hphi_{j}^{2}
		- 16\hphi_{2} (\hphi_{1}+\hphi_{3})^{2}
		\\ & ~~~~
		+ 42 \sum_{j=1}^{3}\hphi_{j}^{4}
		+ 192 \hphi_{1}\hphi_{2}^{2}\hphi_{3}
		+ 48 \hphi_{1}\hphi_{3} (\hphi_{1}^{2}+\hphi_{3}^{2})
		\\ & ~~~~
		+ 144 (\hphi_{1}^{2}\hphi_{2}^{2} + \hphi_{1}^{2}\hphi_{3}^{2}
		+ \hphi_{2}^{2}\hphi_{3}^{2}).
	\end{aligned}
	\label{eq:energy.8.3}
\end{equation}
Moreover, for $m=2$, the optimal primary RLVs of OQC have length scales of $1$
and $\cos(\pi/4)$, which are depicted in the embedded pattern of
\Cref{fig:energy.8}\,(a), with the HC free energy expressed as
\begin{equation}
	\begin{aligned}
		\mcF_{\text{OQC}^2}(\{\hphi\}_{j=1}^{2},\epsilon^{*})
		& = -4\epsilon^{*} \sum_{j=1}^{2}\hphi_{j}^{2}
		- 16\hphi_{1}\hphi_{2}^{2}
		\\
		& ~~~~
		+ 42 \sum_{j=1}^{2}\hphi_{j}^{4}
		+ 120 \hphi_{1}^{2} \hphi_{2}^{2}.
	\end{aligned}
	\label{eq:energy.8.2}
\end{equation}

\begin{figure*}[htbp]
	\centering
	\includegraphics[scale=0.14]{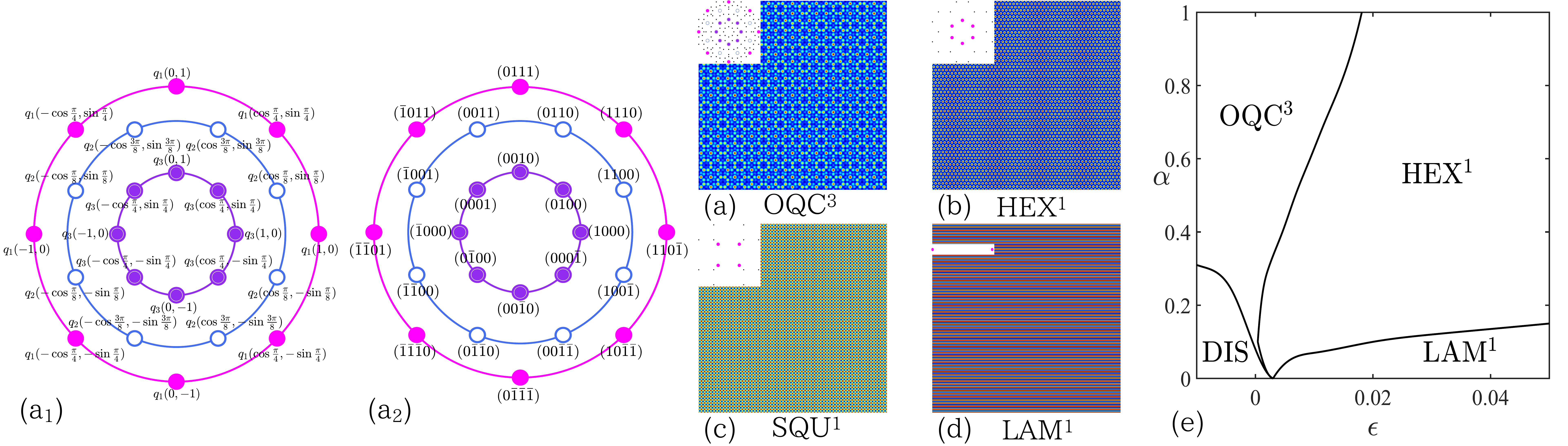}
	\caption{
		Coordinates of optimal primary RLVs of OQC$^3$ in (a$_1$)
		two-dimensional and (a$_2$) four-dimensional reciprocal space.
		Stationary ordered states: 
		(a) OQC$^3$;
		(b) HEX$^1$;
		(c) SQU$^1$;
		(d) LAM$^1$
		calculated by the projection method in the Landau model with three
		length scales $\cos(\pi/8)$, $\cos(\pi/4)$ and $\cos(3\pi/8)$.
		The diffraction pattern embedded in the upper left corner only 
		plots these RLVs with intensities greater than $10^{-6}$.
		(e) SC phase diagram of the three-length-scale model. 
		}
	\label{fig:finite.8}
\end{figure*}

To study the thermodynamic stability of OQCs under HC,
\Cref{fig:energy.8}\,(a) plots the HC free energy of candidate structures as a
function of $\epsilon^*$ for a Landau model with two length scales $1$ and
$\cos(\pi/4)$. We find that SQU$^2$ is favorable when $\epsilon^* \le 0.7688$,
HEX$^1$ for $0.7688 \le \epsilon^* \le 1.9159$, and LAM$^1$ when $\epsilon^*
\gtrsim 1.9159$. Thus the two-length-scale OQC is metastable. For a
Landau model with three length scales $\cos(\pi/8)$, $\cos(\pi/4)$, and
$\cos(3\pi/8)$, we find stable OQCs when $\epsilon^{*} \le 0.0300$, as shown in
\Cref{fig:energy.8}\,(b). The reason could be attributed that the primary RLVs in
the three-length-scale OQC form more three-RLV interactions than those in the
two-length-scale OQC, thereby reducing the HC free energy. Furthermore, the HC
phase diagram in the $\epsilon$-$\alpha$ plane is plotted in
\Cref{fig:energy.8}\,(c). Here, the three-length-scale OQC is expected to be
thermodynamic stable when $-0.1021 \le \epsilon^{*} \le 0.0300$. Therefore,
this three-length-scale Landau model is the minimal model to stabilize OQC under HC.

Based on the HC minimal Landau model, we further study the thermodynamic
stability of OQC under SC to design a SC minimal Landau theory. We apply the
projection method to evaluate OQCs and their free energies accurately.
\Cref{fig:finite.8}\,(a$_1$) depicts the coordinates of the optimal primary RLVs
of OQCs in two-dimensional space. To be consistent with the rescaling model
\eqref{eq:m.model.soft.rescale}, the length scales are measured in units of
$q_3$. We set the four vectors $(1,0)$, $(\sqrt{2}/2,\sqrt{2}/2)$, $(0,1)$,
$(-\sqrt{2}/2,\sqrt{2}/2)$ as basis vectors, allowing the primary RLVs
to be expressed by integer coefficients with the four vectors, as illustrated in
\Cref{fig:finite.8}\,(a$_2$). Accordingly, the projection matrix is
\begin{equation}
	\mcP_{OQC} = 
	\begin{pmatrix}
		1 & \sqrt{2}/2 & 0 & -\sqrt{2}/2 \\
		0 & \sqrt{2}/2 & 1 & \sqrt{2}/2
	\end{pmatrix}.
	\label{eq:pm.8}
\end{equation}
The OQC is embedded into a four-dimensional periodic structure that can be
accurately calculated, and then it is recovered by projecting this four-dimensional
periodic structure into the two-dimensional space.

In \Cref{fig:finite.8}\,(a)-(d), we present the diffraction patterns and
real-space morphologies of stationary candidate structures. \Cref{fig:finite.8}\,(e)
presents the SC phase diagram, where OQC$^3$, HEX$^1$, and LAM$^1$ occupy stable
regions but SQU$^1$ remains metastable. The phase boundaries in the SC phase
diagram are similar to those in the HC phase diagram \Cref{fig:energy.8}\,(c).
It may be attributed to the fact that the primary RLVs play a dominant role in
determining the stability of candidate structures and the contribution of
non-primary RLVs causes slight changes in phase boundaries. Consequently, we
could come to a conclusion that the minimal Landau theory to stabilize OQCs
under SC involves three length scales $\cos(\pi/8)$, $\cos(\pi/4)$, and
$\cos(3\pi/8)$.

In previous work on primary RLVs of OQCs, only a finite number of configurations
were considered\,\cite{jiang2015stability, Savitz2018}. ISM examines nearly all
possible configurations and effectively identifies the optimal configuration.
Therefore, ISM can design a minimal Landau theory to stabilize OQCs.

\subsection{Tetradecagonal (TD), Hexadecagonal (HD) and Octadecagonal (OD) QCs}
\label{subsec:14.16.18fold}

\begin{figure}[htbp]
	\centering
	\includegraphics[scale=0.16]{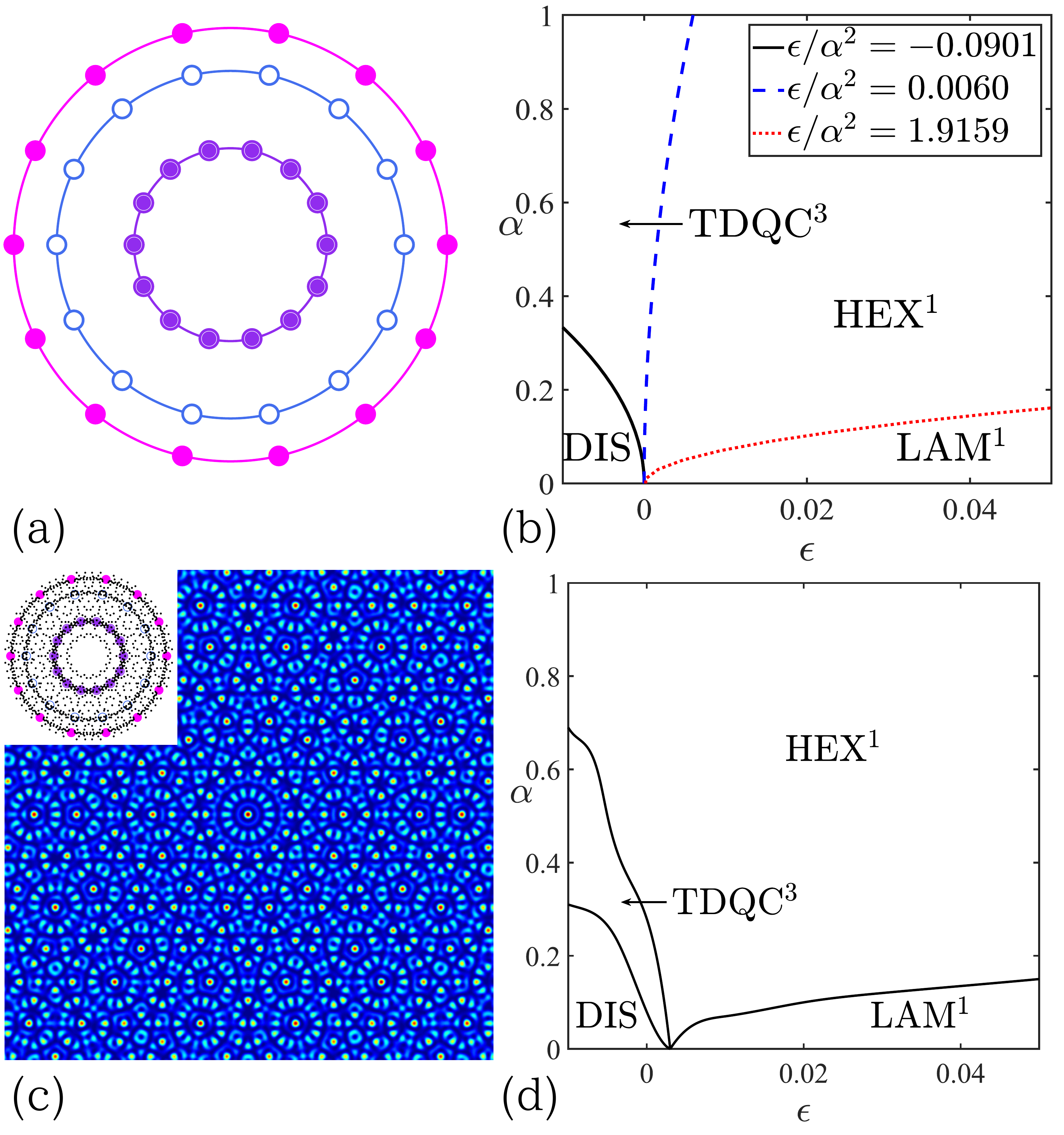}
	\caption{
		(a) Optimal primary RLVs of TDQC$^3$, (b) HC phase diagram,
		(c) stationary patterns of TDQC$^3$ at $\epsilon=-0.01$ and $\alpha=0.5$,
		(d) SC phase diagram. The three length scales are $\cos(\pi/14)$,
		$\cos(3\pi/14)$, and $\cos(5\pi/14)$.
		}
	\label{fig:14}
\end{figure}

\begin{figure}[htbp]
	\centering
	\includegraphics[scale=0.16]{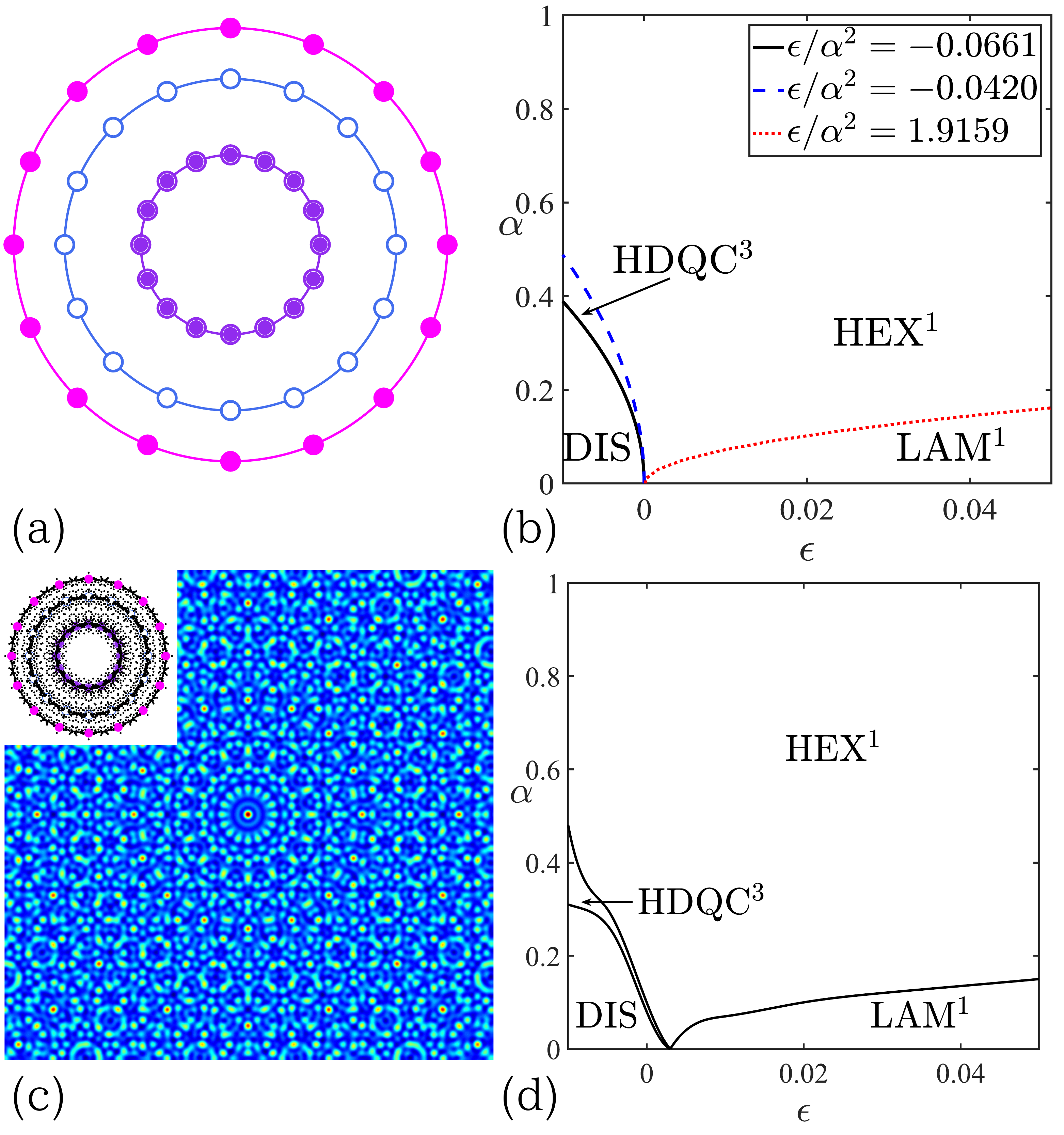}
	\caption{
		(a) Optimal primary RLVs of HDQC$^3$, (b) HC phase diagram,
		(c) stationary patterns at $\epsilon=-0.01$ and $\alpha=0.4$,
		(d) SC phase diagram. The three length scales are $\cos(\pi/8)$,
		$\cos(\pi/4)$, and $\cos(3\pi/8)$.
		}
	\label{fig:16}
\end{figure}

To the best of our knowledge, TDQC and HDQC have not yet been observed in nature
and laboratories. Utilizing ISM, we design minimal Landau theories to stabilize
these structures, which may be helpful for experimental research. Numerical
results demonstrate that the minimal Landau theories both have three length scales. 

For the three-length-scale TDQC, we present its optimal configuration of
primary RLVs in \Cref{fig:14}\,(a). The corresponding HC free energy is
\begin{equation}
	\begin{aligned}
		& \mcF_{\text{TDQC}^3} (\{\hphi\}_{j=1}^{4},\epsilon^{*})
		= -7\epsilon^{*} \sum_{j=1}^{3}\hphi_{j}^{2}
		- 56 \hphi_1 \hphi_2 \hphi_3
		\\ & ~~~~
		- 28 (\hphi_1^2 \hphi_3 + \hphi_1 \hphi_2^2
			+ \hphi_2 \hphi_3^2)
		+ \frac{273}{2} \sum_{j=1}^{3}\hphi_{j}^{4} 
		\\ & ~~~~
		+ 252 \hphi_1 \hphi_2 \hphi_3 \sum_{j=1}^{3} \hphi_j
		+ 84 (\hphi_1^3 \hphi_2 + \hphi_1 \hphi_3^3
			+ \hphi_2^3 \hphi_3)
		\\ & ~~~~
		+ 378 (\hphi_1^2 \hphi_2^2 + \hphi_1^2 \hphi_3^2
			+ \hphi_2^2 \hphi_3^2).
	\end{aligned}
	\label{eq:energy.14.3}
\end{equation}
\Cref{fig:14}\,(b) plots the HC phase diagram, where
TDQC$^3$ is stable when $-0.0901 \le \epsilon^{*} \le 0.0060$. Under SC,
\Cref{fig:14}\,(c) displays the stationary patterns of TDQC$^3$ computed by the
projection method. More details on the projection matrix and high-dimensional
coordinates can be found in SM \Cref{sm.eq:pm.14} and
\Cref{sm.fig:high.14}, respectively. \Cref{fig:14}\,(d) shows the SC phase
diagram, revealing a stable region for TDQC$^3$.

For the three-length-scale HDQC, \Cref{fig:16}\,(a) presents the optimal
configuration of primary RLVs, which results in the HC free energy
\begin{equation}
	\begin{aligned}
		& \mcF_{\text{HDQC}^3} (\{\hphi\}_{j=1}^{4},\epsilon^{*})
		= -8\epsilon^{*} \sum_{j=1}^{3}\hphi_{j}^{2}
		- 32 \hphi_2 (\hphi_1 + \hphi_3)^2
		\\
		&+ 180 \sum_{j=1}^{3}\hphi_{j}^{4}
		+ 96 \hphi_1 \hphi_3 (\hphi_1^2 + \hphi_3^2)
		+ 384 \hphi_1 \hphi_2^2 \hphi_3
		\\
		&+ 480 (\hphi_1^2 \hphi_2^2 + \hphi_1^2 \hphi_3^2
			+ \hphi_2^2 \hphi_3^2).
	\end{aligned}
	\label{eq:energy.16.3}
\end{equation}
We plot the corresponding HC phase diagram in \Cref{fig:16}\,(b), indicating
that HDQC$^3$ is thermodynamic stable for $-0.0661 \le \epsilon^{*} \le -0.0420$.
Under SC, we use the projection method and obtain the stationary pattern of
HDQC$^3$ as shown in \Cref{fig:16}\,(c). We give the projection matrix in SM
\Cref{sm.eq:pm.16} and high-dimensional coordinates of HDQC$^3$ in SM
\Cref{sm.fig:high.16}. \Cref{fig:16}\,(d) plots the SC phase diagram, confirming
the stability of HDQC$^3$ under SC.

\begin{figure}[htbp]
	\centering
	\includegraphics[scale=0.12]{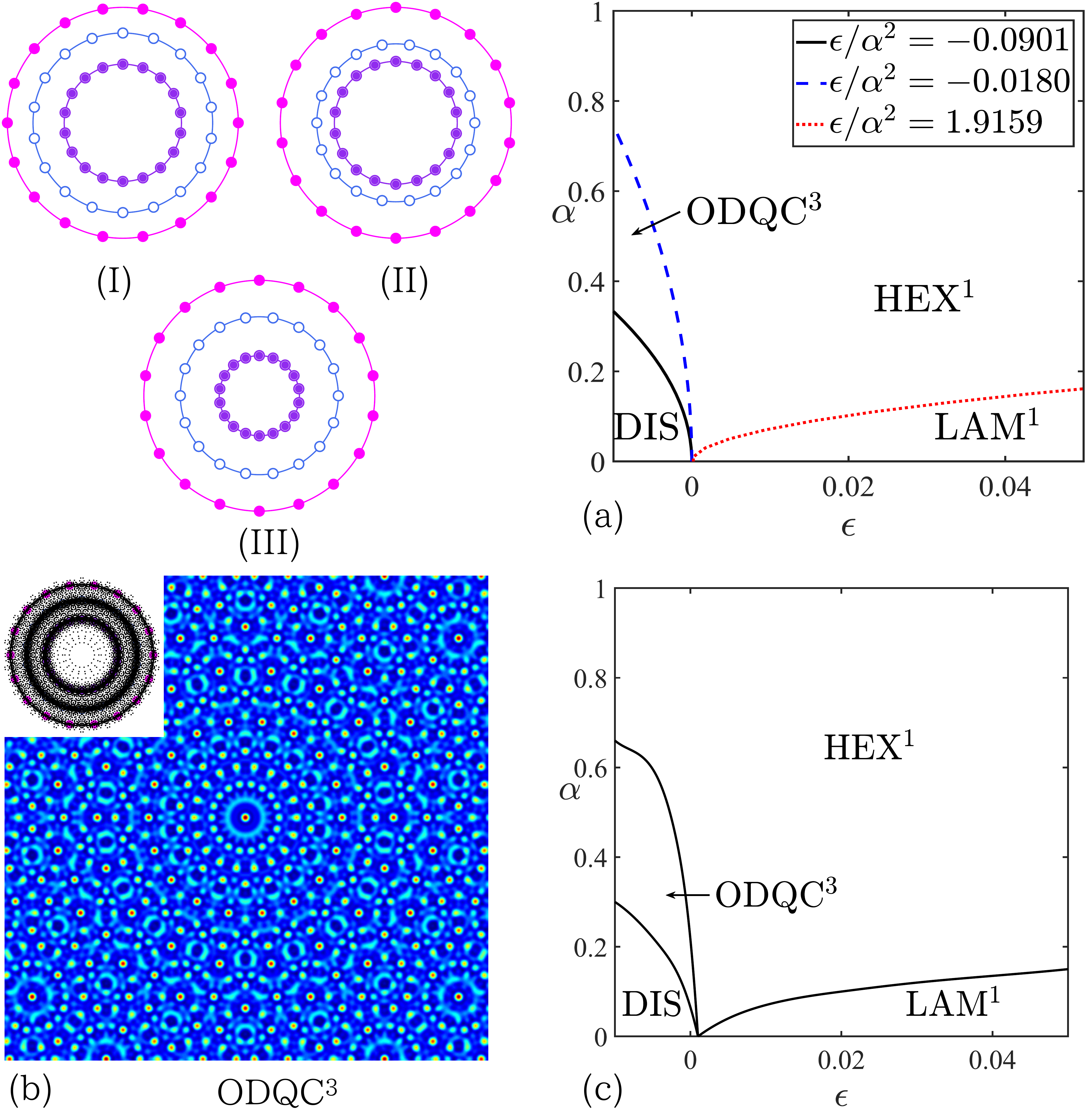}
	\caption{
		Optimal primary RLVs of ODQC$^3$ whose radii of the circles from
		outside to inside are
		(\Rmnum{1}) $\cos(\pi/18)$, $\cos(2\pi/9)$, $1/2$;
		(\Rmnum{2}) $\cos(\pi/9)$, $\cos(5\pi/18)$, $1/2$;
		(\Rmnum{3}) $1/2$, $\cos(7\pi/18)$, $\cos(4\pi/9)$.
		(a) HC phase diagram of the three-length-scale model in $\epsilon$-$\alpha$ plane.
		(b) Stationary patterns of ODQC$^3$ at $\epsilon=-0.01$ and $\alpha=0.5$.
		Non-primary RLVs with intensities greater than $10^{-6}$ are indicated by small dots.
		(c) Phase diagram of the SC minimal Landau model with three length
		scales $\cos(\pi/18)$, $\cos(2\pi/9)$, and $1/2$.
		}
	\label{fig:18}
\end{figure}

ODQC has been discovered in soft colloidal systems, as evidenced by their
diffraction patterns that exhibit multiple length scales\,\cite{Fischer2011}.
Stable ODQCs in theoretical study have been obtained by a Landau model with four
length scales\,\cite{Savitz2018}. In our study, we design three minimal Landau
models by ISM, each with three different length scales:
(\Rmnum{1}) $\cos(\pi/18)$, $\cos(2\pi/9)$, $1/2$; (\Rmnum{2}) $\cos(\pi/9)$,
$\cos(5\pi/18)$, $1/2$; (\Rmnum{3}) $1/2$, $\cos(7\pi/18)$, $\cos(4\pi/9)$. 
\Cref{fig:18}\,(\Rmnum{1})-(\Rmnum{3}) show the optimal configurations of
primary RLVs of ODQCs, corresponding to the minimal models (\Rmnum{1})-(\Rmnum{3})
respectively. These configurations have the same number of three- and four-RLV
interactions, resulting in the same HC free energy
\begin{equation}
	\begin{aligned}
		& \mcF_{\text{ODQC}^3} (\{\hphi\}_{j=1}^{3},\epsilon^{*})
		= -9\epsilon^{*} \sum_{j=1}^{3}\hphi_{j}^{2}
		- 12 \sum_{j=1}^{3}\hphi_{j}^{3} 
		\\ &
		- 36 \hphi_{1} \sum_{j=2}^{3}\hphi_{j}^{2}
		- 36 \hphi_{2}\hphi_{3} (2\hphi_{1}+\hphi_{3})
		+ \frac{459}{2}\sum_{j=1}^{3}\hphi_{j}^{4} 
		\\ &
		+ 108 \hphi_2^3 (2\hphi_1 + \hphi_3)
		+ 216 \hphi_3^3 (\hphi_1 + \hphi_2)
		+ 702 \hphi_2^2 \hphi_3^2
		\\ &
		+ 702 \hphi_1^2 (\hphi_2^2 + \hphi_3^2)
		+ 756 \hphi_1^2 \hphi_2 \hphi_3
		+ 432 \hphi_1 \hphi_2 \hphi_3 (\hphi_2 + \hphi_3).
	\end{aligned}
	\label{eq:energy.18.3}
\end{equation}
\Cref{fig:18}\,(a) plots the HC phase diagram, where ODQC$^3$ is expected to be
thermodynamic stable in $-0.0901 \le \epsilon^{*} \le -0.0180$. Moreover, we
take the minimal model (\Rmnum{1}) as an example to examine the thermodynamic
stability of ODQCs under SC. By use of the projection method (see SM
\Cref{sm.eq:pm.18} for projection matrix and \Cref{sm.fig:high.18} for
high-dimensional coordinates), we obtain the stationary ODQC$^3$ phase. Its
diffraction pattern and real-space morphology are shown in \Cref{fig:18}\,(b).
\Cref{fig:18}\,(c) presents the SC phase diagram, revealing a stable region for
ODQC$^3$. The phase boundaries exhibit a slight shift compared to the HC phase
diagram, which may be attributed to the non-negligible influence of non-primary
RLVs and the predominant contribution of primary RLVs. Moreover, we find stable
ODQC$^3$ in the models (\Rmnum{2}) and (\Rmnum{3}) with SC.

\section{Conclusion}
\label{sec:conclu}

In this paper, we propose an efficient method (ISM) to design a minimal Landau
theory to stabilize desired QCs. ISM evaluates almost all possible
configurations of RLVs for the target QC, allowing us to identify the optimal
configuration with the lowest free energy, as the free energy functional can be
expressed as a polynomial under HC. With this optimal configuration, ISM then
constructs phase diagrams to assess the thermodynamic stability of the target
QC. Generally, configurations with more length scales contain more primary RLVs,
which can form more three-RLV interactions to lower the free energy. Thus, we
can always design a minimal Landau theory to stabilize desired QCs by gradually
increasing the number of length scales.

Using ISM, we design minimal Landau theories to stabilize $2n$-fold QCs
($n=4,\ldots,9$). Concretely, two-length-scale Landau models can stabilize
$10$- and $12$-fold QCs, which is consistent with previous results.
Moreover, we obtain stable $8$- and $18$-fold QCs in three-length-scale
Landau models, reducing the number of length scales compared to earlier
studies. Our findings also indicate that three-length-scale models can
stabilize $14$- and $16$-fold QCs. We believe that these minimal models with
relatively simple potentials could be helpful to control the synthesis of QCs.



%




\section{Supplementary Material}

\graphicspath{{swfigEnd/supp/}}

\Cref{sm.tab:hc} gives the optimal primary RLVs, the formula of HC free energies,
and HC phase diagrams for $2n$-fold quasicrystals (QCs) ($n=4,\ldots,9$).
From these HC phase diagrams, one can conclude that the HC minimal models of
$10$- and $12$-fold QCs have two length scales and the HC minimal models of the
rest QCs require three length scales.
\begin{longtable*}{|m{0.12\linewidth}<{\centering}|
	m{0.28\linewidth}<{\centering}|
	m{0.29\linewidth}<{\centering}|
	m{0.30\linewidth}<{\centering}|}
	\caption{
		\label{sm.tab:hc}
		Optimal primary RLVs, HC free energies, and HC phase diagrams of
		$2n$-fold QCs ($n=4,\ldots,9$).
		Length scales satisfy $q_{j}=\cos(w_{j}\pi/(2n))$, 
		and offset angles are $\theta_{j}=s_{j}\pi/(2n)$,
		where $j=1,2,\ldots,m$, $m$ is the number of length scales.
		$s_{1}$ defaults to $0$.
		The second to fourth columns correspond to $m=2,3,4$.
		The parameters of optimal primary RLVs $w_{j}$ and $s_{j}$ for different
		configurations are separated by semicolons.
		The primary RLVs located on four length scales are marked by 
		magenta \textcolor{magenta}{$\bullet$}, 
		royal blue \textcolor{RoyalBlue2}{$\circ$}, 
		purple \textcolor{Purple2}{$\odot$},
		and dark turquoise \textcolor{DarkTurquoise}{$\otimes$},
		respectively.
		The specific values of length scales and offset angles are listed 
		at the bottom of the RLV patterns.
		We also write the HC free energies of desired QCs based on the
		optimal primary RLVs.
		In phase diagrams, we use superscripts to denote the number of 
		valid length scales.
	} \\*
	\endfirsthead
	\multicolumn{4}{c}{\Cref{sm.tab:hc} (continued): optimal
	primary RLVs, HC free energies and HC phase diagrams,}%
	\endhead
	\endfoot
	\endlastfoot
	\hline
	Desired QCs & \multicolumn{3}{c|}{\textbf{Octagonal (O) QCs} ($8$-fold)} \\
	\hline
	$\{w_j\}_{j=1}^{m}$ & 0, 2 & 1, 2, 3 & 0, 1, 2, 3 \\
	\hline
	$\{s_j\}_{j=1}^{m}$ & 0, 0 & 0, 1, 0 & 0, 1, 0, 1 \\
	\hline
	Optimal primary RLVs \rule[45pt]{100pt}{0pt} &
	\includegraphics[scale=0.10]{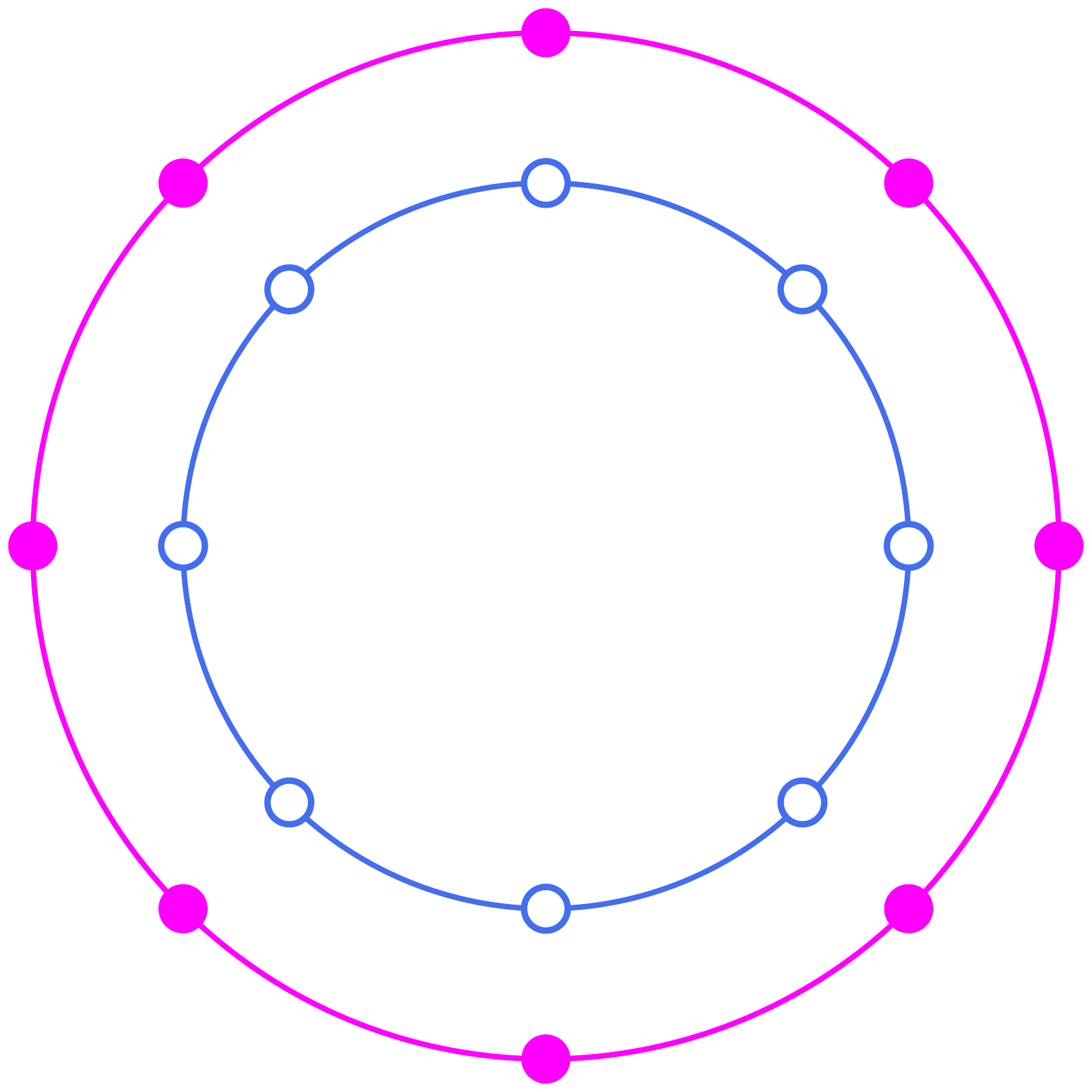} 
	\centerline{$q: 1, \cos\frac{\pi}{4}$}
	\centerline{$\theta: 0, 0$}
	& 
	\includegraphics[scale=0.10]{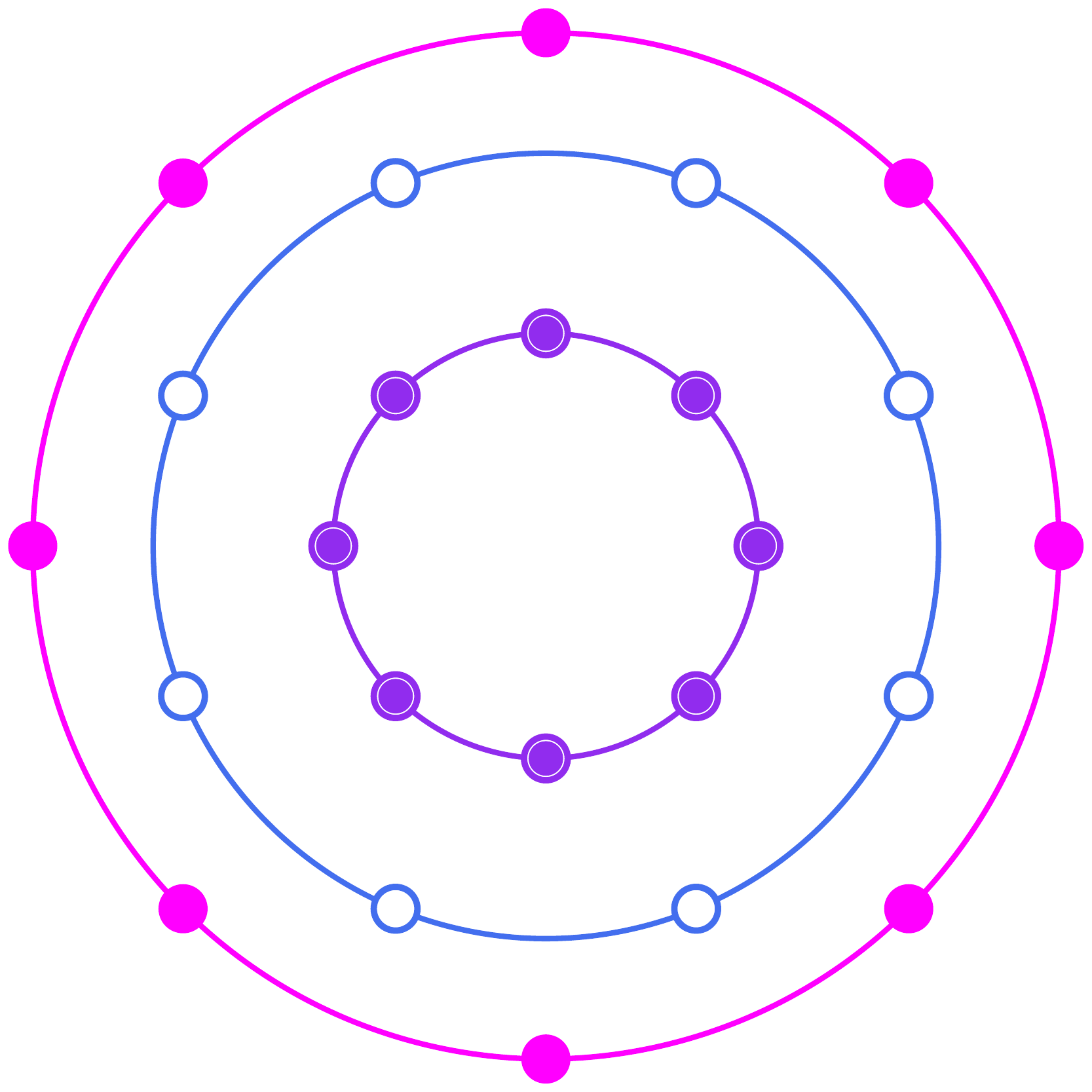} 
	\centerline{$q: \cos\frac{\pi}{8}, \cos\frac{\pi}{4}, 
	\cos\frac{3\pi}{8}$}
	\centerline{$\theta: 0, \pi/8, 0$}
	& 
	\includegraphics[scale=0.10]{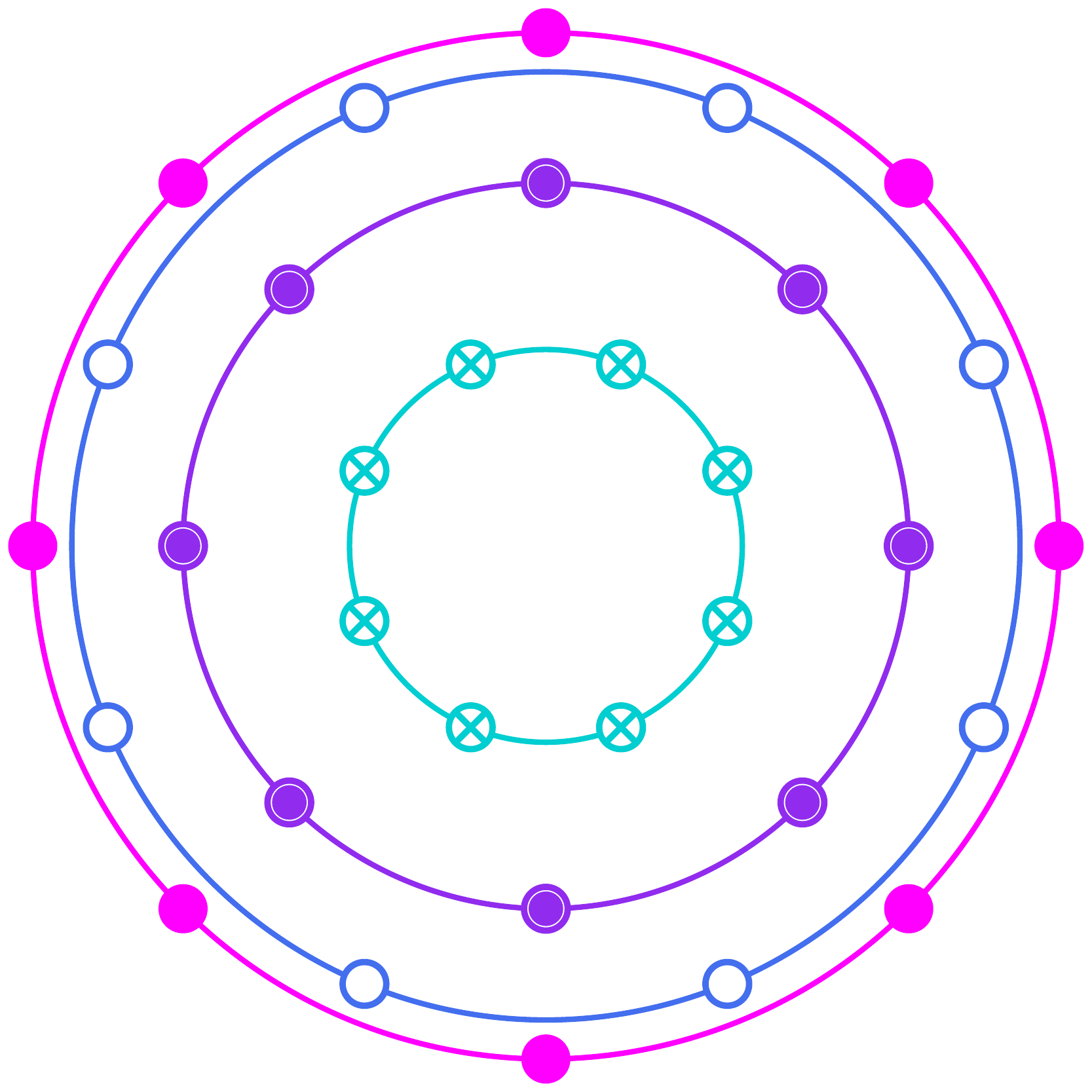}
	\centerline{$q: 1, \cos\frac{\pi}{8}, \cos\frac{\pi}{4}, 
	\cos\frac{3\pi}{8}$}
	\centerline{$\theta: 0, \pi/8, 0, \pi/8$}
	\\
	\hline
	HC free energies \rule[70pt]{100pt}{0pt} &
	{
	\begin{equation*}
		\begin{aligned}
			&-4\epsilon^{*} \sum_{j=1}^{2}\hphi_{j}^{2}
			- 16\hphi_{1}\hphi_{2}^{2}
			\\
			&+ 42 \sum_{j=1}^{2}\hphi_{j}^{4}
			+ 120 \hphi_{1}^{2} \hphi_{2}^{2}
		\end{aligned}
		\label{sm.eq:energy.8.2}
	\end{equation*}
	}
	\rule[45pt]{100pt}{0pt}
	&
	{
	\begin{equation*}
		\begin{aligned}
			&-4\epsilon^{*} \sum_{j=1}^{3}\hphi_{j}^{2}
			- 16\hphi_{2} (\hphi_{1}+\hphi_{3})^{2}
			\\
			&+ 42 \sum_{j=1}^{3}\hphi_{j}^{4}
			+ 192 \hphi_{1}\hphi_{2}^{2}\hphi_{3}
			\\
			&+ 144 (\hphi_{1}^{2}\hphi_{2}^{2} + \hphi_{1}^{2}\hphi_{3}^{2}
			+ \hphi_{2}^{2}\hphi_{3}^{2})
			\\
			&+ 48 \hphi_{1}\hphi_{3} (\hphi_{1}^{2}+\hphi_{3}^{2})
		\end{aligned}
		\label{sm.eq:energy.8.3}
	\end{equation*}
	}
	\rule[25pt]{100pt}{0pt}
	&
	{
	\begin{equation*}
		\begin{aligned}
			&-4\epsilon^{*} \sum_{j=1}^{4}\hphi_{j}^{2}
			- 16 \hphi_1 (\hphi_3^2 + 2\hphi_2\hphi_4)
			\\
			&- 16 \hphi_3 (\hphi_2 + \hphi_4)^2
			+ 42 \sum_{j=1}^{4} \hphi_j^4
			\\
			&+ 48 \hphi_2 \hphi_4 (\hphi_2^2 + \hphi_4^2)
			+ 120 \hphi_1^2 \sum_{j=2}^{4} \hphi_j^2
			\\
			&+ 144 ( \hphi_3 (\hphi_1 + \hphi_3) (\hphi_2^2 + \hphi_4^2)
			+ \hphi_2^2 \hphi_4^2 )
			\\
			&+ 192 \hphi_2 \hphi_3^2 \hphi_4
			+ 288 \hphi_1 \hphi_2 \hphi_3 \hphi_4
		\end{aligned}
		\label{sm.eq:energy.8.4}
	\end{equation*}
	}
	\\
	\hline
	HC phase diagrams \rule[45pt]{100pt}{0pt} & 
	\includegraphics[scale=0.16]{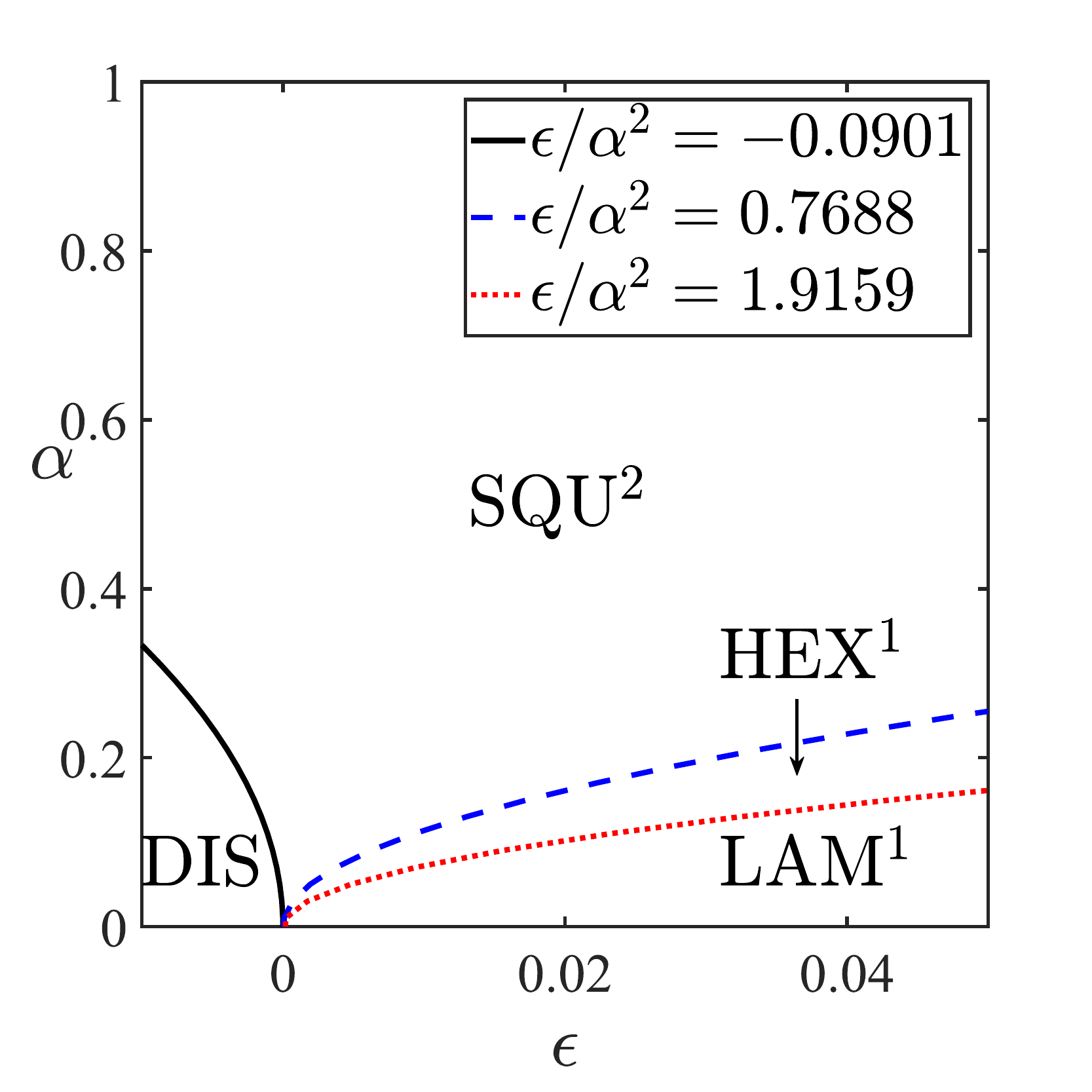} & 
	\includegraphics[scale=0.16]{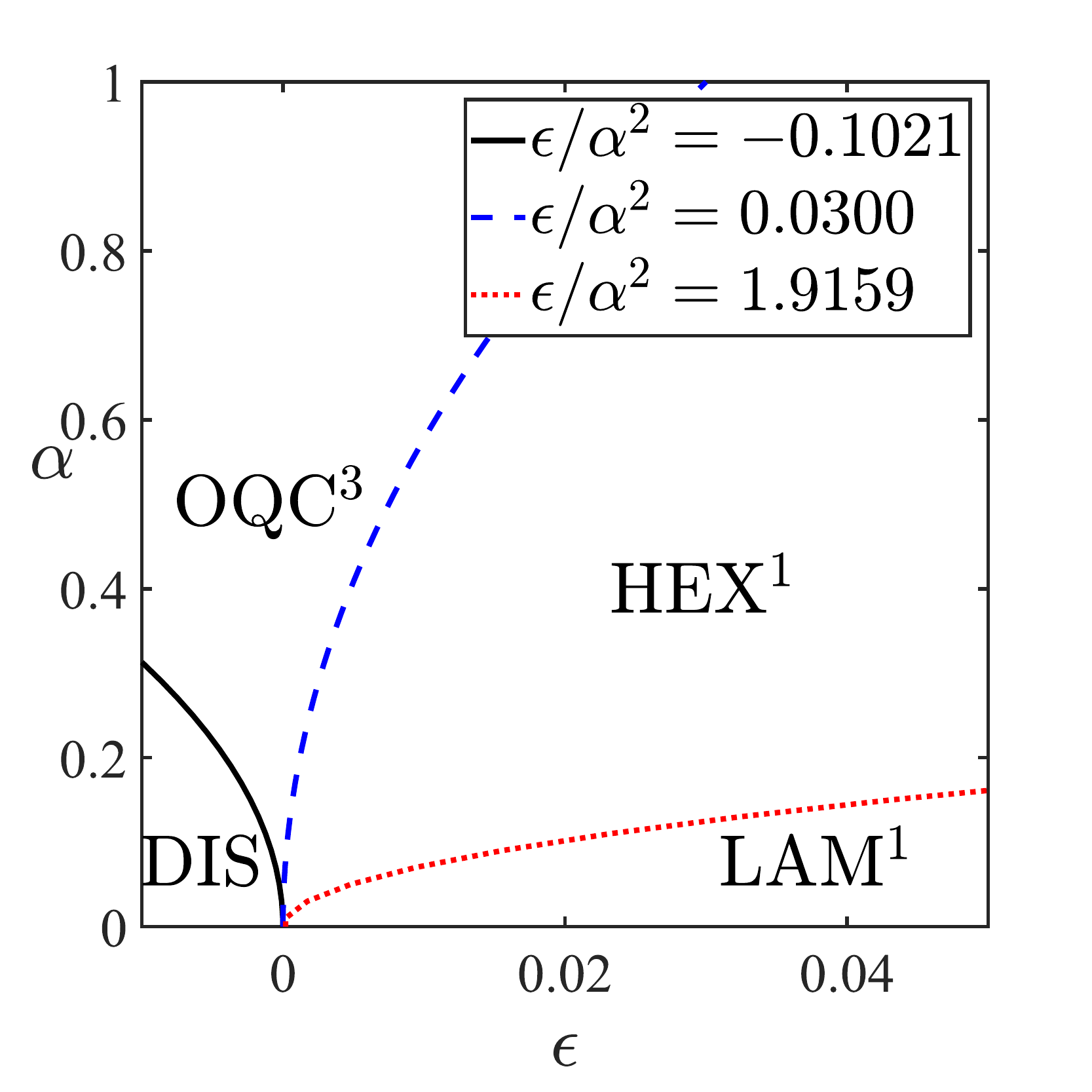} & 
	\includegraphics[scale=0.16]{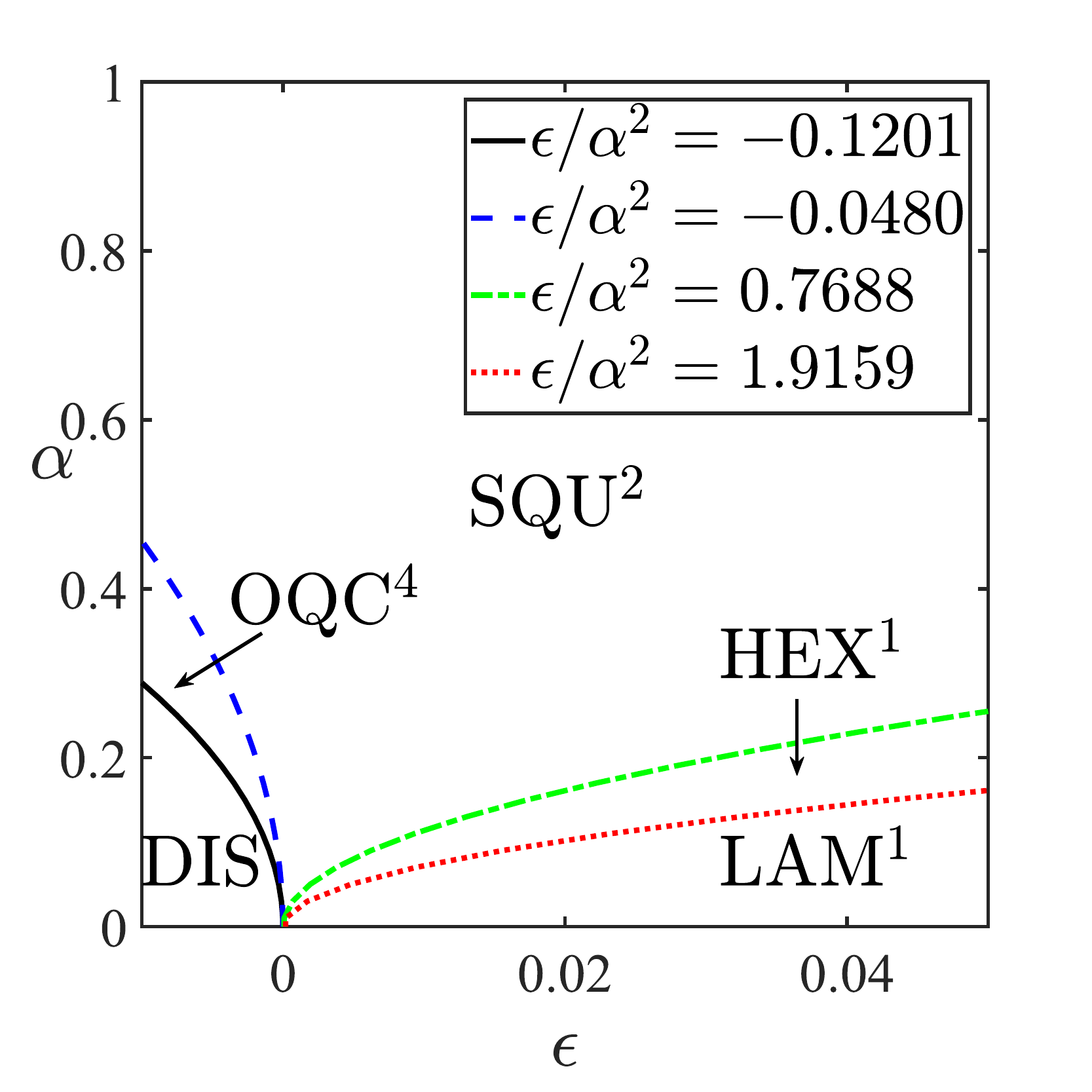} \\
	\hline
	\multicolumn{4}{c}{} \\
	\multicolumn{4}{c}{} \\
	\hline
	Desired QCs & \multicolumn{3}{c|}{\textbf{Decagonal (D) QCs} ($10$-fold)} \\
	\hline
	$\{w_j\}_{j=1}^{m}$ & 1, 3 & 1, 2, 4; 2, 3, 4 & 0, 1, 2, 4; 0, 2, 3, 4 \\
	\hline
	$\{s_j\}_{j=1}^{m}$ & 0, 0 & 0, 1, 1; 0, 1, 0 & 0, 1, 0, 0; 0, 0, 1, 0 \\
	\hline
	Optimal primary RLVs \rule[85pt]{100pt}{0pt} &
	\includegraphics[scale=0.10]{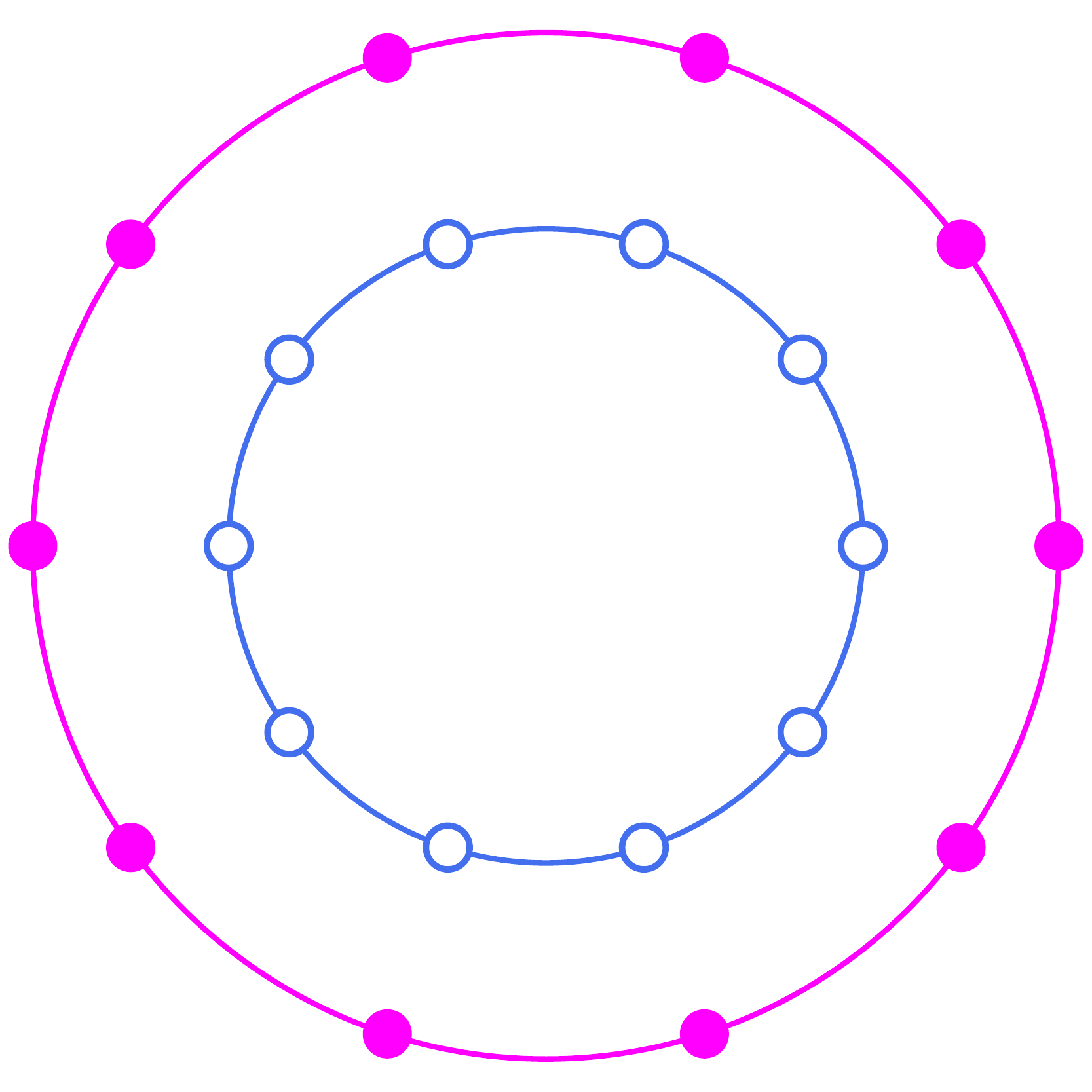} 
	\centerline{$q: \cos\frac{\pi}{10}, \cos\frac{3\pi}{10}$}
	\centerline{$\theta: 0, 0$}
	\rule[45pt]{100pt}{0pt}
	& 
	\includegraphics[scale=0.10]{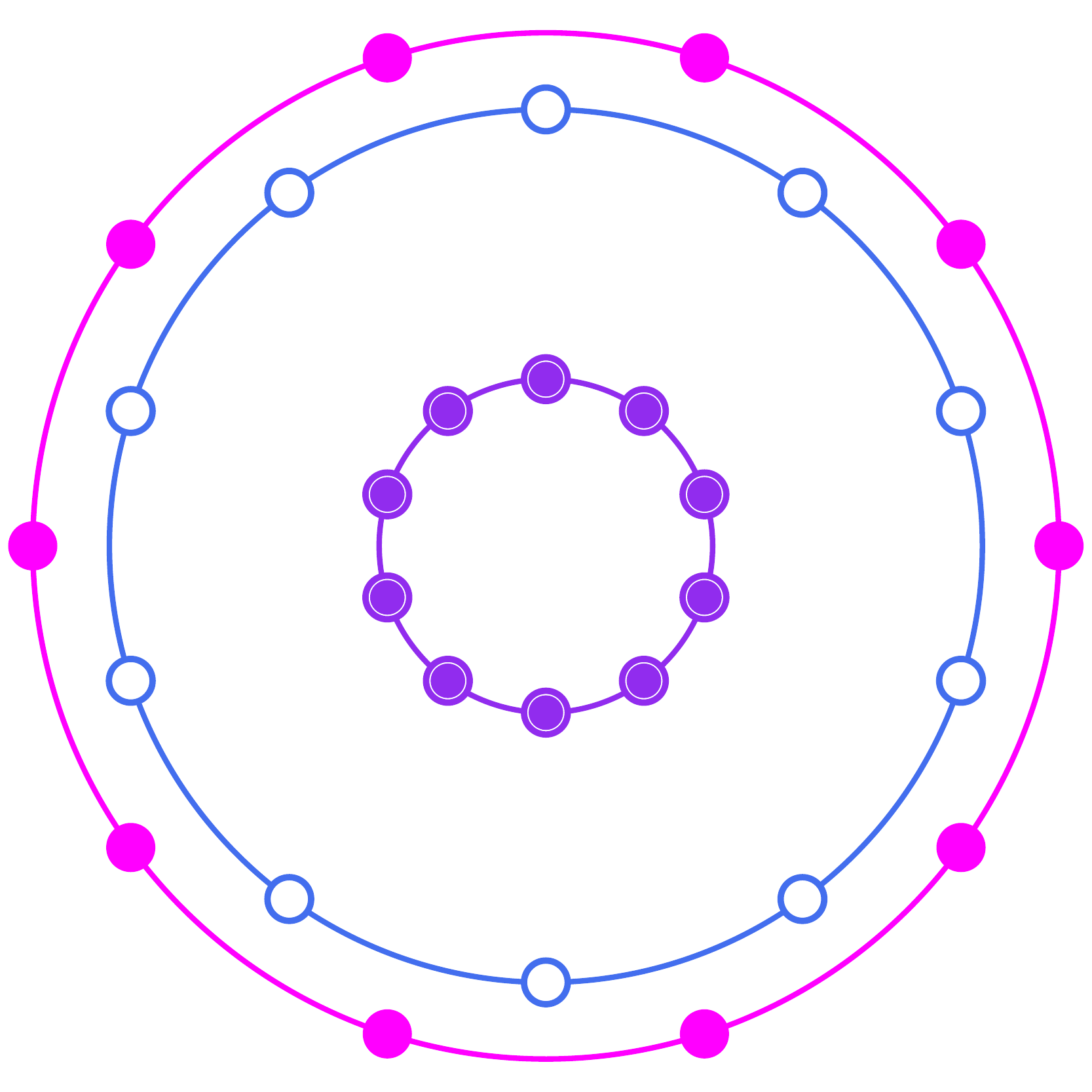} 
	\centerline{$q: \cos\frac{\pi}{10}, \cos\frac{\pi}{5}, 
	\cos\frac{2\pi}{5}$}
	\centerline{$\theta: 0, \pi/10, \pi/10$}
	\includegraphics[scale=0.10]{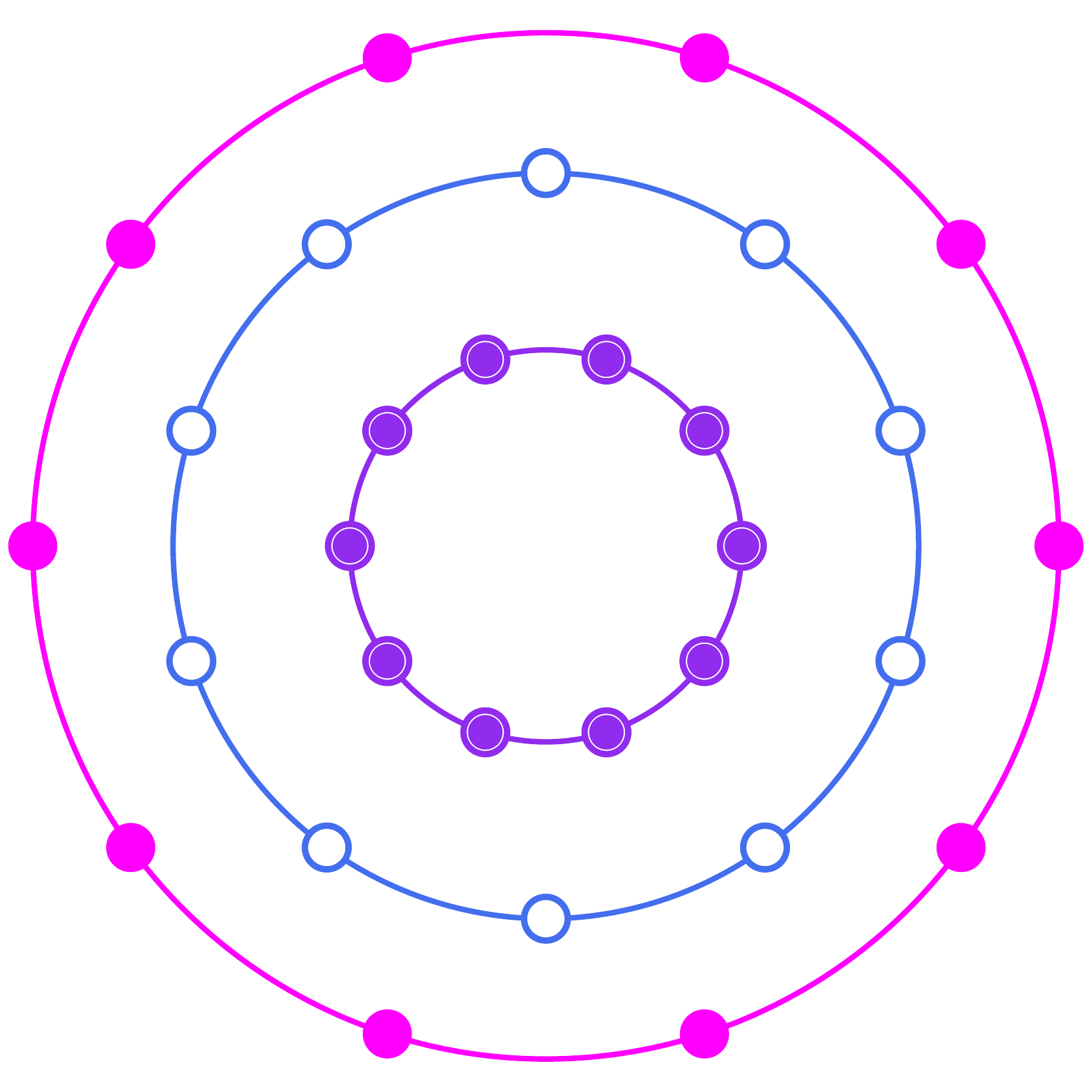} 
	\centerline{$q: \cos\frac{\pi}{5}, \cos\frac{3\pi}{10}, 
	\cos\frac{2\pi}{5}$}
	\centerline{$\theta: 0, \pi/10, 0$}
	& 
	\includegraphics[scale=0.10]{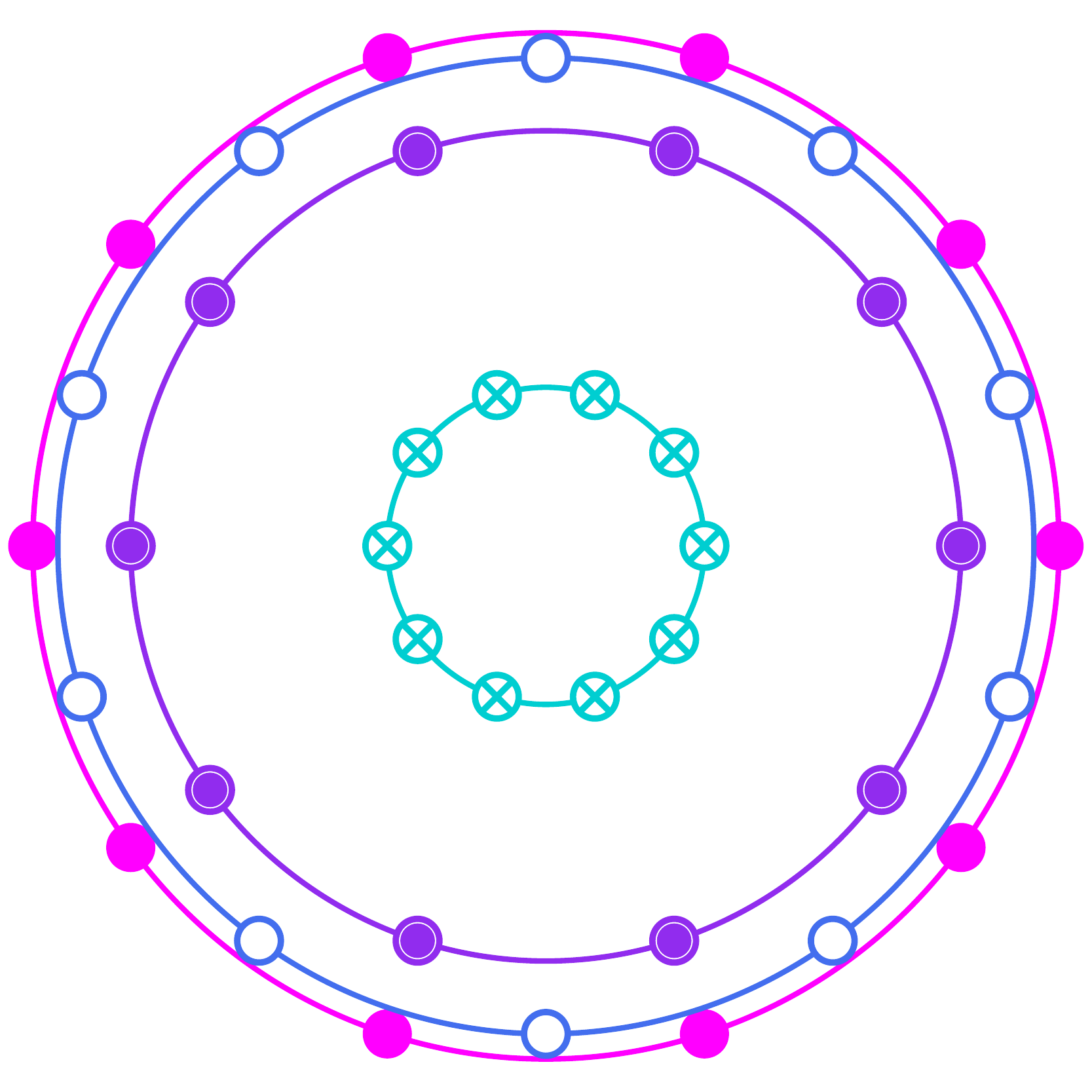} 
	\centerline{$q: 1, \cos\frac{\pi}{10}, \cos\frac{\pi}{5}, 
	\cos\frac{2\pi}{5}$}
	\centerline{$\theta: 0, \pi/10, 0, 0$}
	\includegraphics[scale=0.10]{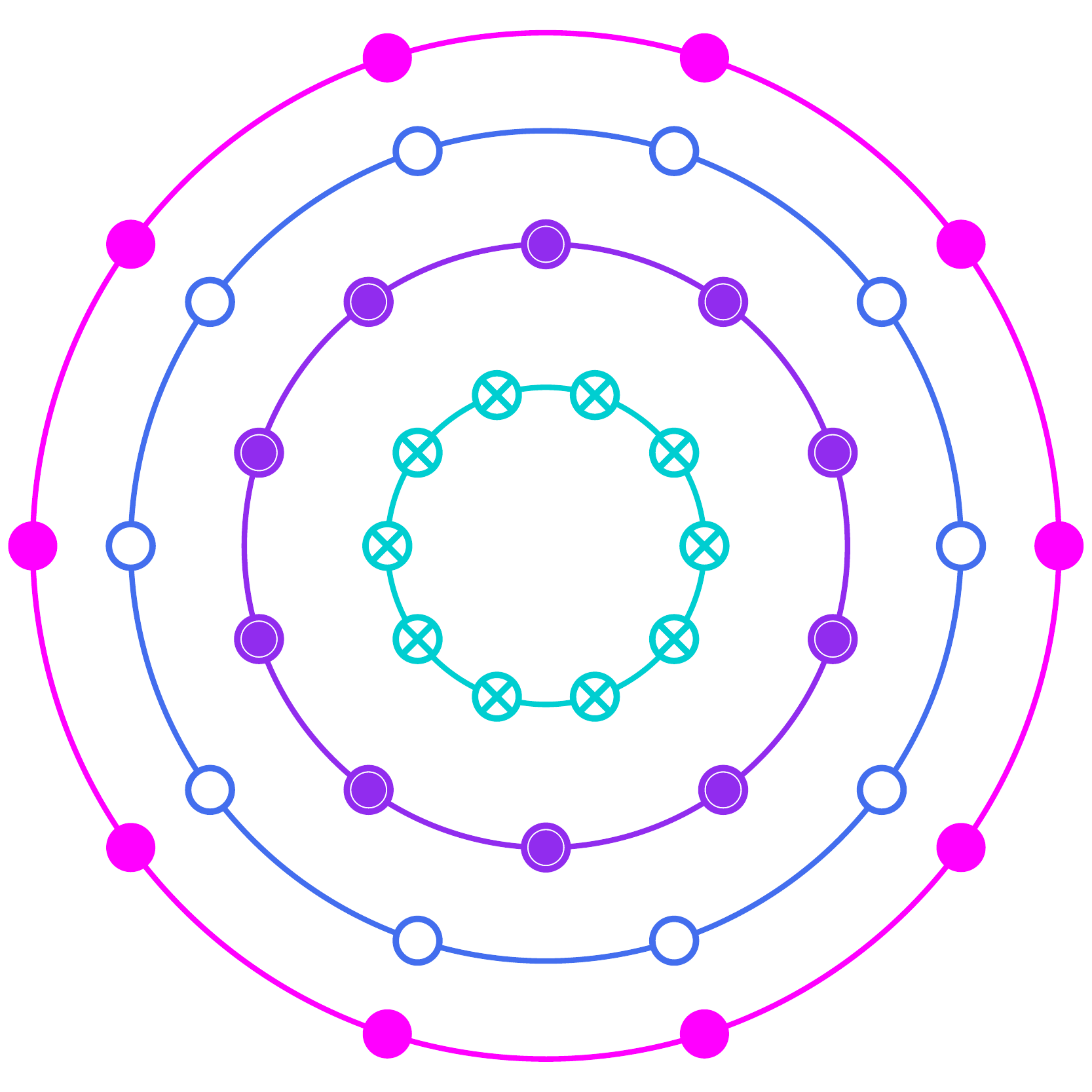} 
	\centerline{$q: 1, \cos\frac{\pi}{5}, \cos\frac{3\pi}{10}, 
	\cos\frac{2\pi}{5}$}
	\centerline{$\theta: 0, 0, \pi/10, 0$}
	\\
	\hline
	HC free energies \rule[80pt]{100pt}{0pt} &
	{
	\begin{equation*}
		\begin{aligned}
			&-5\epsilon^{*} \sum_{j=1}^{2}\hphi_{j}^{2} 
			- 20\hphi_{1}\hphi_{2}(\hphi_{1}+\hphi_{2})
			\\
			&+ \frac{135}{2}\sum_{j=1}^{2}\hphi_{j}^{4} 
			+ 60\hphi_{1}\hphi_{2}\sum_{j=1}^{2}\hphi_{j}^{2}
			\\
			&+ 210\hphi_{1}^{2}\hphi_{2}^{2}
		\end{aligned}
		\label{sm.eq:energy.10.2}
	\end{equation*}
	}
	\rule[45pt]{100pt}{0pt}
	&
	{
	\begin{equation*}
		\begin{aligned}
			&-5\epsilon^{*} \sum_{j=1}^{3}\hphi_{j}^{2}
			- 20 \hphi_1 \hphi_2 (\hphi_2 + 2\hphi_3)
			\\
			&+ \frac{135}{2}\sum_{j=1}^{3}\hphi_{j}^{4} 
			+ 60 \hphi_2 \hphi_3 \sum_{j=2}^{3} \hphi_j^2
			\\
			&+ 210 ( \hphi_1^2 \hphi_2^2 + \hphi_1^2 \hphi_3^2
				+ \hphi_2^2 \hphi_3^2 )
			\\
			&+ 180 \hphi_1^2 \hphi_2 \hphi_3
		\end{aligned}
		\label{sm.eq:energy.10.3}
	\end{equation*}
	}
	\rule[35pt]{100pt}{0pt}
	&
	{
	\begin{equation*}
		\begin{aligned}
			&-5\epsilon^{*} \sum_{j=1}^{4}\hphi_{j}^{2}
			-40 \hphi_2 \hphi_4 (\hphi_1+\hphi_3)
			\\
			&-20 \hphi_2 \hphi_3^2
			+ \frac{135}{2} \sum_{j=1}^{4}\hphi_{j}^{4} 
			\\
			&+ 60 \hphi_3 \hphi_4 \sum_{j=3}^{4} \hphi_j^2
			+ 180 \hphi_1^2 \sum_{j=2}^{4} \hphi_j^2
			\\
			&+ 180 \hphi_1 \hphi_3 (\hphi_2^2 + \hphi_4 (\hphi_3+\hphi_4))
			\\
			&+ 210 (\hphi_2^2 \hphi_3^2 + \hphi_2^2 \hphi_4^2
				+ \hphi_3^2 \hphi_4^2)
			\\
			&+ 180 \hphi_2^2 \hphi_3 \hphi_4
		\end{aligned}
		\label{sm.eq:energy.10.4}
	\end{equation*}
	}
	\\
	\hline
	HC phase diagrams \rule[45pt]{100pt}{0pt} & 
	\includegraphics[scale=0.16]{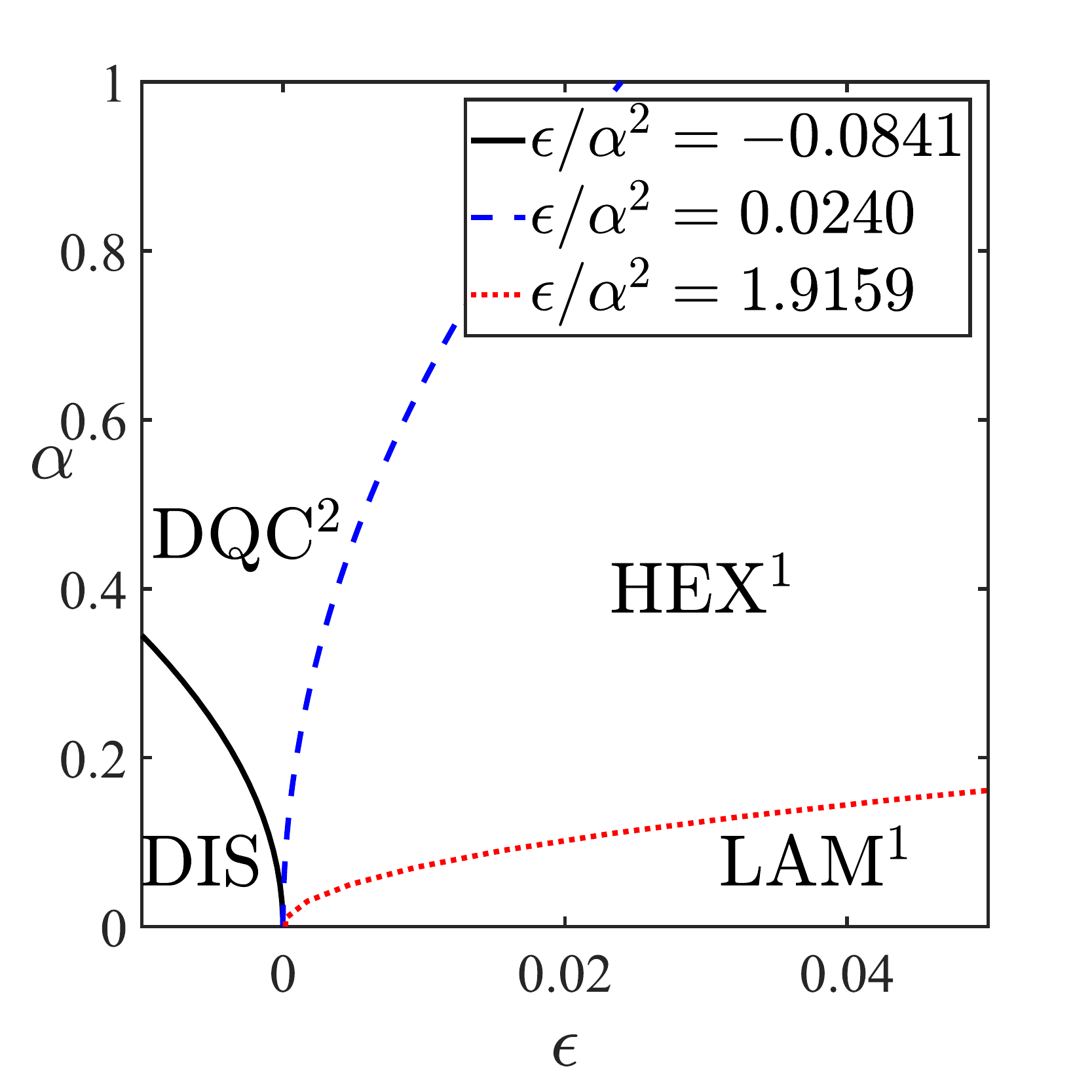} & 
	\includegraphics[scale=0.16]{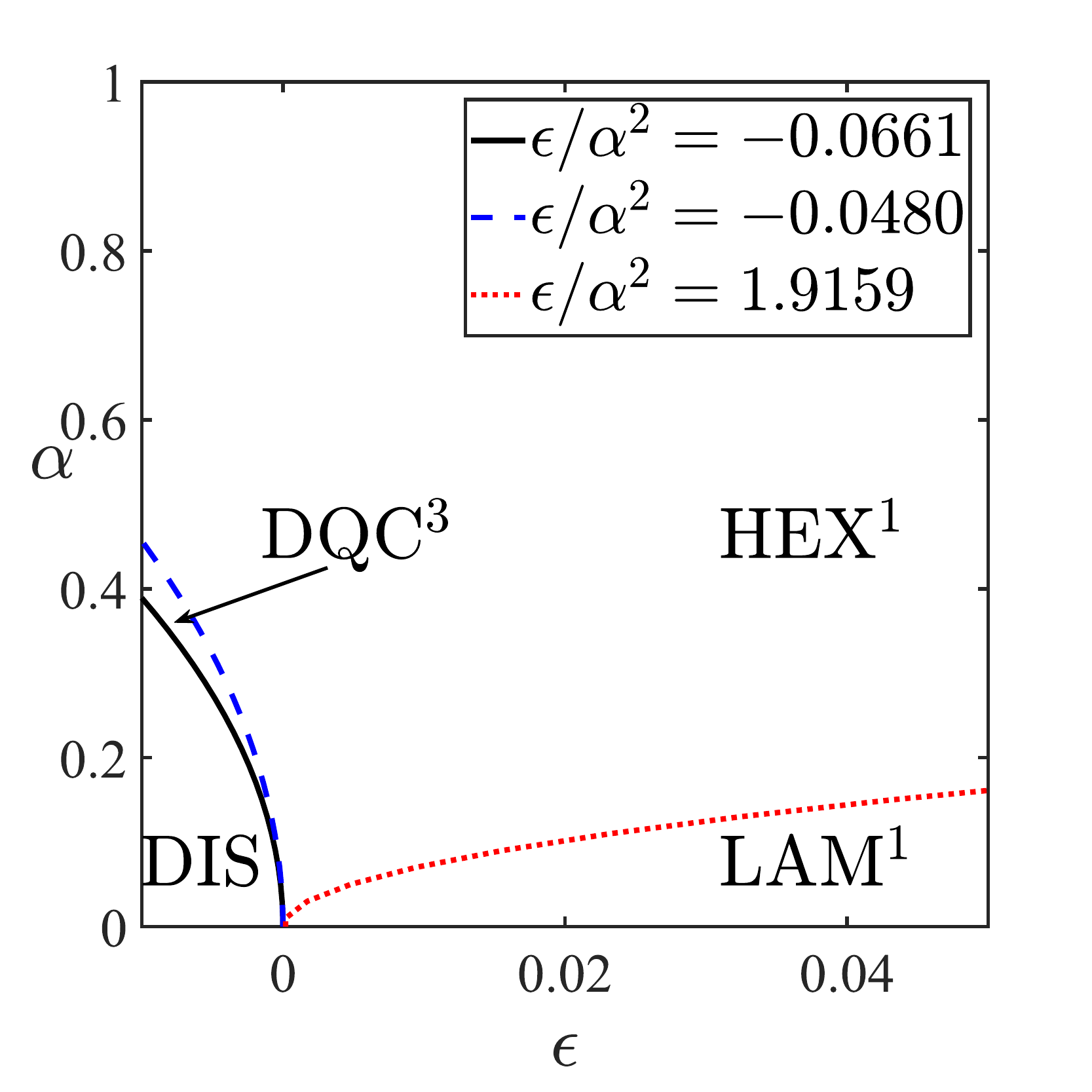} & 
	\includegraphics[scale=0.16]{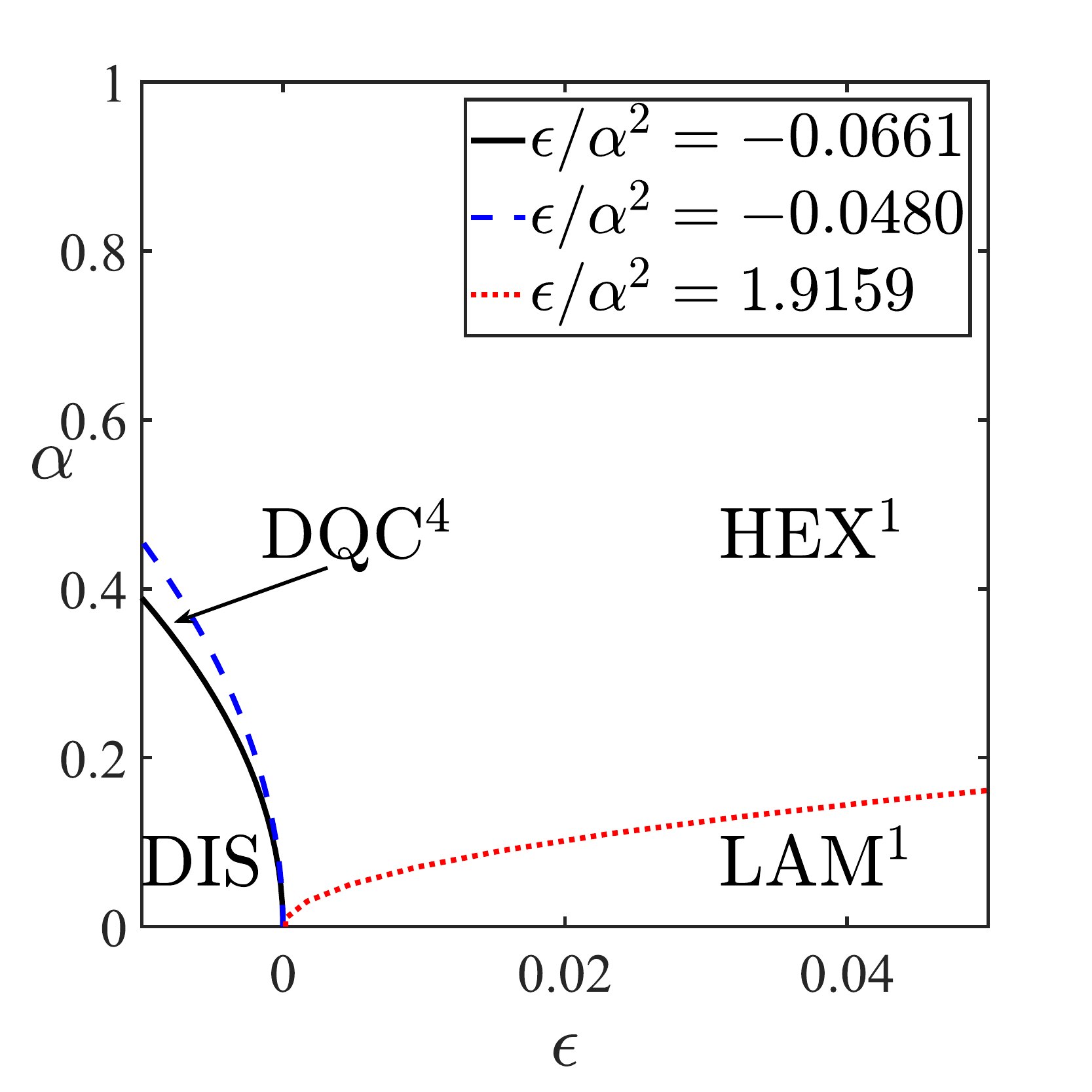} \\
	\hline
	\multicolumn{4}{c}{} \\
	\multicolumn{4}{c}{} \\
	\hline
	Desired QCs & \multicolumn{3}{c|}{\textbf{Dodecagonal (DD) QCs} ($12$-fold)} \\
	\hline
	$\{w_j\}_{j=1}^{m}$ & 1, 4; 4, 5 & 1, 3, 4 & 1, 2, 3, 4 \\
	\hline
	$\{s_j\}_{j=1}^{m}$ & 0, 1; 0, 1 & 0, 0, 1 & 0, 1, 0, 1 \\
	\hline
	Optimal primary RLVs \rule[100pt]{100pt}{0pt} &
	\includegraphics[scale=0.10]{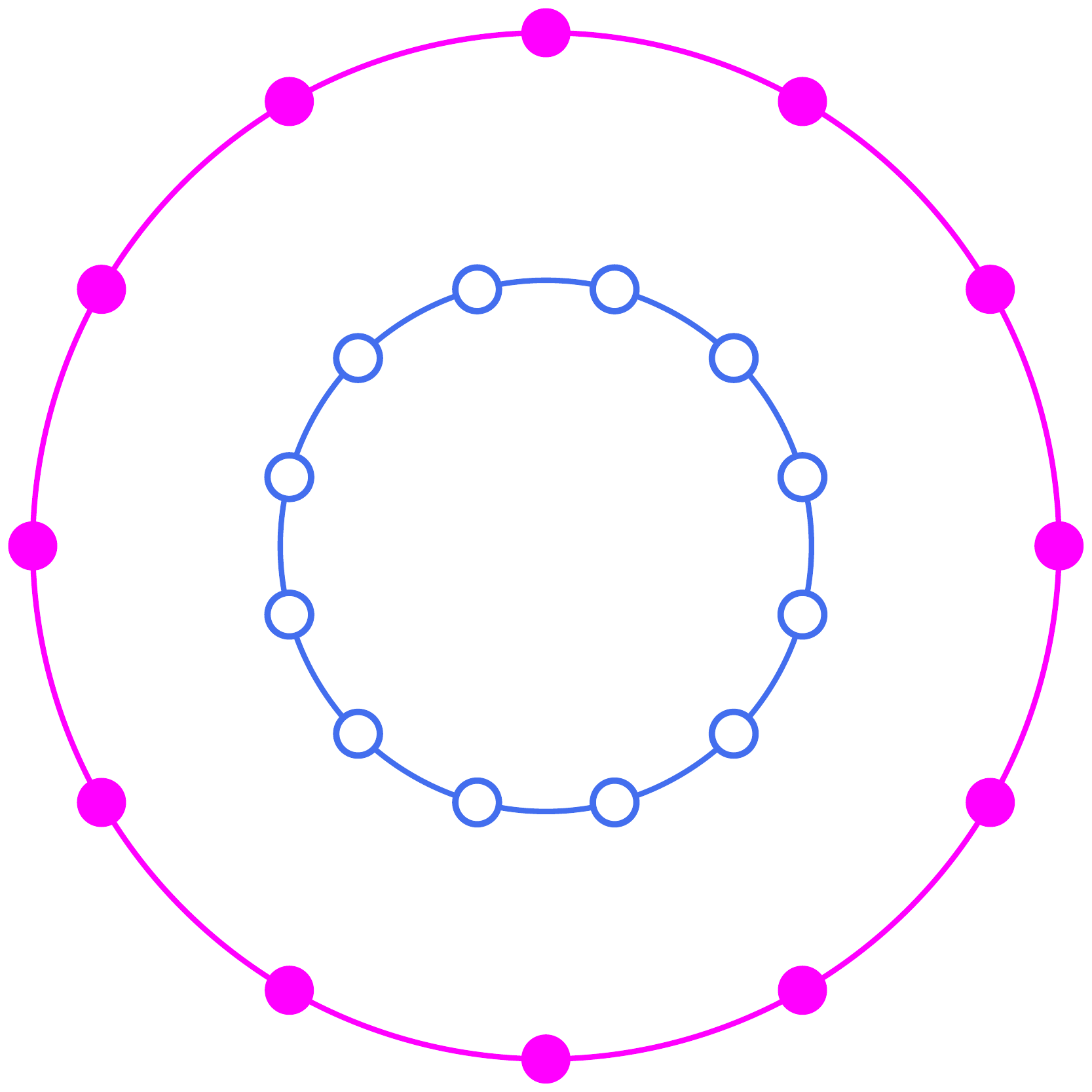}
	\centerline{$q: \cos\frac{\pi}{12}, \frac{1}{2}$}
	\centerline{$\theta: 0, \pi/12$}
	\includegraphics[scale=0.10]{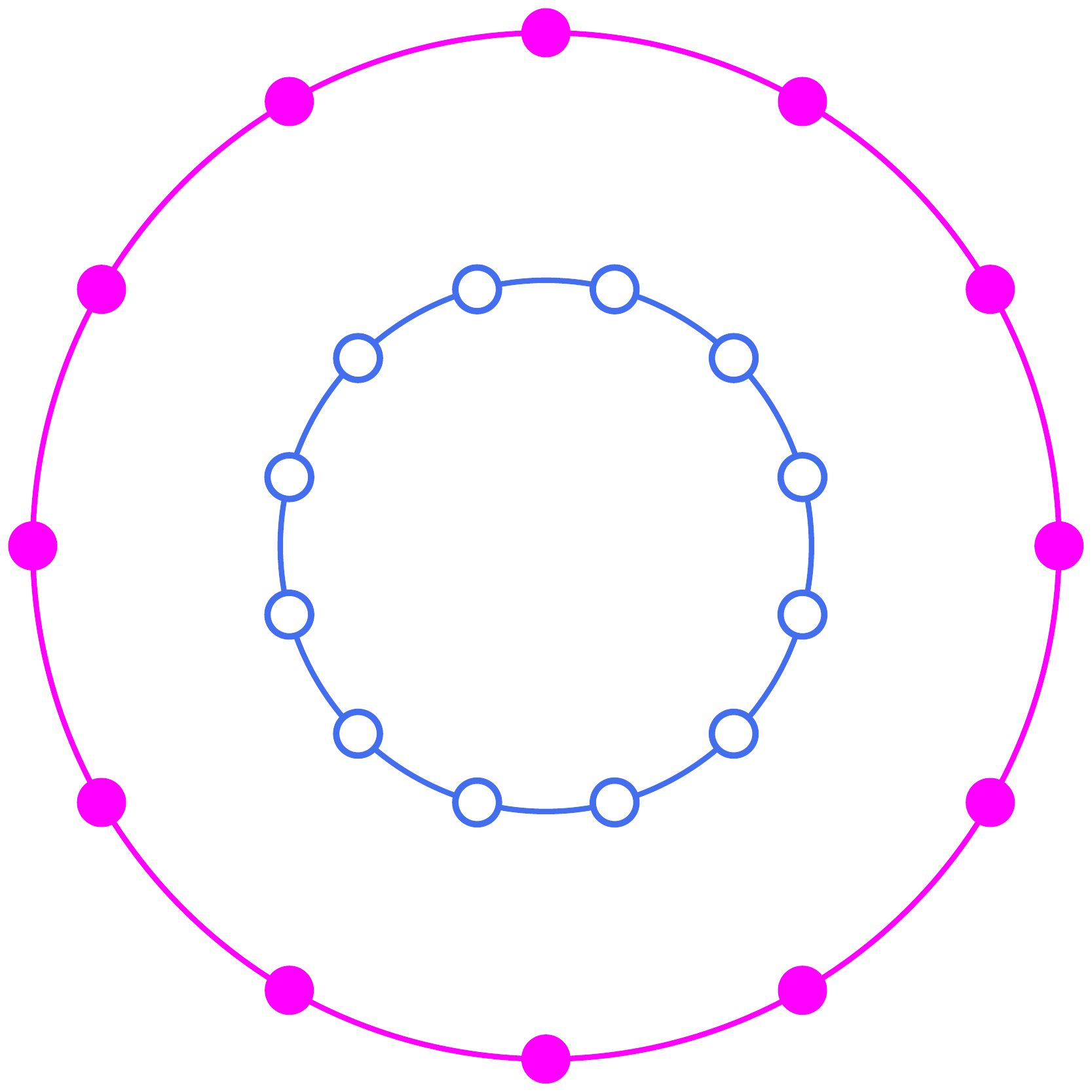}
	\centerline{$q: \frac{1}{2}, \cos\frac{5\pi}{12}$}
	\centerline{$\theta: 0, \pi/12$}
	& 
	\includegraphics[scale=0.10]{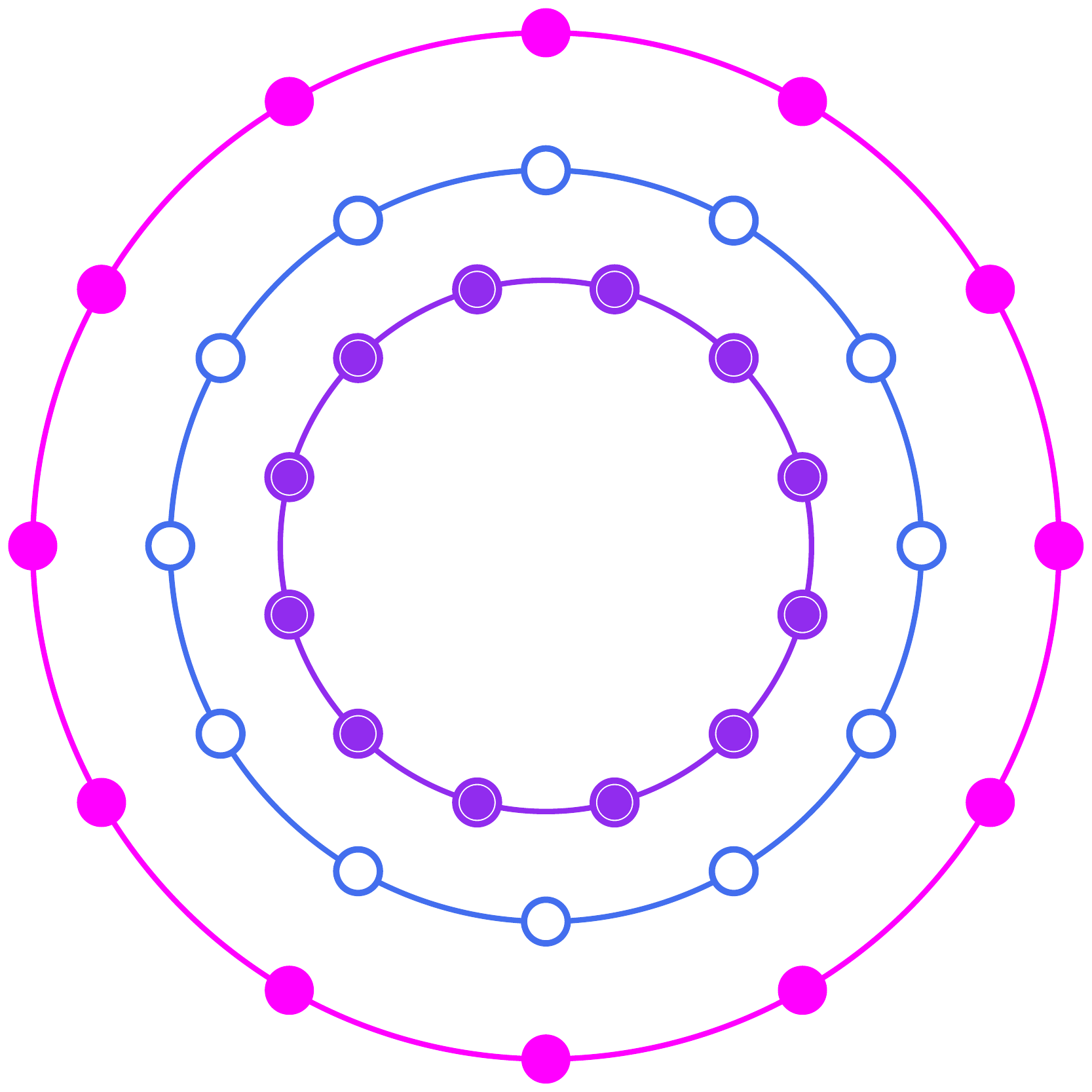}
	\centerline{$q: \cos\frac{\pi}{12}, \cos\frac{\pi}{4}$, 
	$\frac{1}{2}$}
	\centerline{$\theta: 0, 0, \pi/12$}
	\rule[45pt]{100pt}{0pt}
	& 
	\includegraphics[scale=0.10]{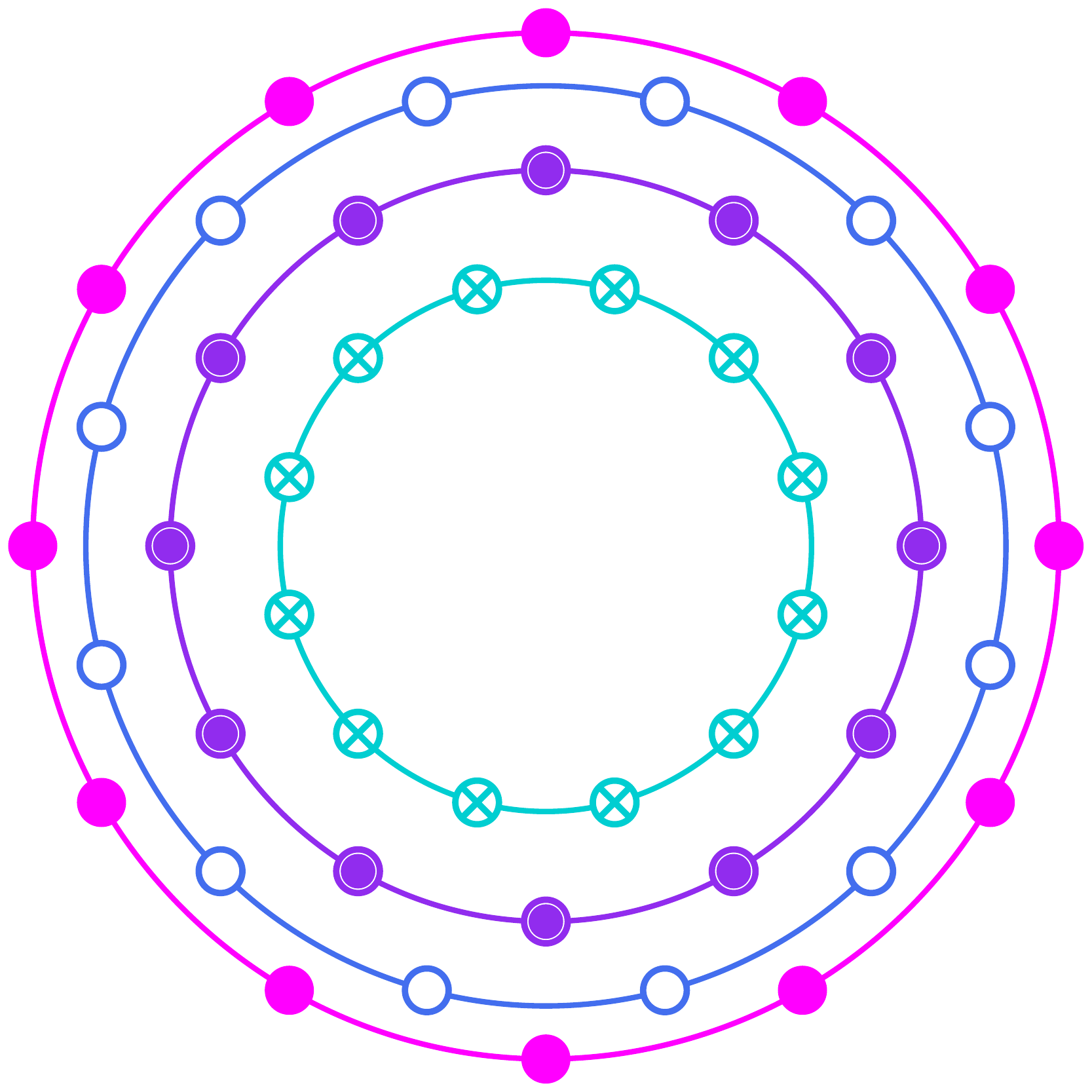}
	\centerline{$q: \cos\frac{\pi}{12}, \cos\frac{\pi}{6},
	\cos\frac{\pi}{4}, \frac{1}{2}$}
	\centerline{$\theta: 0, \pi/12, 0, \pi/12$}
	\rule[45pt]{100pt}{0pt}
	\\
	\hline
	HC free energies \rule[125pt]{100pt}{0pt} &
	{
	\begin{equation*}
		\begin{aligned}
			&-6\epsilon^{*} \sum_{j=1}^{2}\hphi_{j}^{2}
			- 8(\hphi_{1}+\hphi_{2})^{3}
			\\
			&+ 99 \sum_{j=1}^{2}\hphi_{j}^{4}
			+ 144 \hphi_{1}\hphi_{2}\sum_{j=1}^{2}\hphi_{j}^{2}
			\\
			&+ 360 \hphi_{1}^{2}\hphi_{2}^{2}
		\end{aligned}
		\label{sm.eq:energy.12.2}
	\end{equation*}
	}
	\rule[80pt]{100pt}{0pt}
	&
	{
	\begin{equation*}
		\begin{aligned}
			&-6\epsilon^{*} \sum_{j=1}^{3}\hphi_{j}^{2}
			- 8 \sum_{j=1}^{3} \hphi_j^3
			-48 \hphi_1 \hphi_2 \hphi_3
			\\
			&-24 \hphi_3 ( \hphi_1^2 + (\hphi_1+\hphi_2) \hphi_3 )
			\\
			&+ 99 \sum_{j=1}^{3}\hphi_{j}^{4}
			+ 72 \hphi_1^3 (\hphi_2 + 2\hphi_3)
			\\
			&+ 288 \hphi_1 \hphi_2 ( \hphi_1 (\hphi_2+\hphi_3) 
				+ \hphi_2 \hphi_3 )
			\\
			&+ 324 \hphi_2^2 \hphi_3^2
			+ 360 \hphi_1 \hphi_3^2 (\hphi_1+\hphi_2)
			\\
			&+ 144 \hphi_3^3 (\hphi_1 + \hphi_2)
		\end{aligned}
		\label{sm.eq:energy.12.3}
	\end{equation*}
	}
	\rule[45pt]{100pt}{0pt}
	& 
	{
	\begin{equation*}
		\begin{aligned}
			&-6\epsilon^{*} \sum_{j=1}^{4}\hphi_{j}^{2}
			-8 \sum_{j=1}^{4} \hphi_j^3
			- 24 \hphi_1^2 \hphi_4
			\\
			&- 24 \hphi_4^2 \sum_{j=1}^{3} \hphi_j
			- 48 \hphi_1 \hphi_3 (\hphi_2 + \hphi_4)
			\\
			&+ 99 \sum_{j=1}^{4} \hphi_{j}^{4}
			+ 72 \hphi_1^3 (\hphi_3 + 2\hphi_4)
			\\
			&+ 72 \hphi_4^3 (2\hphi_1 + \hphi_2 + 2\hphi_3)
			\\
			&+ 144 \hphi_1 \hphi_2 \hphi_3 (\hphi_1+\hphi_3)
			\\
			&+ 216 \hphi_1 \hphi_2 ( \hphi_1\hphi_4 + \hphi_2\hphi_3)
			\\
			&+ 288 \hphi_1^2 (\hphi_2^2 + \hphi_3^2)
			+ 288 \hphi_2^2 (\hphi_3^2 + \hphi_4^2)
			\\
			&+ 288 \hphi_1 \hphi_3 \hphi_4 (\hphi_1+\hphi_3)
			+ 288 \hphi_1 \hphi_2 \hphi_4^2
			\\
			&+ 288 \hphi_2 \hphi_3 \hphi_4 (\hphi_3+\hphi_4)
			+ 324 \hphi_3^2 \hphi_4^2
			\\
			&+ 360 \hphi_1 \hphi_4^2 (\hphi_1+\hphi_3)
			+ 432 \hphi_1 \hphi_2 \hphi_3 \hphi_4
		\end{aligned}
		\label{sm.eq:energy.12.4}
	\end{equation*}
	}
	\\
	\hline
	HC phase diagrams \rule[45pt]{100pt}{0pt} & 
	\includegraphics[scale=0.16]{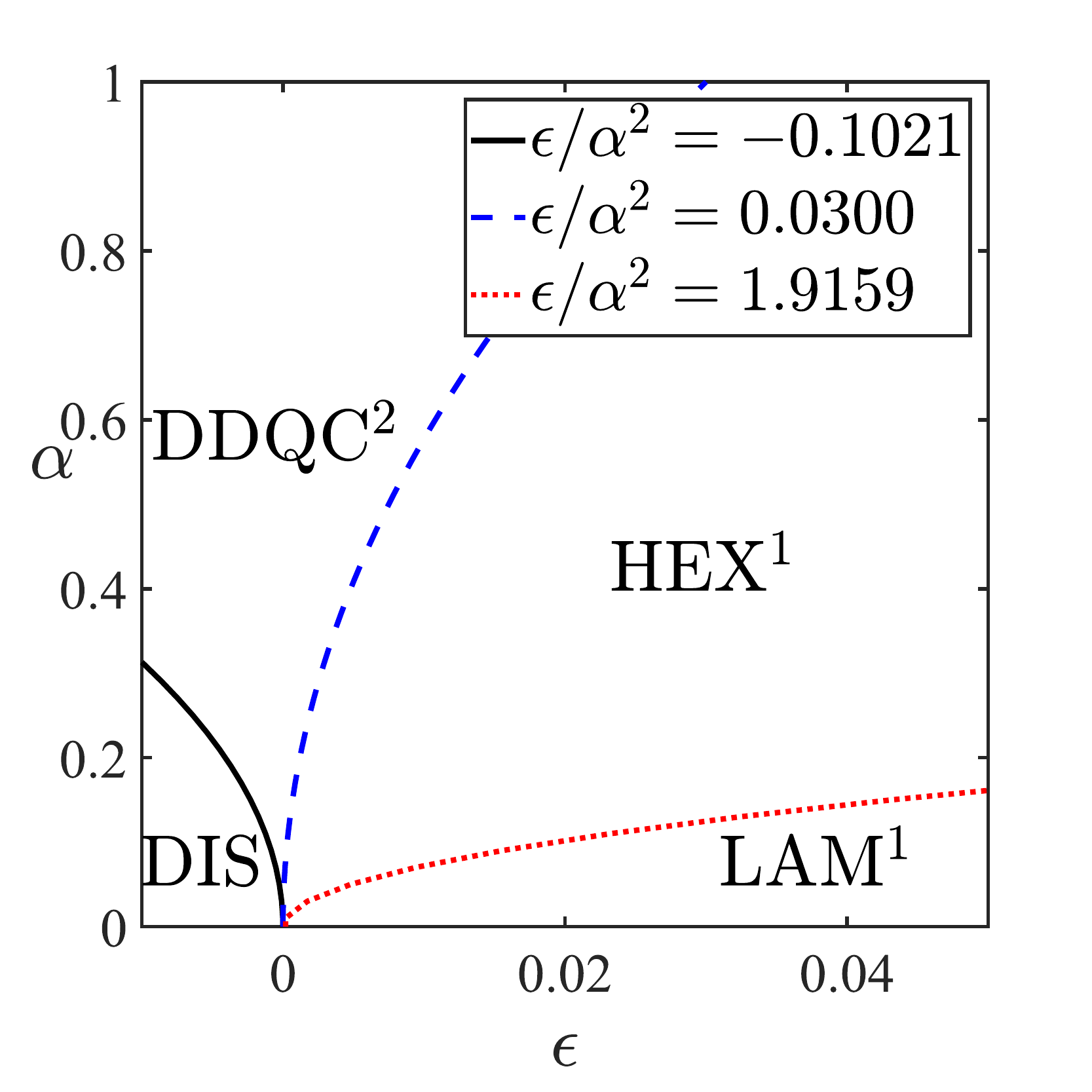} & 
	\includegraphics[scale=0.16]{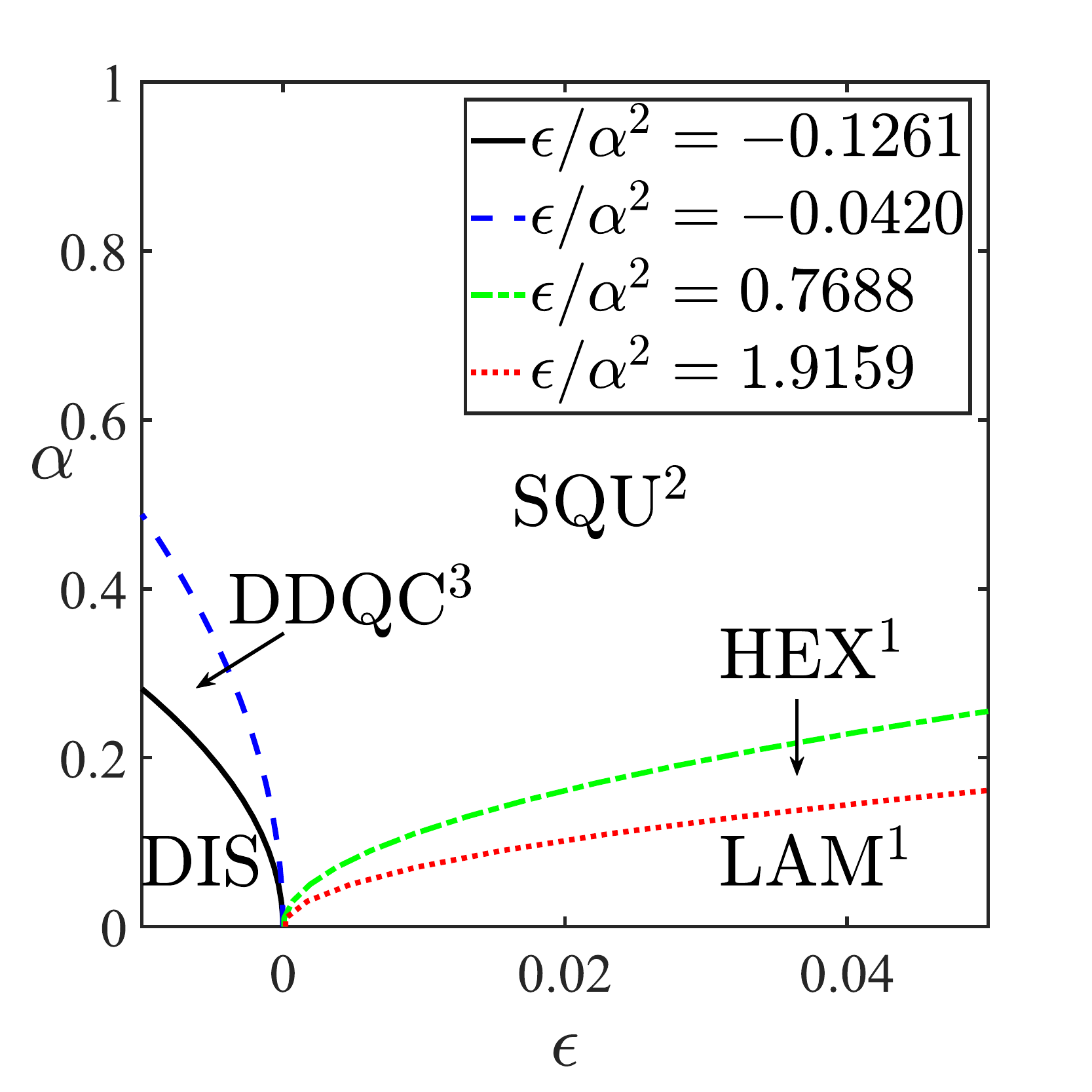} & 
	\includegraphics[scale=0.16]{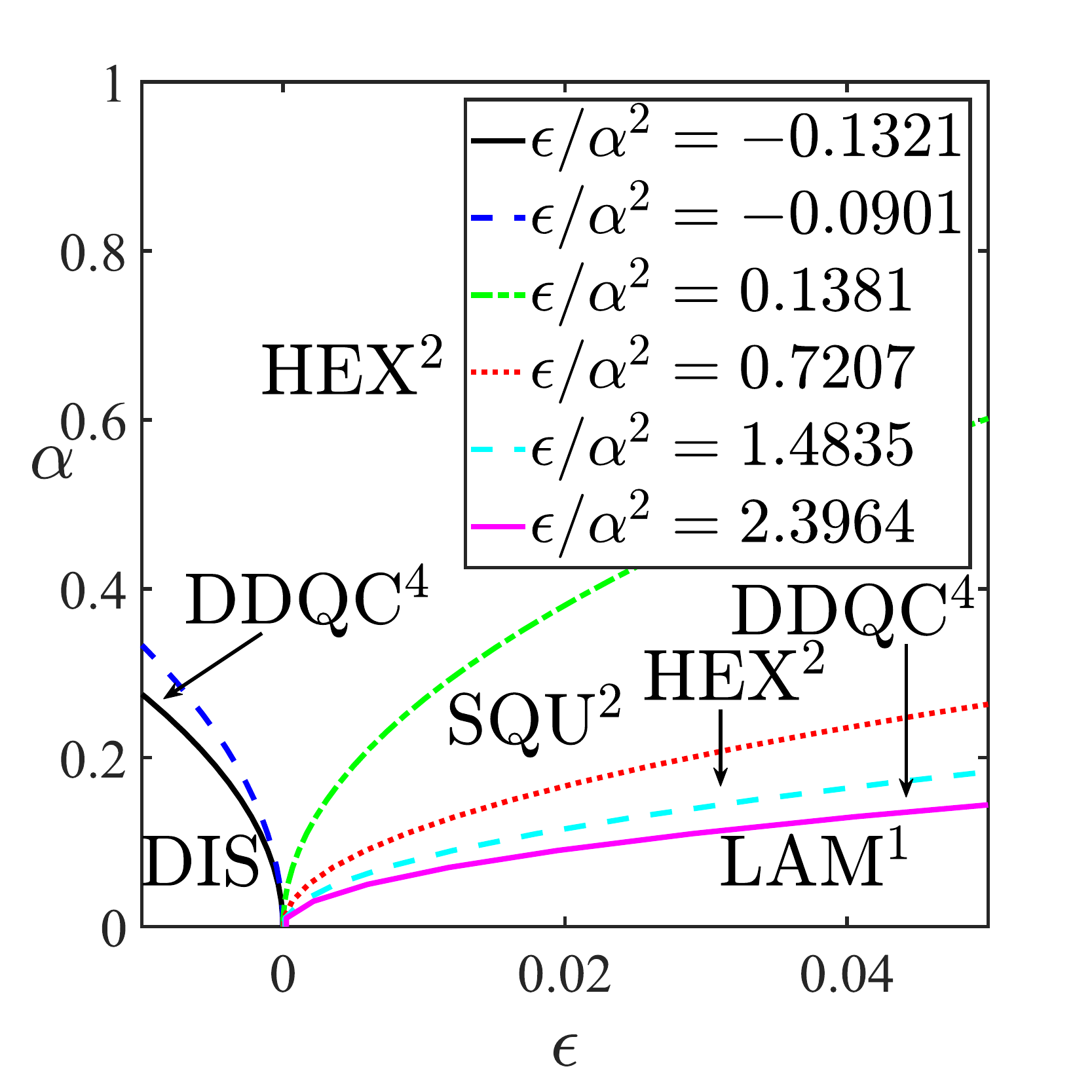} \\
	\hline
	\multicolumn{4}{c}{} \\
	\multicolumn{4}{c}{} \\
	\hline
	Desired QCs & \multicolumn{3}{c|}{\textbf{Tetradecagonal (TD) QCs} ($14$-fold)} \\
	\hline
	$\{w_j\}_{j=1}^{m}$ & 1, 3; 1, 5; 3, 5 & 1, 3, 5 & 1, 2, 3, 6; 1, 2, 4, 5;
	3, 4, 5, 6 \\
	\hline                                                               
	$\{s_j\}_{j=1}^{m}$ & 0, 0; 0, 0; 0, 0 & 0, 0, 0 & 0, 1, 0, 1; 0, 1, 1, 0;
	0, 1, 0, 1 \\
	\hline
	Optimal primary RLVs \rule[165pt]{100pt}{0pt} &
	\includegraphics[scale=0.10]{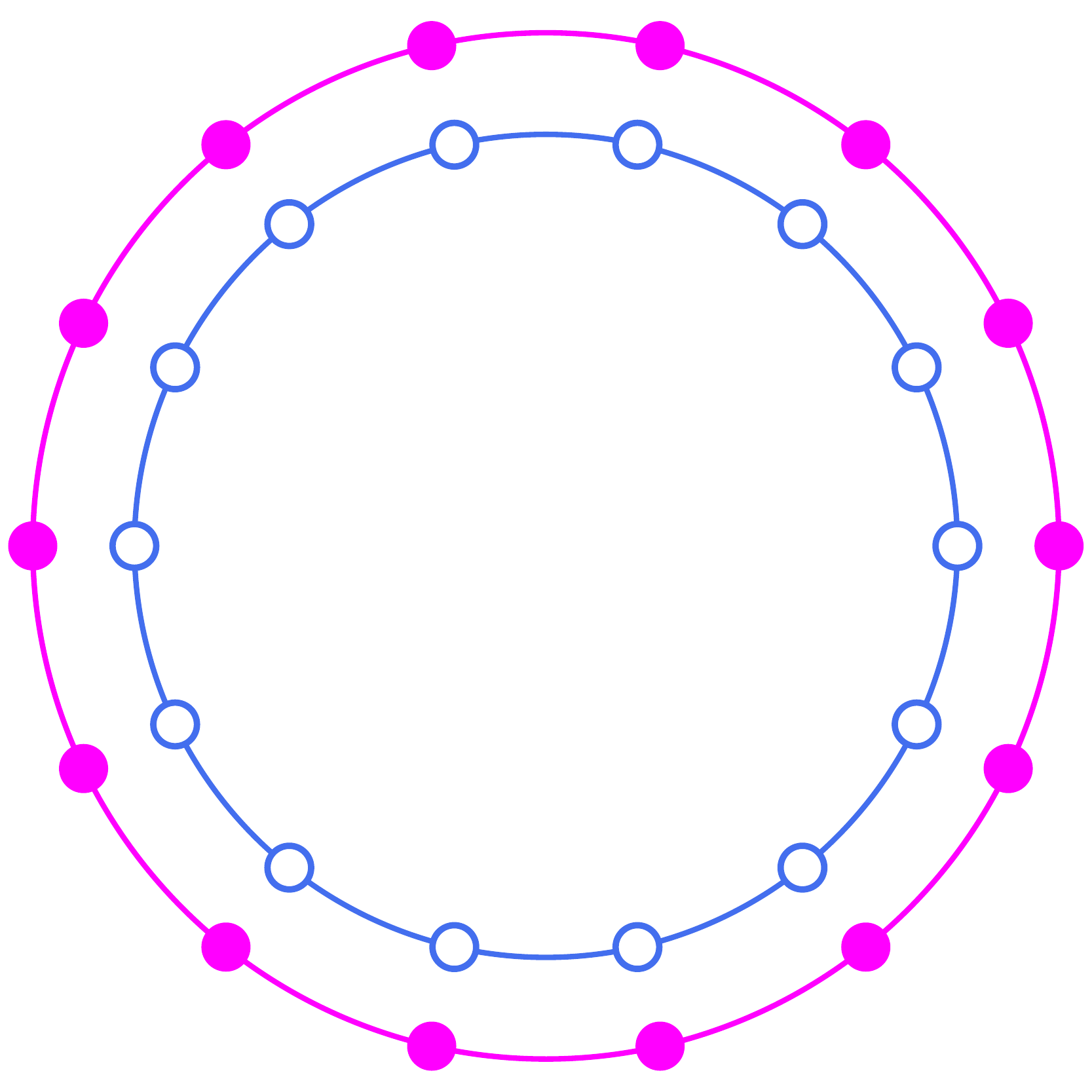}
	\centerline{$q: \cos\frac{\pi}{14}, \cos\frac{3\pi}{14}$}
	\centerline{$\theta: 0, 0$}
	\includegraphics[scale=0.10]{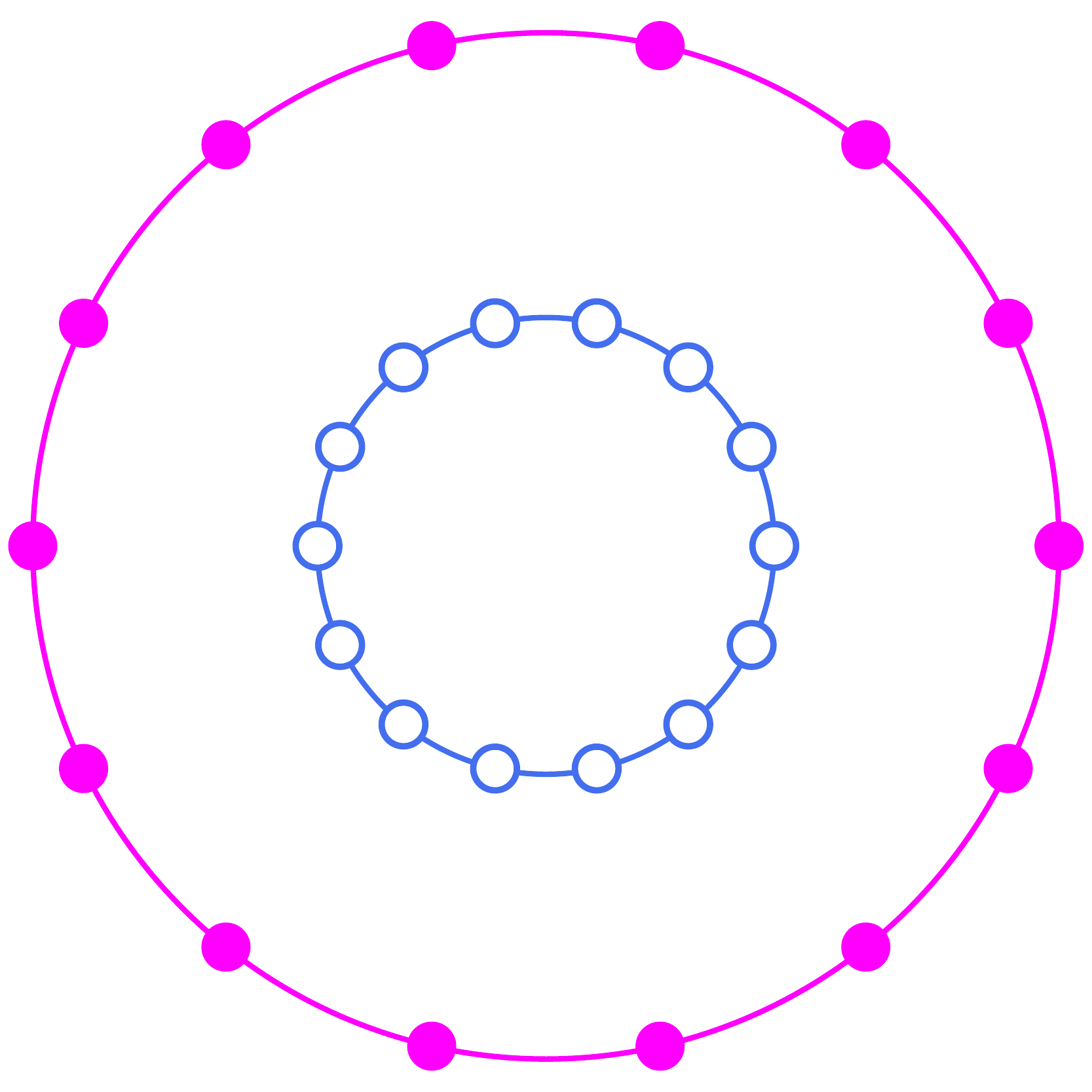}
	\centerline{$q: \cos\frac{\pi}{14}, \cos\frac{5\pi}{14}$}
	\centerline{$\theta: 0, 0$}
	\includegraphics[scale=0.10]{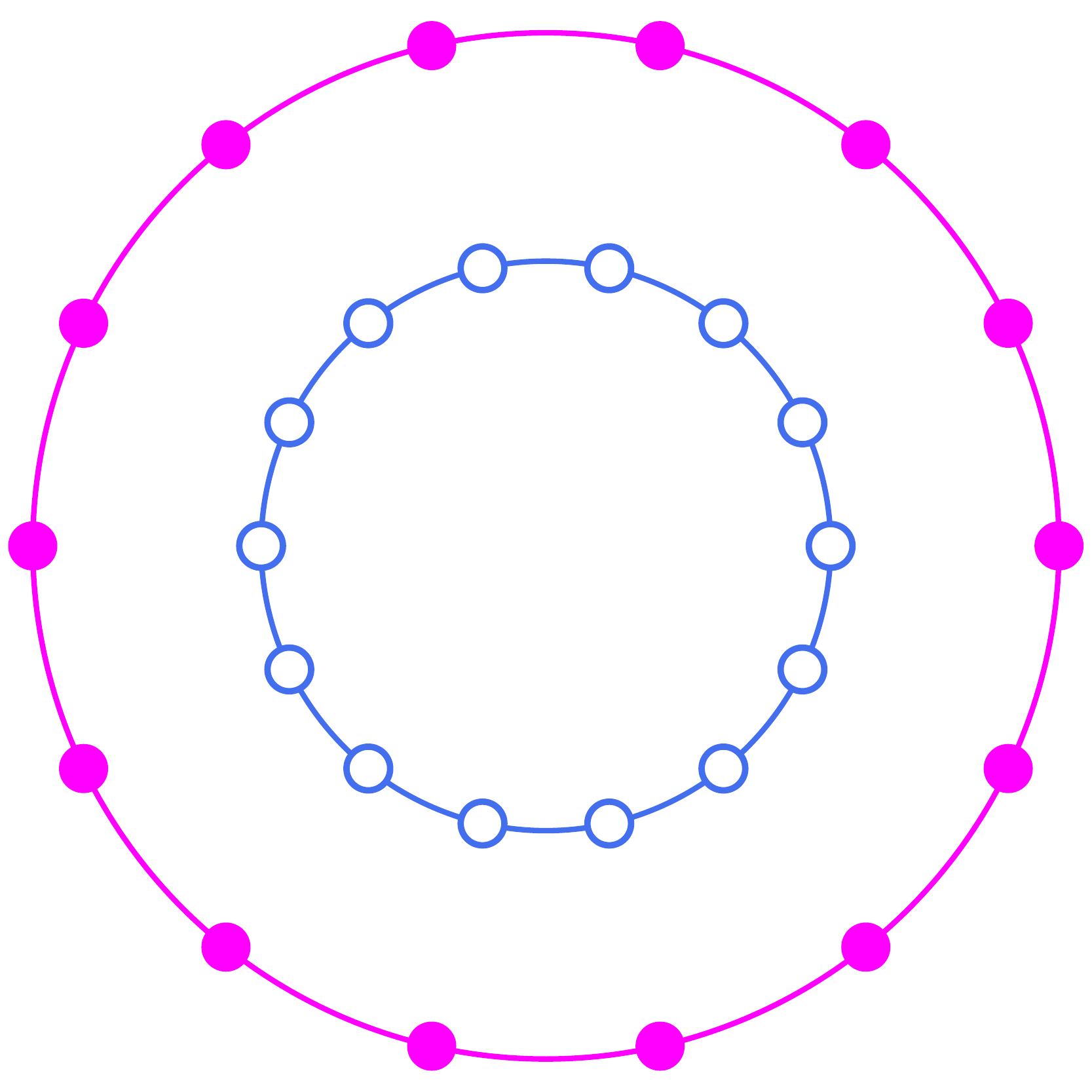}
	\centerline{$q: \cos\frac{3\pi}{14}, \cos\frac{5\pi}{14}$}
	\centerline{$\theta: 0, 0$}
	& 
	\includegraphics[scale=0.10]{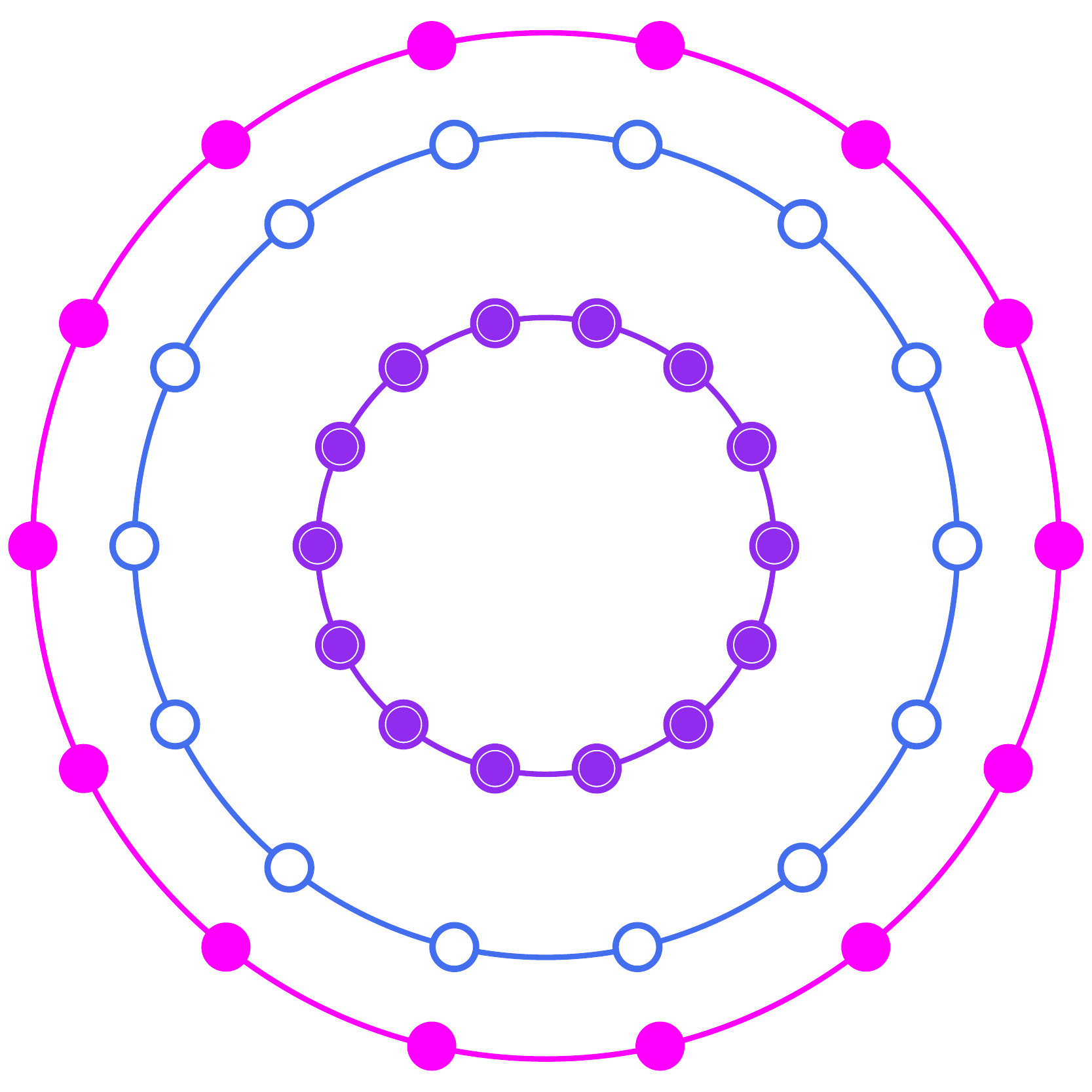}
	\centerline{$q: \cos\frac{\pi}{14}, \cos\frac{3\pi}{14},
	\cos\frac{5\pi}{14}$}
	\centerline{$\theta: 0, 0, 0$}
	\rule[110pt]{100pt}{0pt}
	& 
	\includegraphics[scale=0.10]{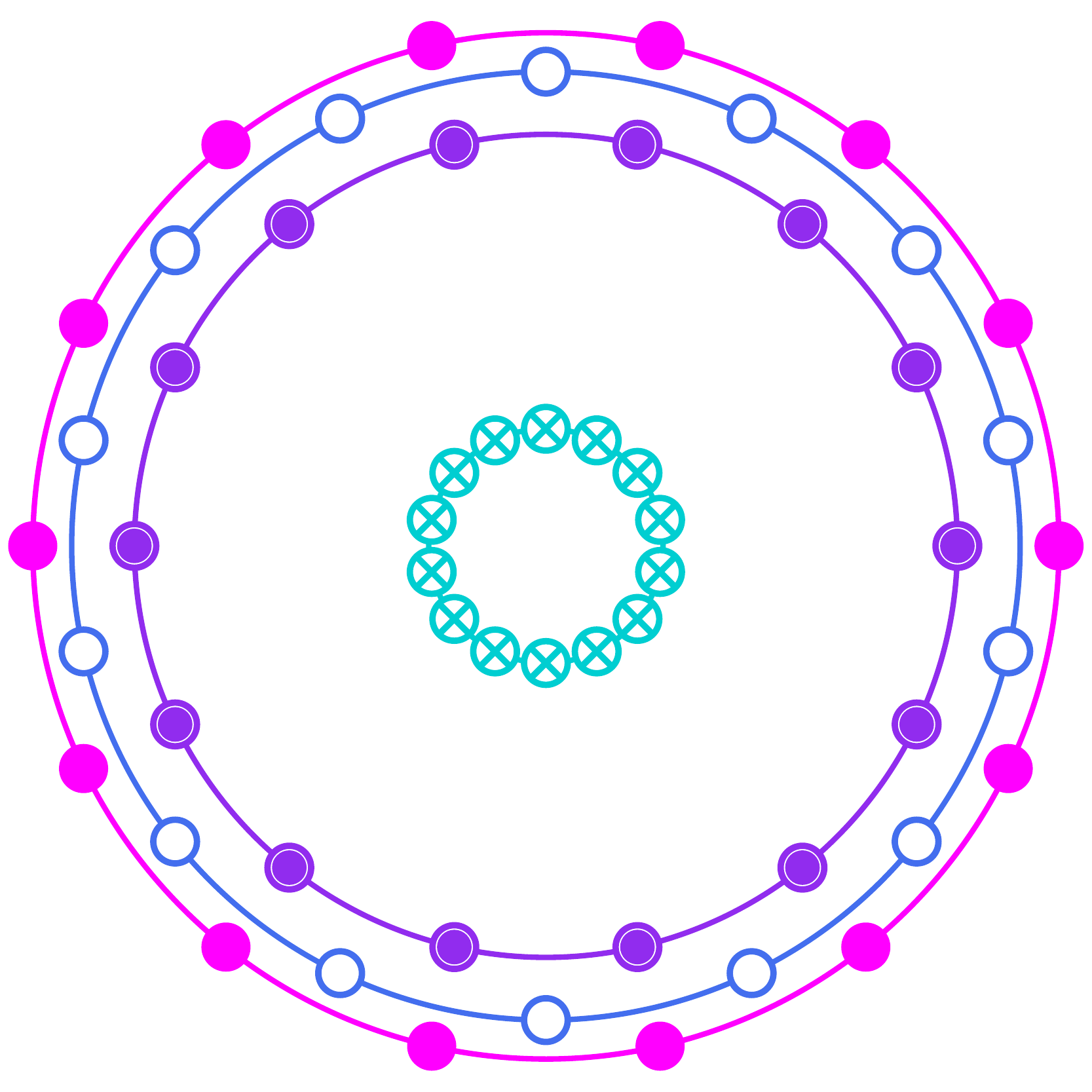}
	\centerline{$q: \cos\frac{\pi}{14}, \cos\frac{\pi}{7},
	\cos\frac{3\pi}{14}, \cos\frac{3\pi}{7}$}
	\centerline{$\theta: 0, \pi/14, 0, \pi/14$}
	\includegraphics[scale=0.10]{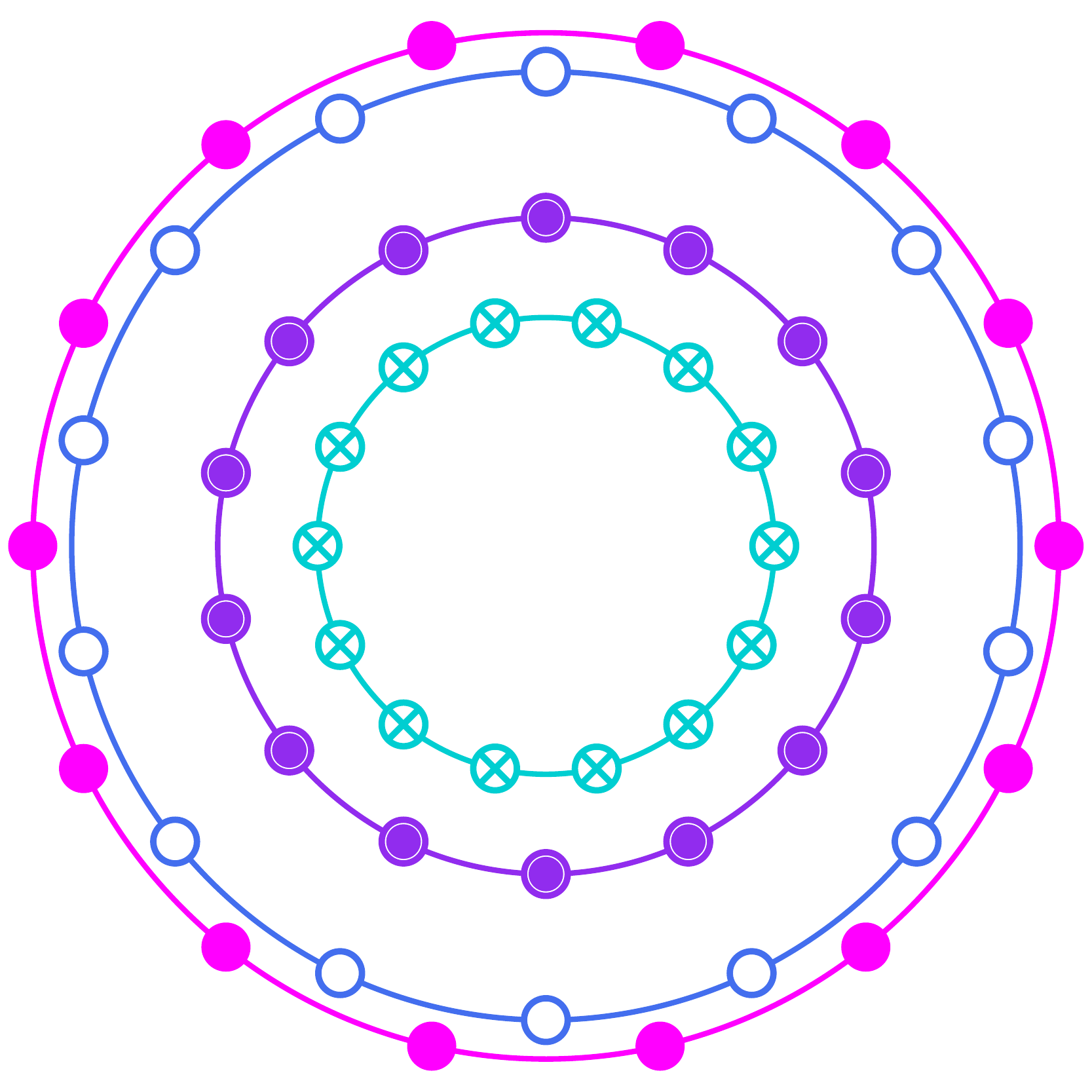}
	\centerline{$q: \cos\frac{\pi}{14}, \cos\frac{\pi}{7},
	\cos\frac{2\pi}{7}, \cos\frac{5\pi}{14}$}
	\centerline{$\theta: 0, \pi/14, \pi/14, 0$}
	\includegraphics[scale=0.10]{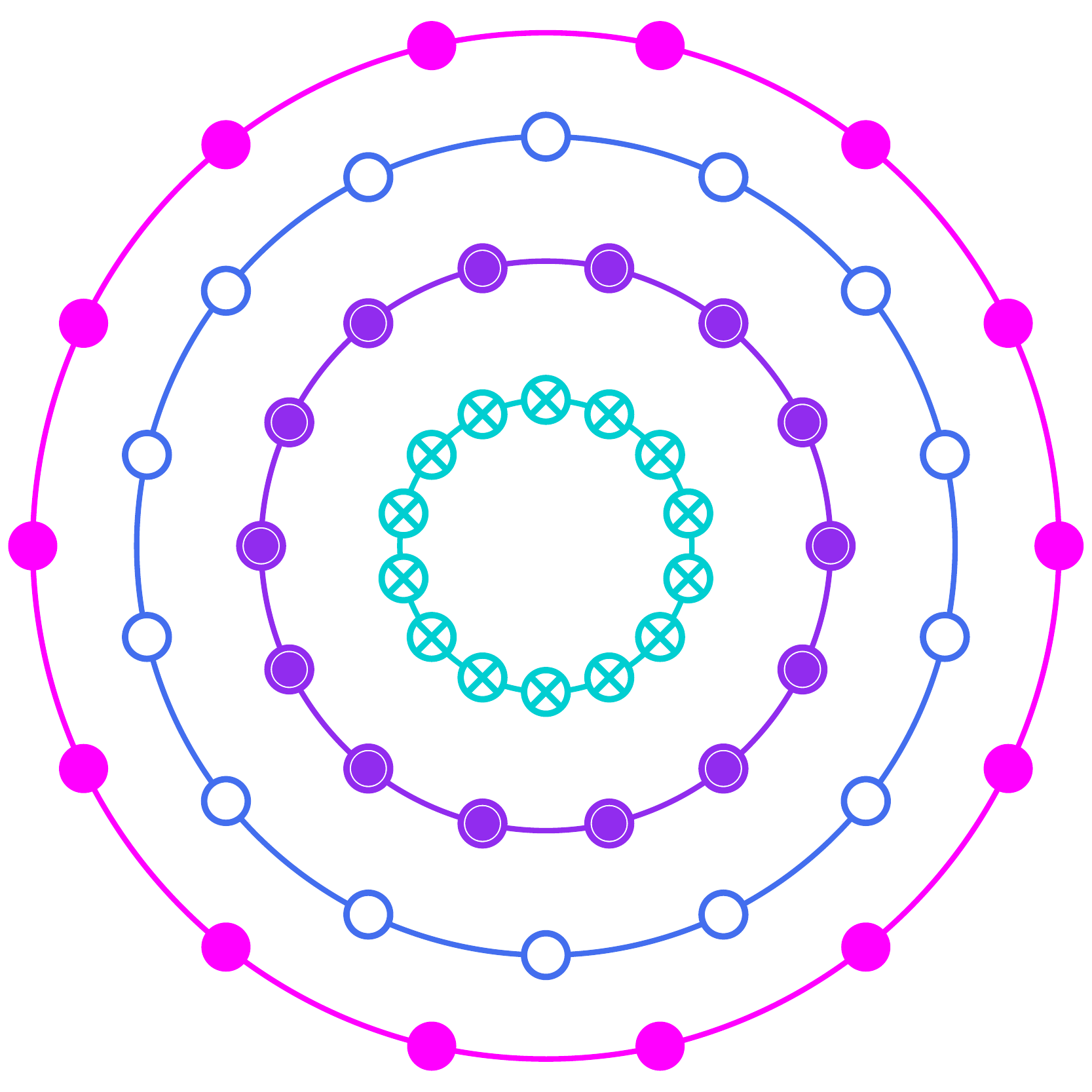}
	\centerline{$q: \cos\frac{3\pi}{14}, \cos\frac{2\pi}{7},
	\cos\frac{5\pi}{14}, \cos\frac{3\pi}{7}$}
	\centerline{$\theta: 0, \pi/14, 0, \pi/14$}
	\\
	\hline
	HC free energies \rule[85pt]{100pt}{0pt} &
	{
	\begin{equation*}
		\begin{aligned}
			&-7\epsilon^{*} \sum_{j=1}^{2}\hphi_{j}^{2}
			- 28 \hphi_{1} \hphi_{2}^{2}
			+ 84 \hphi_{1}^{3} \hphi_{2}
			\\
			&+ \frac{273}{2} \sum_{j=1}^{2}\hphi_{j}^{4} 
			+ 378 \hphi_{1}^{2} \hphi_{2}^{2}
		\end{aligned}
		\label{sm.eq:energy.14.2}
	\end{equation*}
	}
	\rule[45pt]{100pt}{0pt}
	&
	{
	\begin{equation*}
		\begin{aligned}
			&-7\epsilon^{*} \sum_{j=1}^{3}\hphi_{j}^{2}
			- 56 \hphi_1 \hphi_2 \hphi_3
			\\
			&- 28 (\hphi_1^2 \hphi_3 + \hphi_1 \hphi_2^2
				+ \hphi_2 \hphi_3^2)
			\\
			&+ \frac{273}{2} \sum_{j=1}^{3}\hphi_{j}^{4} 
			+ 252 \hphi_1 \hphi_2 \hphi_3 \sum_{j=1}^{3} \hphi_j
			\\
			&+ 84 (\hphi_1^3 \hphi_2 + \hphi_1 \hphi_3^3
				+ \hphi_2^3 \hphi_3)
			\\
			&+ 378 (\hphi_1^2 \hphi_2^2 + \hphi_1^2 \hphi_3^2
				+ \hphi_2^2 \hphi_3^2)
		\end{aligned}
		\label{sm.eq:energy.14.3}
	\end{equation*}
	}
	\rule[25pt]{100pt}{0pt}
	&
	{
	\begin{equation*}
		\begin{aligned}
			&-7\epsilon^{*} \sum_{j=1}^{4}\hphi_{j}^{2}
			- 28 \hphi_3 (\hphi_1 \hphi_3 + \hphi_2^2)
			\\
			&- 56 \hphi_2 \hphi_4 (\hphi_1 + \hphi_3)
			\\
			&+ \frac{273}{2}\sum_{j=1}^{4}\hphi_{j}^{4} 
			+ 84 (\hphi_1^3 \hphi_3 + \hphi_2^3 \hphi_4)
			\\
			&+ 252 \hphi_3 (\hphi_1 \hphi_2^2 + \hphi_1 \hphi_4^2
				+ \hphi_2 \hphi_3 \hphi_4)
			\\
			&+ 378 \hphi_1^2 \sum_{j=2}^{4} \hphi_j^2
			+ 378 \hphi_2^2 \sum_{j=3}^{4} \hphi_j^2
			\\
			&+ 378 \hphi_3^2 \hphi_4^2
			+ 504 \hphi_1 \hphi_2 \hphi_3 \hphi_4
		\end{aligned}
		\label{sm.eq:energy.14.4}
	\end{equation*}
	}
	\\
	\hline
	HC phase diagrams \rule[45pt]{100pt}{0pt} & 
	\includegraphics[scale=0.16]{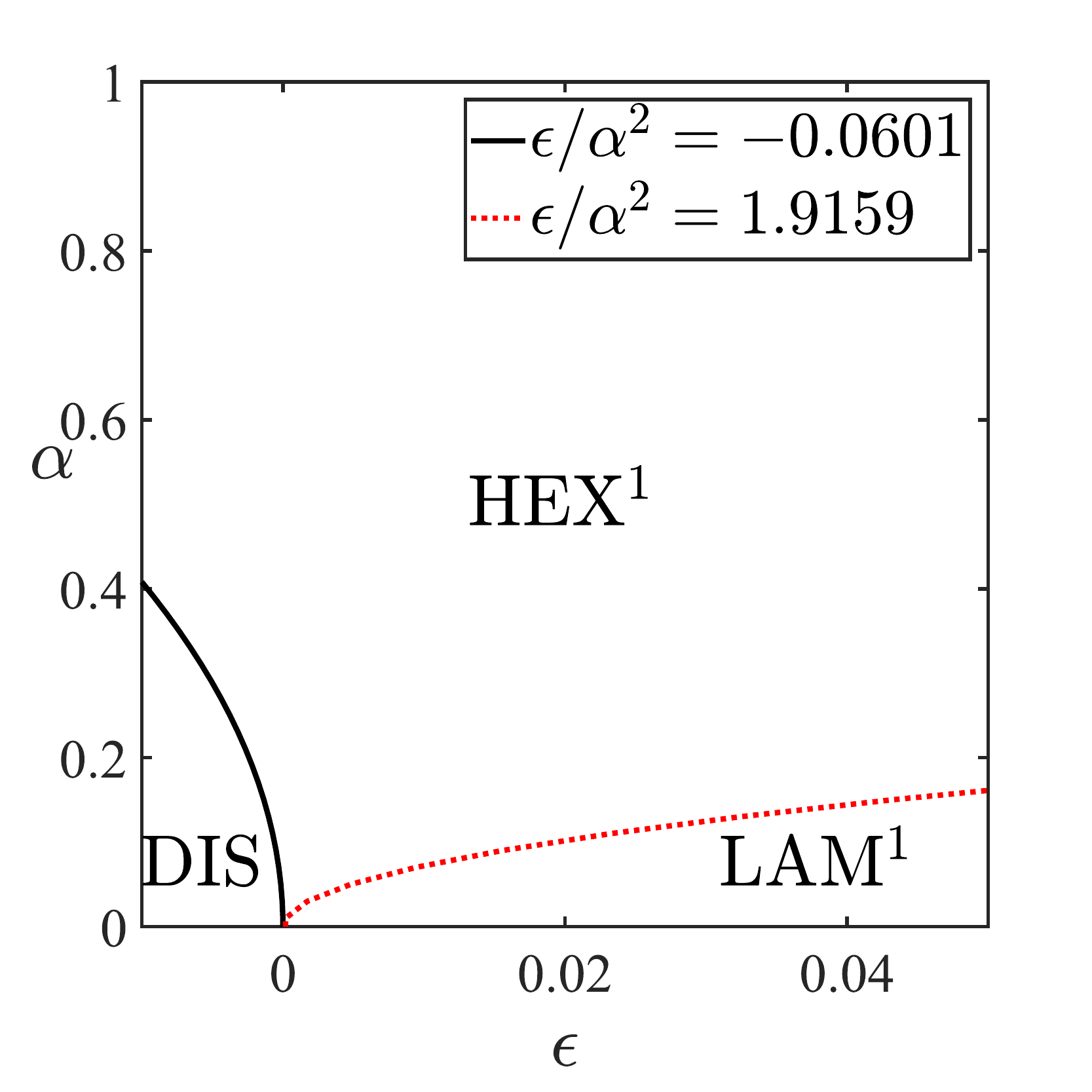} & 
	\includegraphics[scale=0.16]{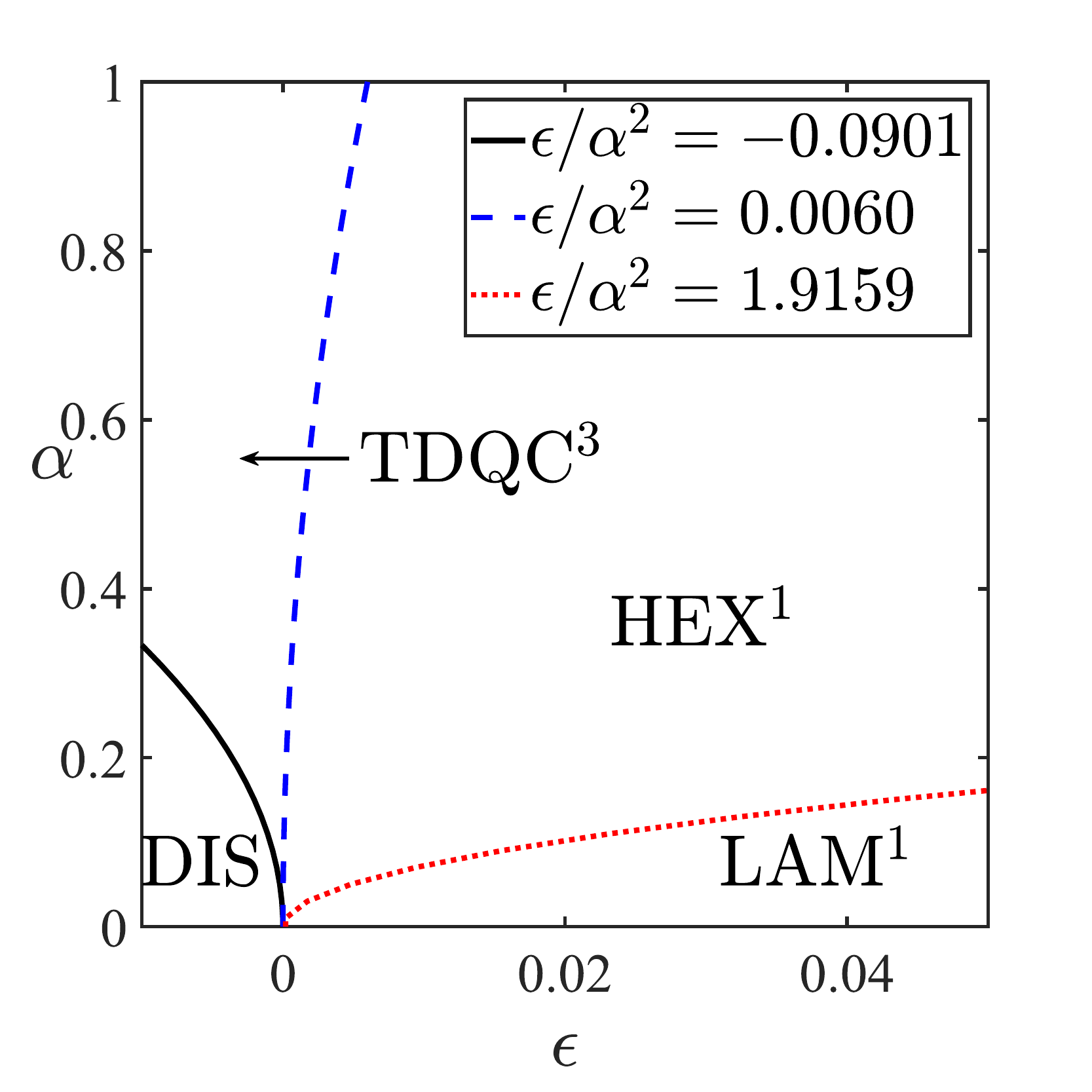} & 
	\includegraphics[scale=0.16]{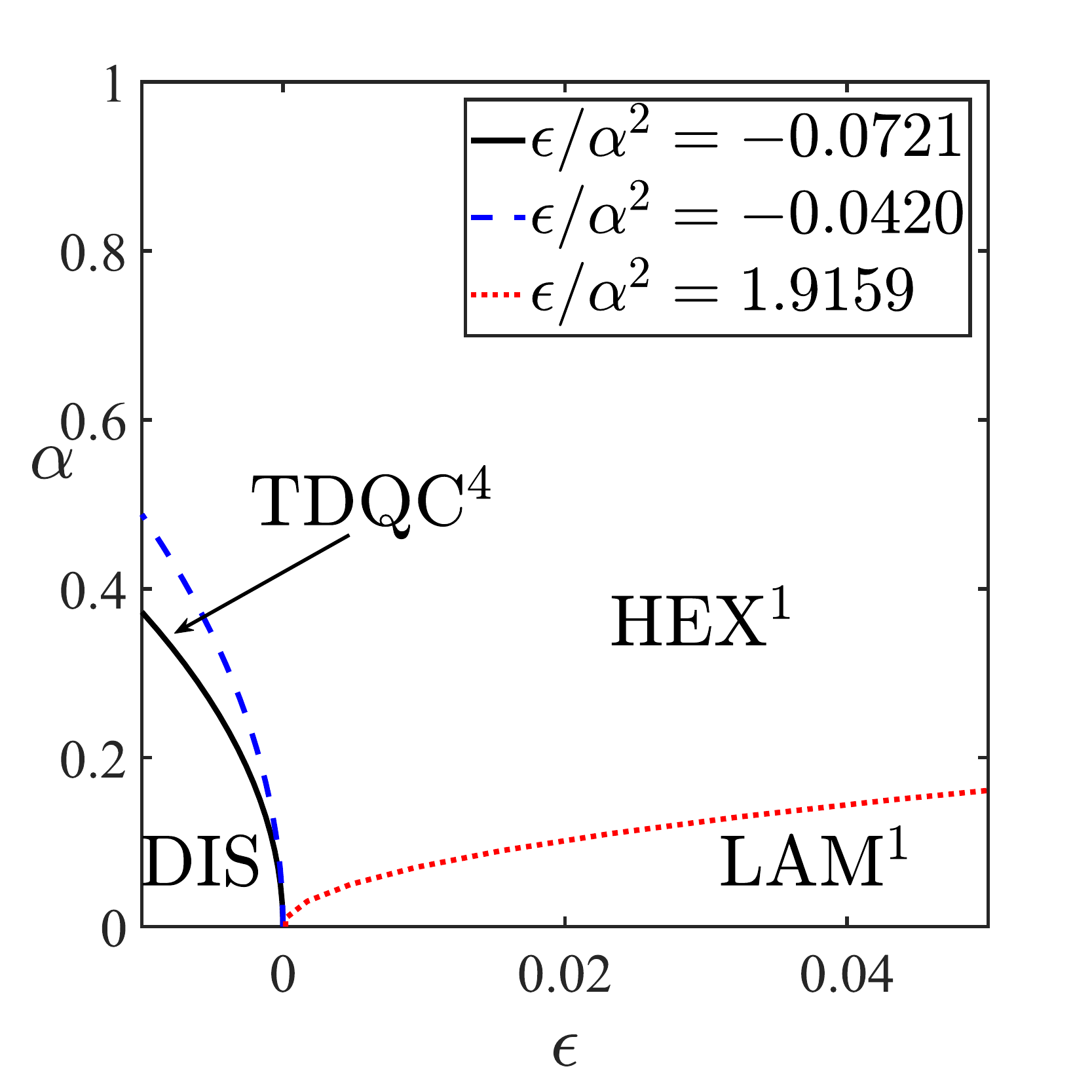} \\
	\hline
	\multicolumn{4}{c}{} \\
	\multicolumn{4}{c}{} \\
	\hline
	Desired QCs & \multicolumn{3}{c|}{\textbf{Hexadecagonal (HD) QCs} ($16$-fold)} \\
	\hline
	$\{w_j\}_{j=1}^{m}$ & 0, 4 & 2, 4, 6 & 1, 2, 4, 5; 1, 3, 4, 6 \\
	\hline
	$\{s_j\}_{j=1}^{m}$ & 0, 0 & 0, 0, 0 & 0, 1, 1, 0; 0, 0, 1, 1 \\
	\hline
	Optimal primary RLVs \rule[105pt]{100pt}{0pt} &
	\includegraphics[scale=0.10]{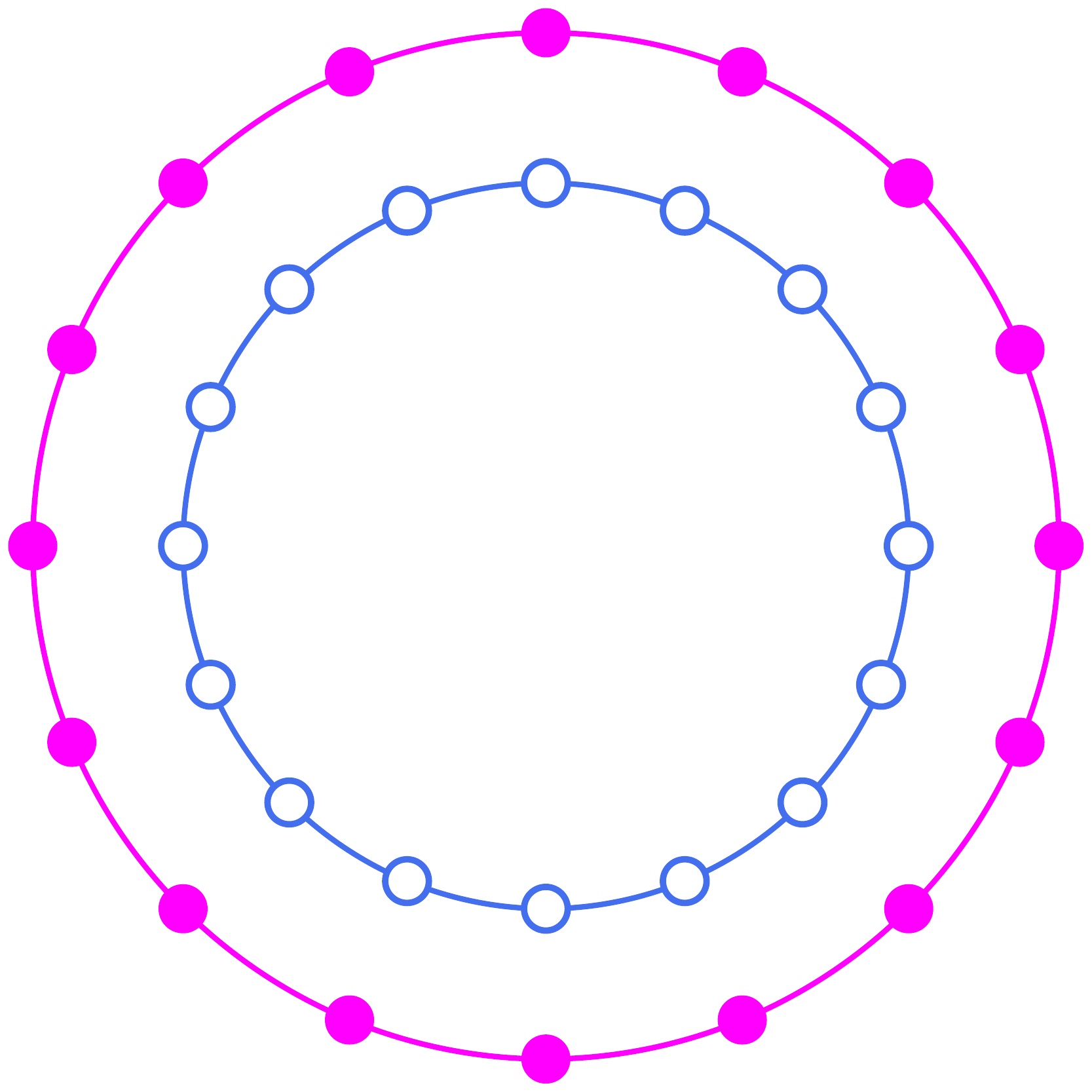}
	\centerline{$q: 1, \cos\frac{\pi}{4}$}
	\centerline{$\theta: 0, 0$}
	\rule[45pt]{100pt}{0pt}
	& 
	\includegraphics[scale=0.10]{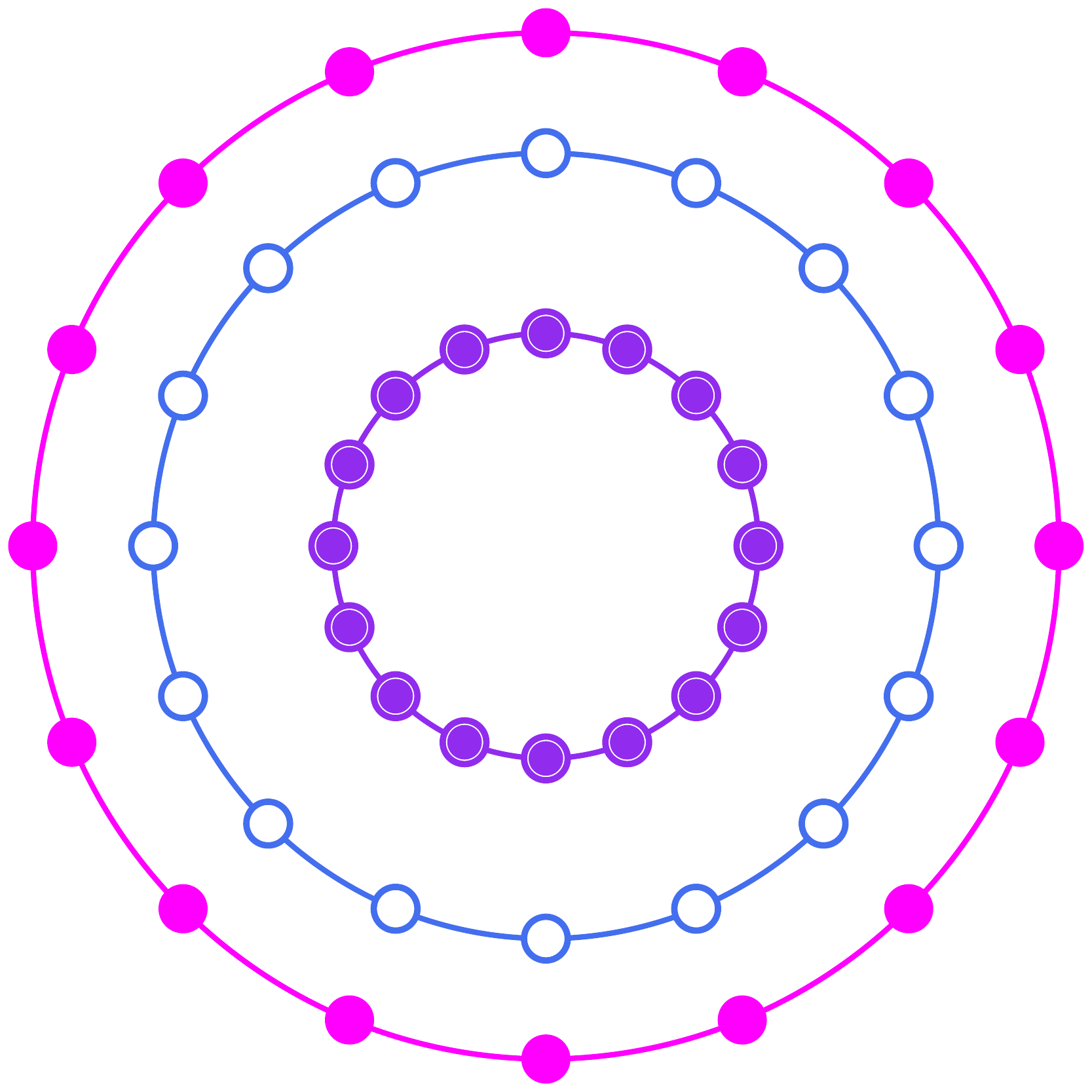}
	\centerline{$q: \cos\frac{\pi}{8}, \cos\frac{\pi}{4}, 
	\cos\frac{3\pi}{8}$}
	\centerline{$\theta: 0, 0, 0$}
	\rule[45pt]{100pt}{0pt}
	& 
	\includegraphics[scale=0.10]{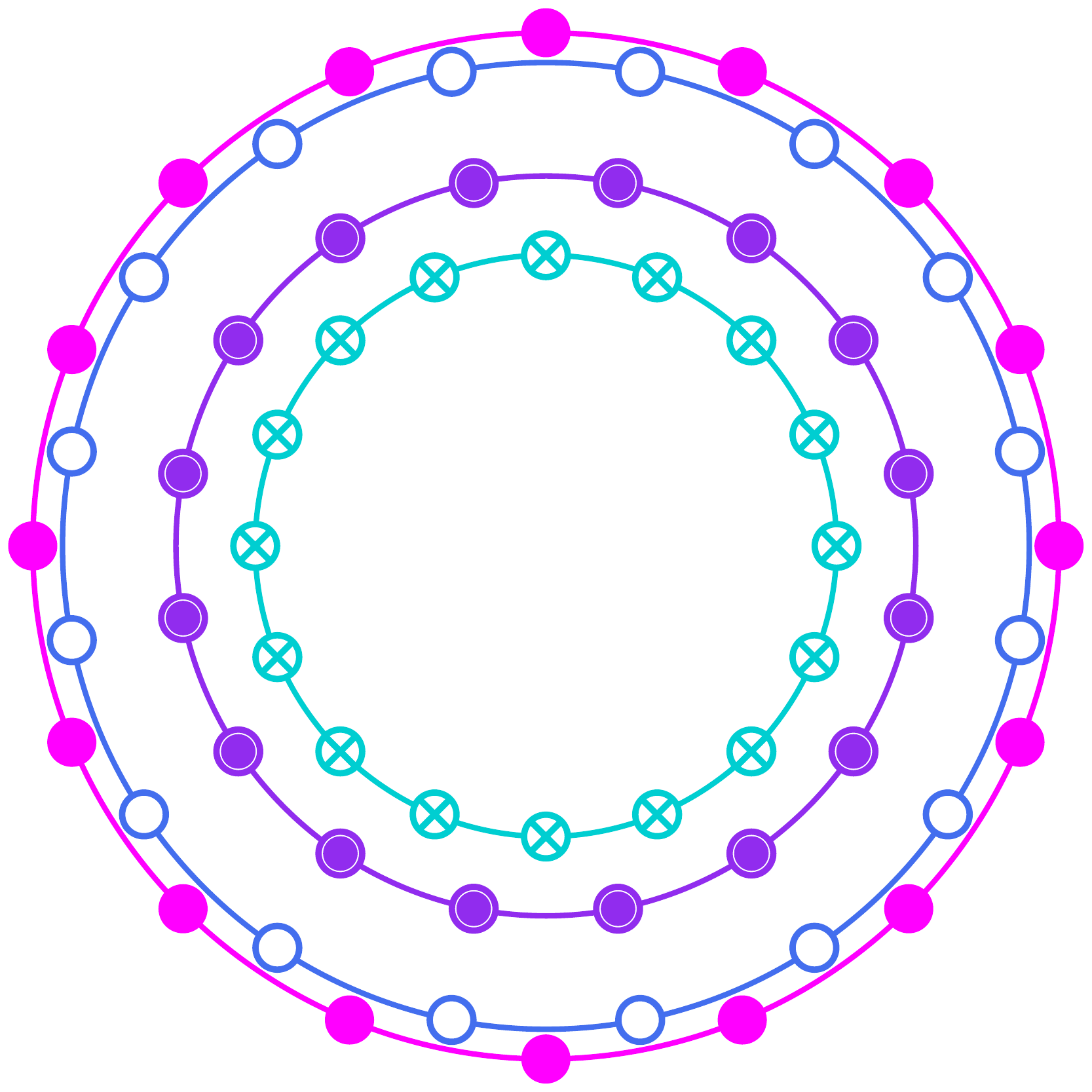}
	\centerline{$q: \cos\frac{\pi}{16}, \cos\frac{\pi}{8}, 
	\cos\frac{\pi}{4}, \cos\frac{5\pi}{16}$}
	\centerline{$\theta: 0, \pi/16, \pi/16, 0$}
	\includegraphics[scale=0.10]{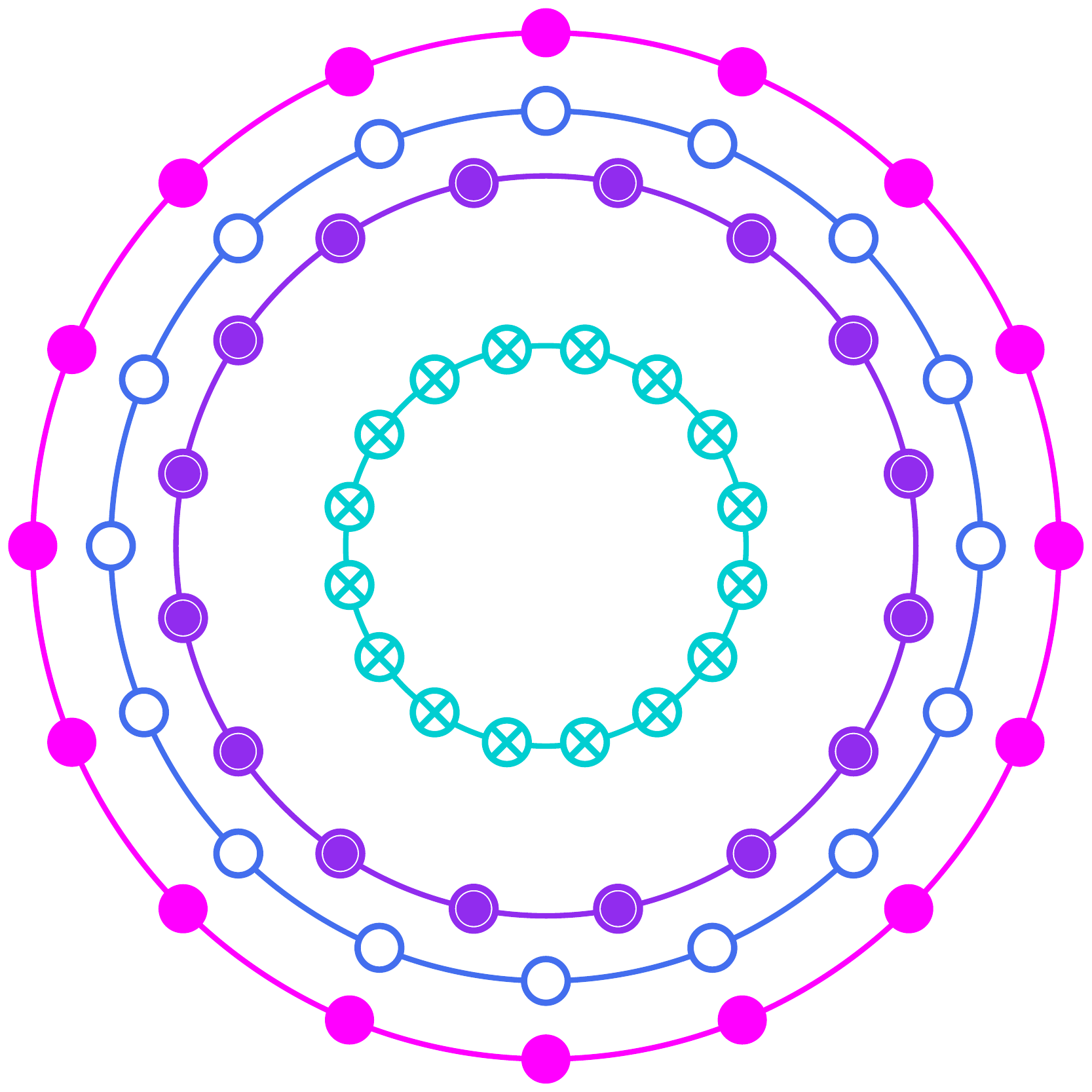}
	\centerline{$q: \cos\frac{\pi}{16}, \cos\frac{3\pi}{16},
	\cos\frac{\pi}{4}, \cos\frac{3\pi}{8}$}
	\centerline{$\theta: 0, 0, \pi/16, \pi/16$}
	\\
	\hline
	HC free energies \rule[65pt]{100pt}{0pt} &
	{
	\begin{equation*}
		\begin{aligned}
			&-8\epsilon^{*} \sum_{j=1}^{2}\hphi_{j}^{2}
			- 32 \hphi_{1} \hphi_{2}^{2}
			\\
			&+ 180 \sum_{j=1}^{2}\hphi_{j}^{4}
			+ 432 \hphi_{1}^{2} \hphi_{2}^{2}
		\end{aligned}
		\label{sm.eq:energy.16.2}
	\end{equation*}
	}
	\rule[45pt]{100pt}{0pt}
	&
	{
	\begin{equation*}
		\begin{aligned}
			&-8\epsilon^{*} \sum_{j=1}^{3}\hphi_{j}^{2}
			- 32 \hphi_2 (\hphi_1 + \hphi_3)^2
			\\
			&+ 180 \sum_{j=1}^{3}\hphi_{j}^{4}
			+ 96 \hphi_1 \hphi_3 (\hphi_1^2 + \hphi_3^2)
			\\
			&+ 480 (\hphi_1^2 \hphi_2^2 + \hphi_1^2 \hphi_3^2
				+ \hphi_2^2 \hphi_3^2)
			\\
			&+ 384 \hphi_1 \hphi_2^2 \hphi_3
		\end{aligned}
		\label{sm.eq:energy.16.3}
	\end{equation*}
	}
	\rule[25pt]{100pt}{0pt}
	&
	{
	\begin{equation*}
		\begin{aligned}
			&-8\epsilon^{*} \sum_{j=1}^{4}\hphi_{j}^{2}
			- 32 \hphi_2 (\hphi_2 \hphi_3 + \hphi_4^2)
			\\
			&- 64 \hphi_1 \hphi_4 (\hphi_2 + \hphi_3)
			\\
			&+ 180\sum_{j=1}^{4}\hphi_{j}^{4} 
			+ 96 \hphi_1 \hphi_4^3
			\\
			&+ 288 \hphi_2 (\hphi_1^2 \hphi_3 + \hphi_1 \hphi_2 \hphi_4
				+ \hphi_3 \hphi_4^2 )
			\\
			&+ 480 \hphi_1^2 \sum_{j=2}^{4} \hphi_j^2
			+ 480 \hphi_2^2 (\hphi_3^2 + \hphi_4^2)
			\\
			&+ 480 \hphi_3^2 \hphi_4^2
			+ 576 \hphi_1 \hphi_2 \hphi_3 \hphi_4
		\end{aligned}
		\label{sm.eq:energy.16.4}
	\end{equation*}
	}
	\\
	\hline
	HC phase diagrams \rule[45pt]{100pt}{0pt} & 
	\includegraphics[scale=0.16]{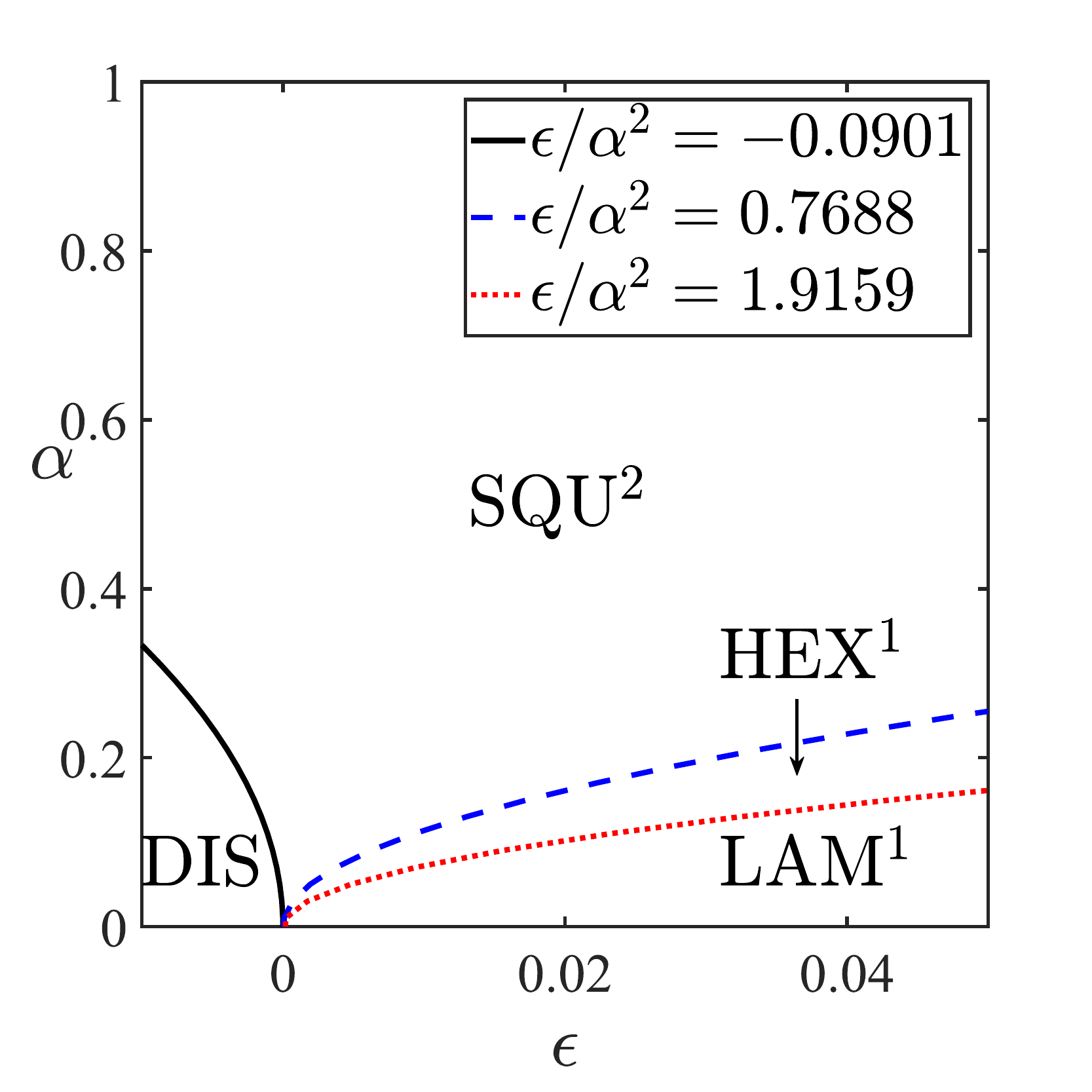} & 
	\includegraphics[scale=0.16]{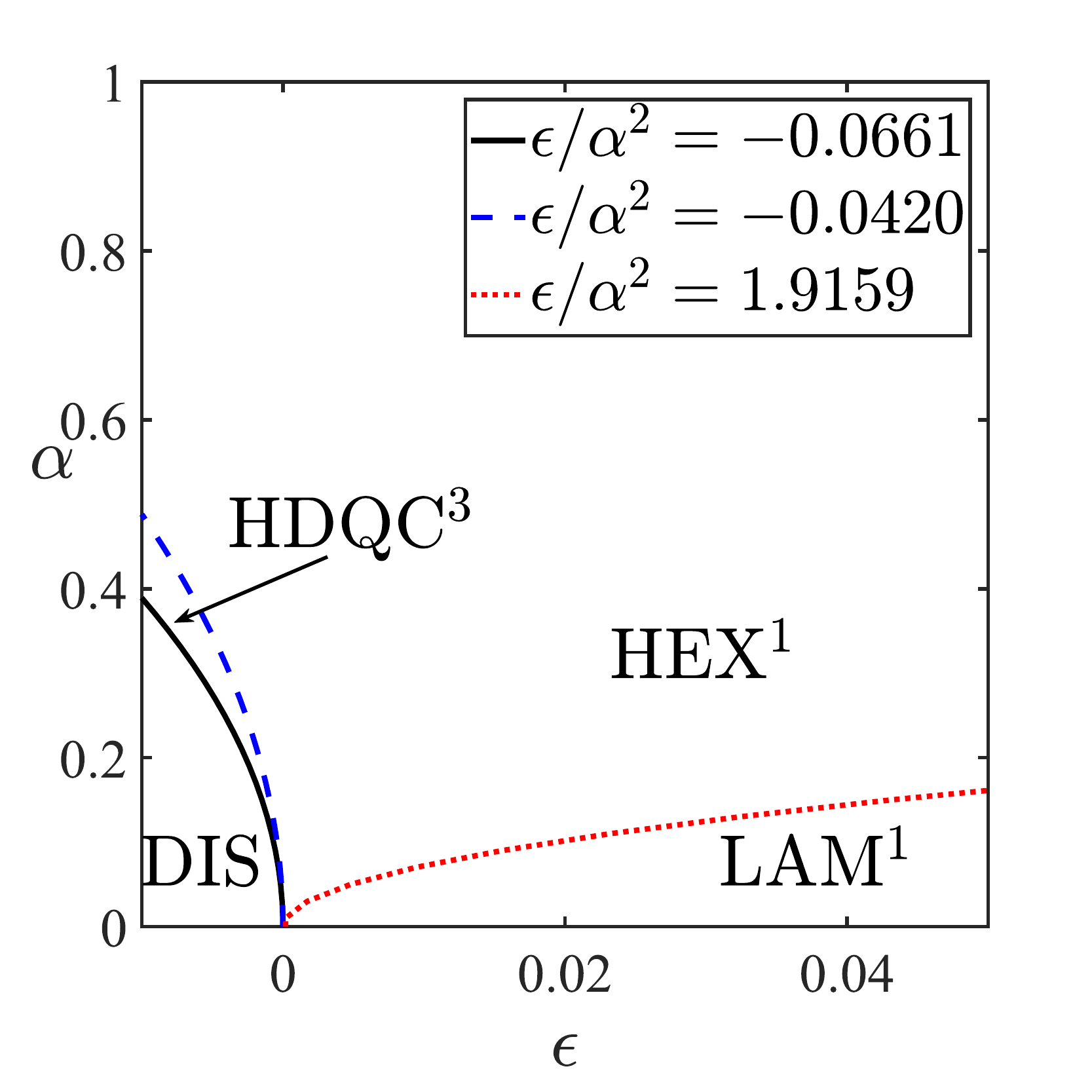} & 
	\includegraphics[scale=0.16]{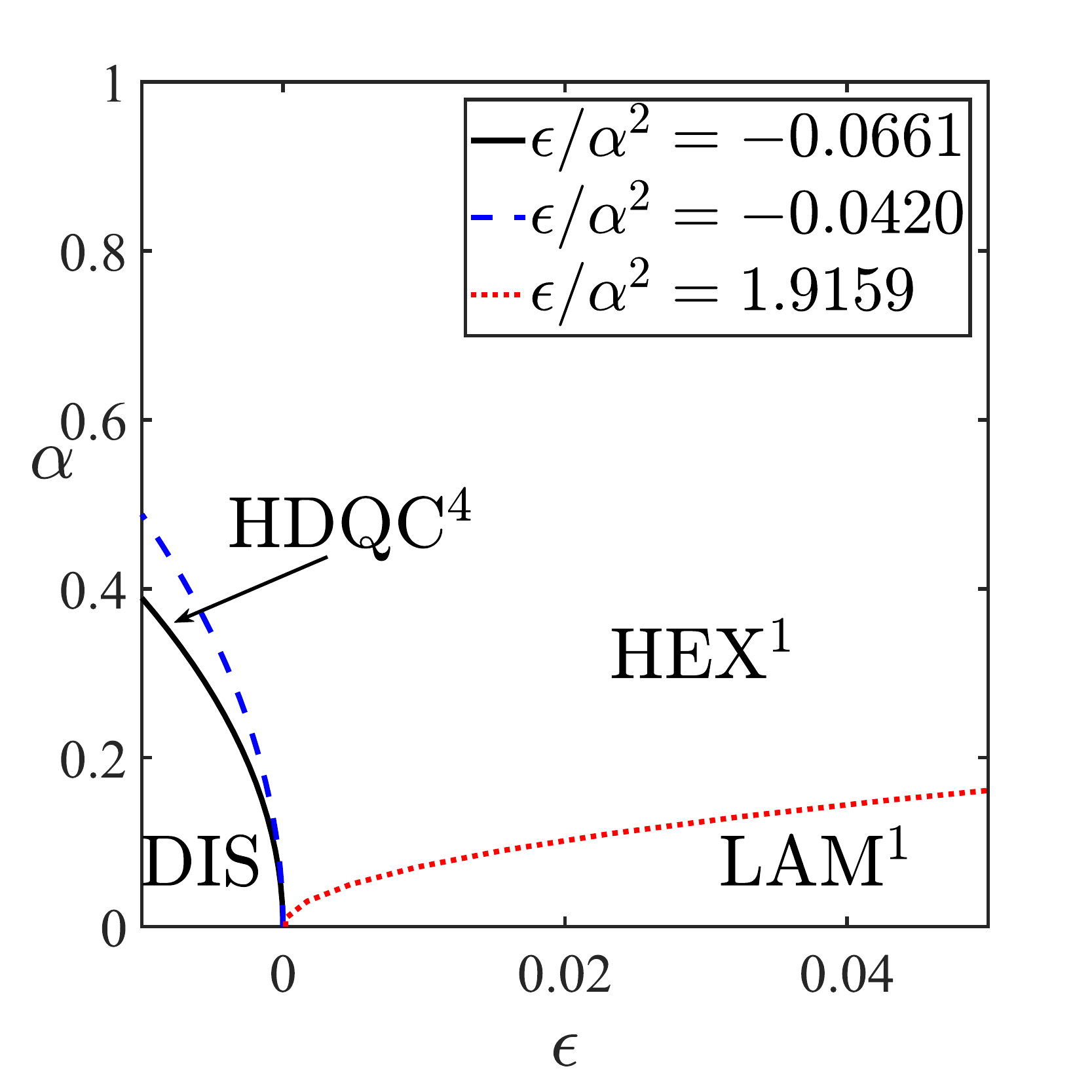} \\
	\hline
	\multicolumn{4}{c}{} \\
	\multicolumn{4}{c}{} \\
	\hline
	Desired QCs & \multicolumn{3}{c|}{\textbf{Octadecagonal (OD) QCs} ($18$-fold)} \\
	\hline
	$\{w_j\}_{j=1}^{m}$ & 3, 6 & 1, 4, 6; 2, 5, 6; 6, 7, 8 & 
	1, 2, 5, 6; 1, 4, 6, 7; 5, 6, 7, 8 \\
	\hline                                                         
	$\{s_j\}_{j=1}^{m}$ & 0, 1 & 0, 1, 1; 1, 0, 1; 1, 0, 1 & 
	0, 1, 0, 1; 0, 1, 1, 0; 0, 1, 0, 1 \\
	\hline
	Optimal primary RLVs \rule[165pt]{100pt}{0pt} &
	\includegraphics[scale=0.10]{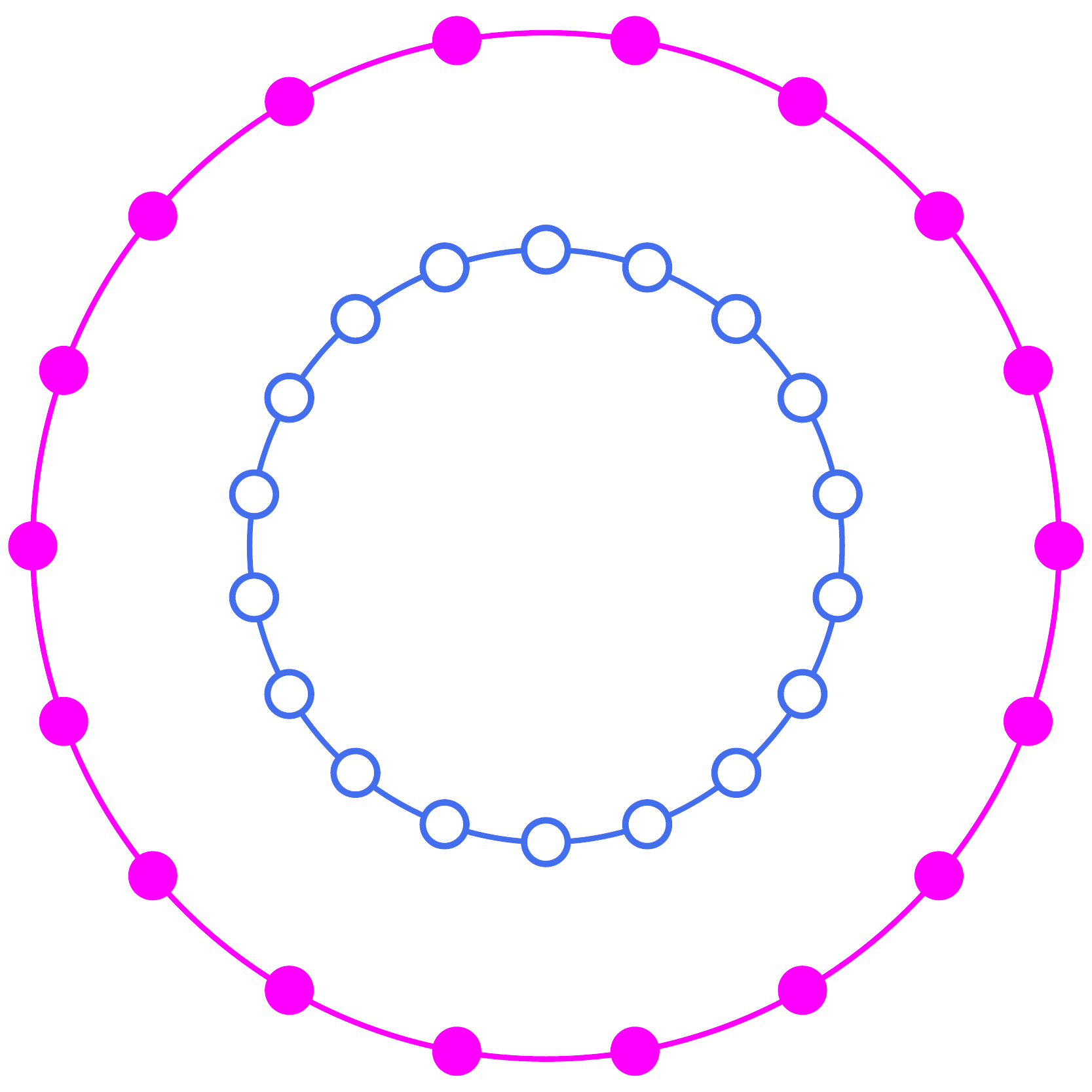}
	\centerline{$q: \cos\frac{\pi}{6}, \frac{1}{2}$}
	\centerline{$\theta: 0, \pi/18$}
	\rule[110pt]{100pt}{0pt}
	& 
	\includegraphics[scale=0.10]{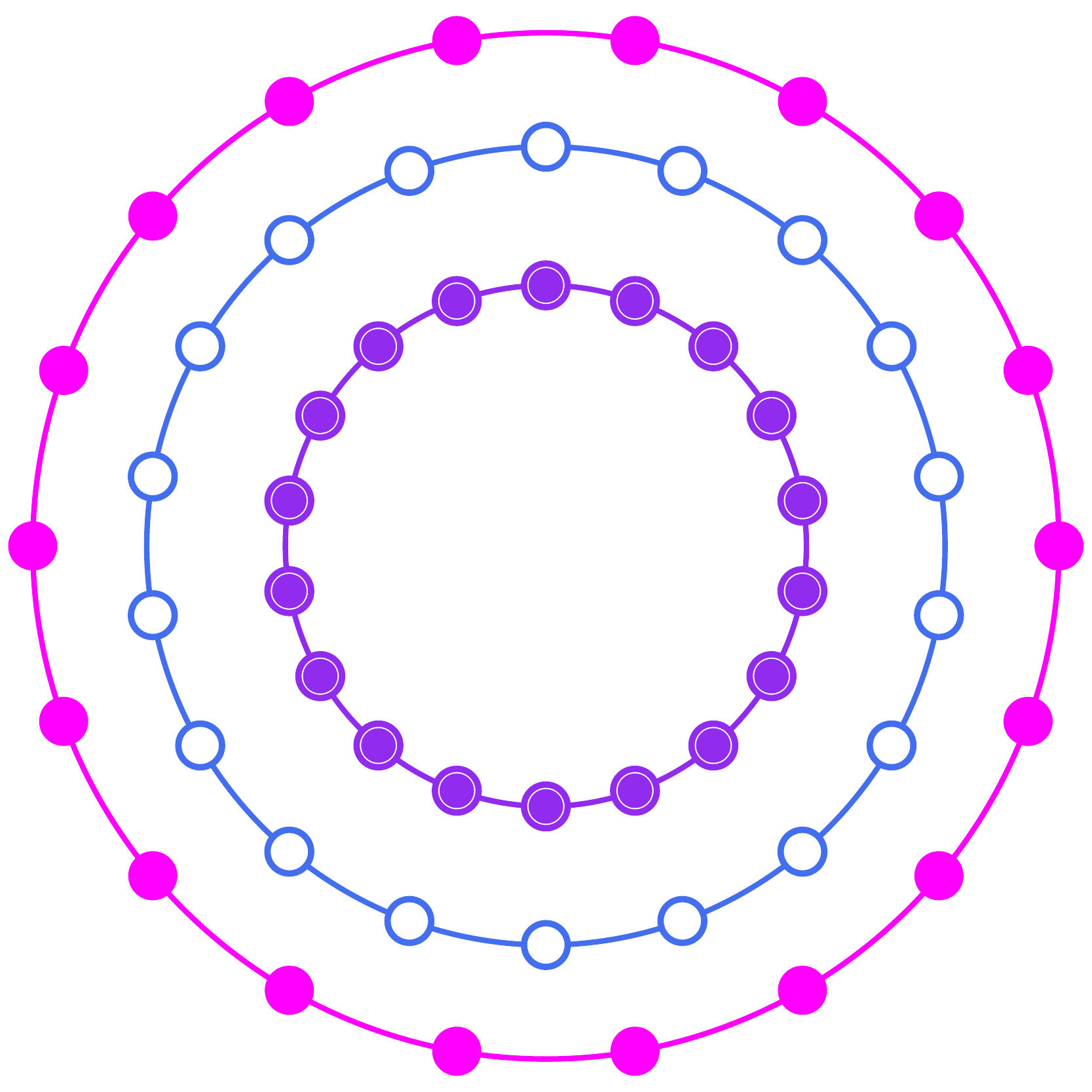}
	\centerline{$q: \cos\frac{\pi}{18}, \cos\frac{2\pi}{9}, \frac{1}{2}$}
	\centerline{$\theta: 0, \pi/18, \pi/18$}
	\includegraphics[scale=0.10]{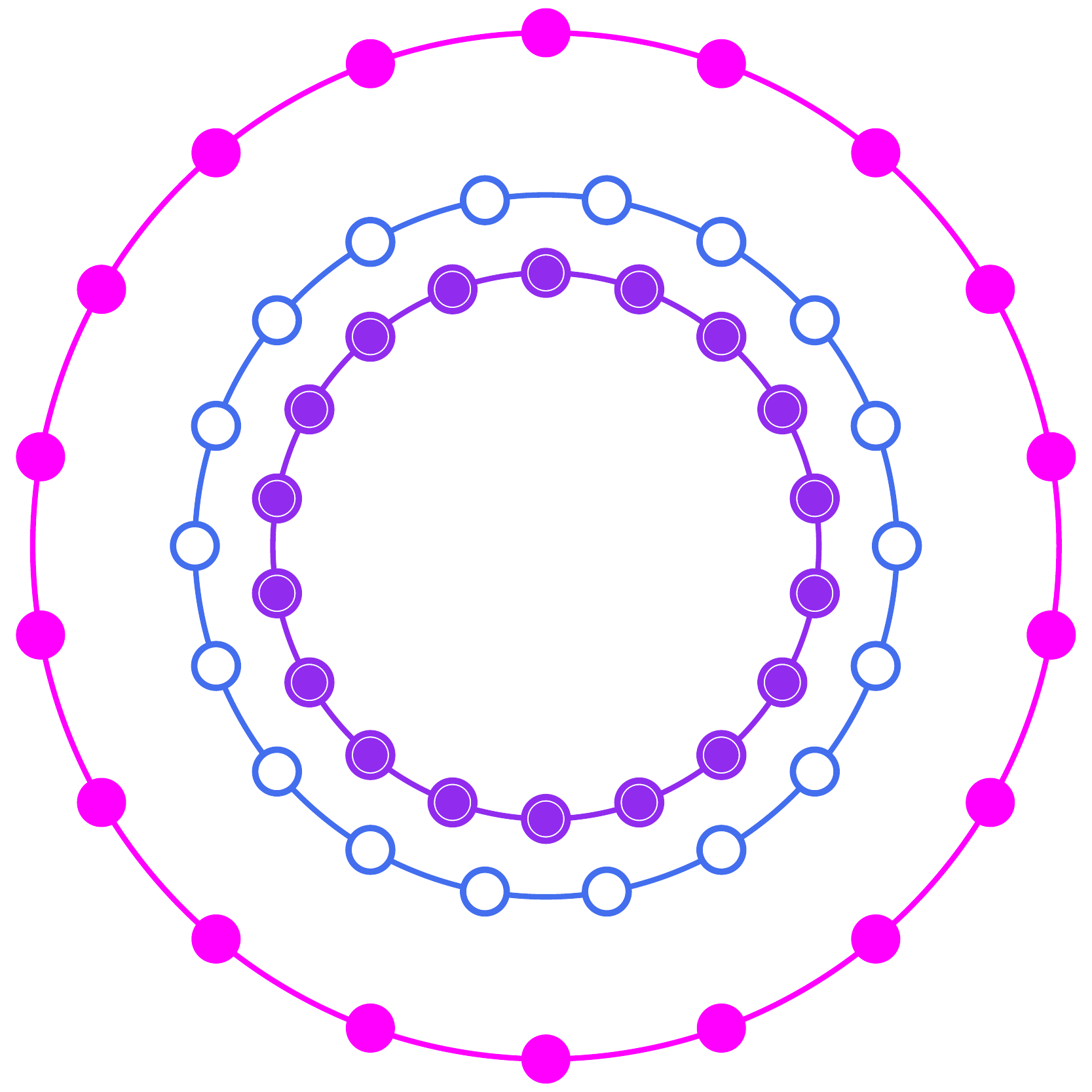}
	\centerline{$q: \cos\frac{\pi}{9}, \cos\frac{5\pi}{18}, \frac{1}{2}$}
	\centerline{$\theta: \pi/18, 0, \pi/18$}
	\includegraphics[scale=0.10]{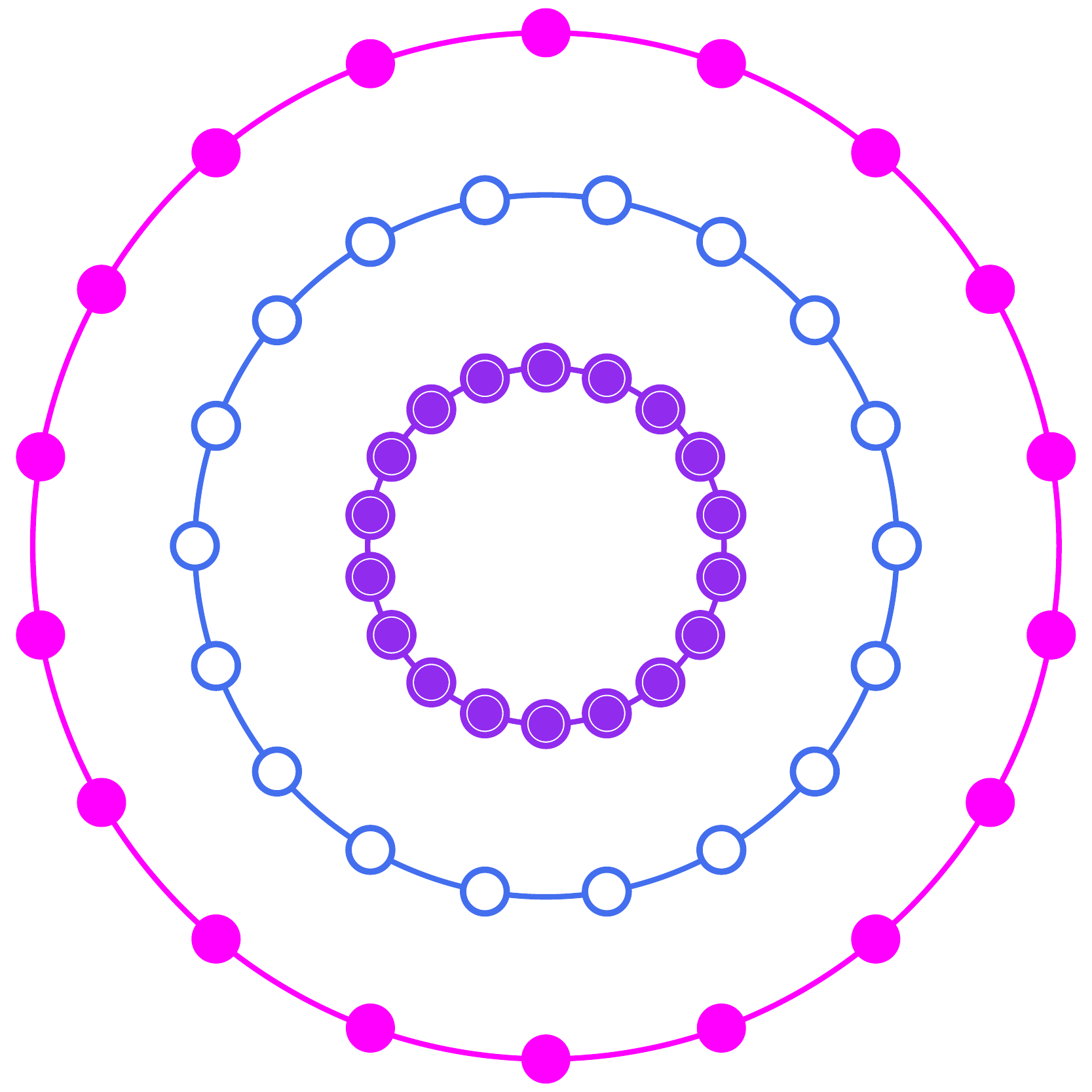}
	\centerline{$q: \frac{1}{2}, \cos\frac{7\pi}{18}, \cos\frac{4\pi}{9}$}
	\centerline{$\theta: \pi/18, 0, \pi/18$}
	& 
	\includegraphics[scale=0.10]{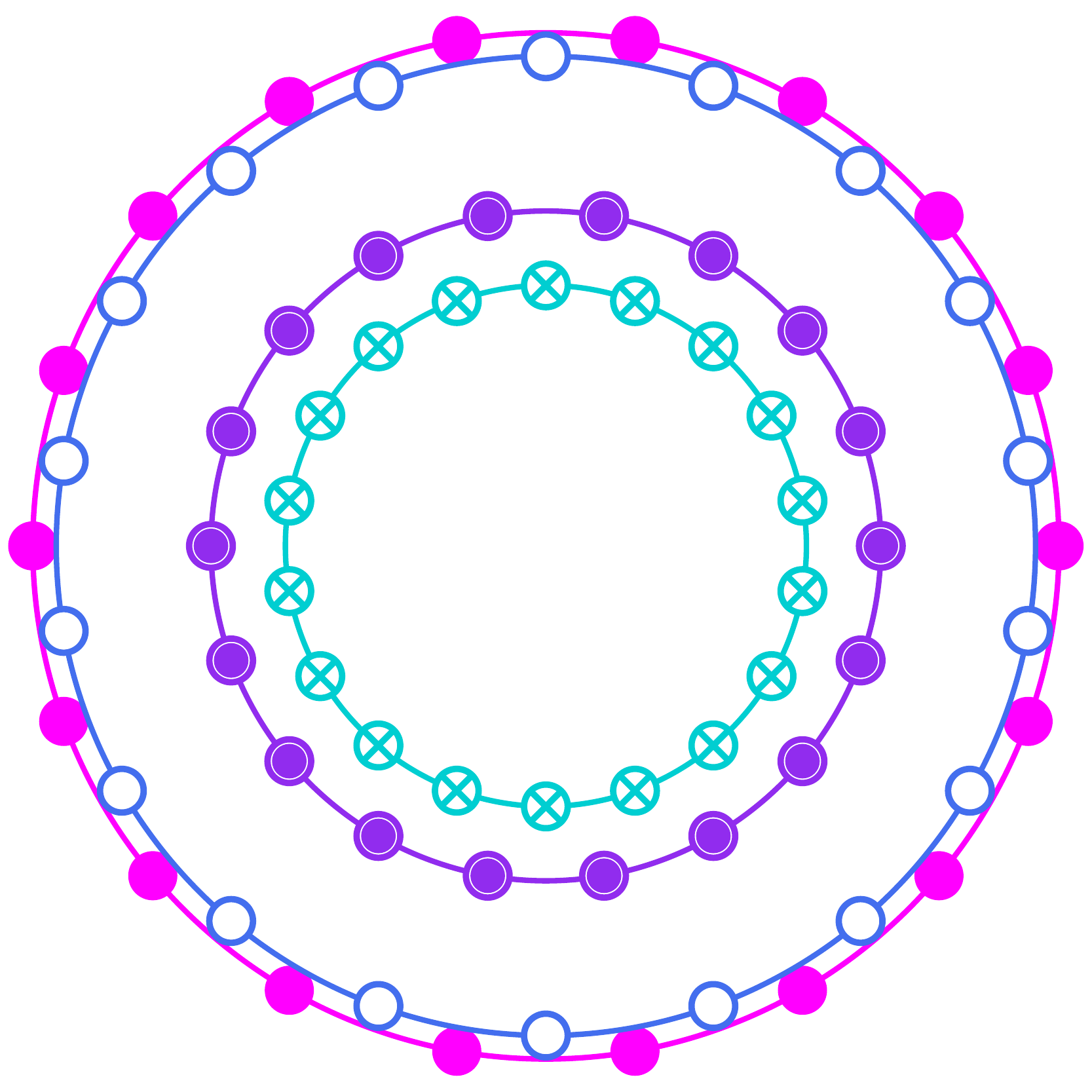}
	\centerline{$q: \cos\frac{\pi}{18}, \cos\frac{\pi}{9}, 
	\cos\frac{5\pi}{18}, \frac{1}{2}$}
	\centerline{$\theta: 0, \pi/18, 0, \pi/18$}
	\includegraphics[scale=0.10]{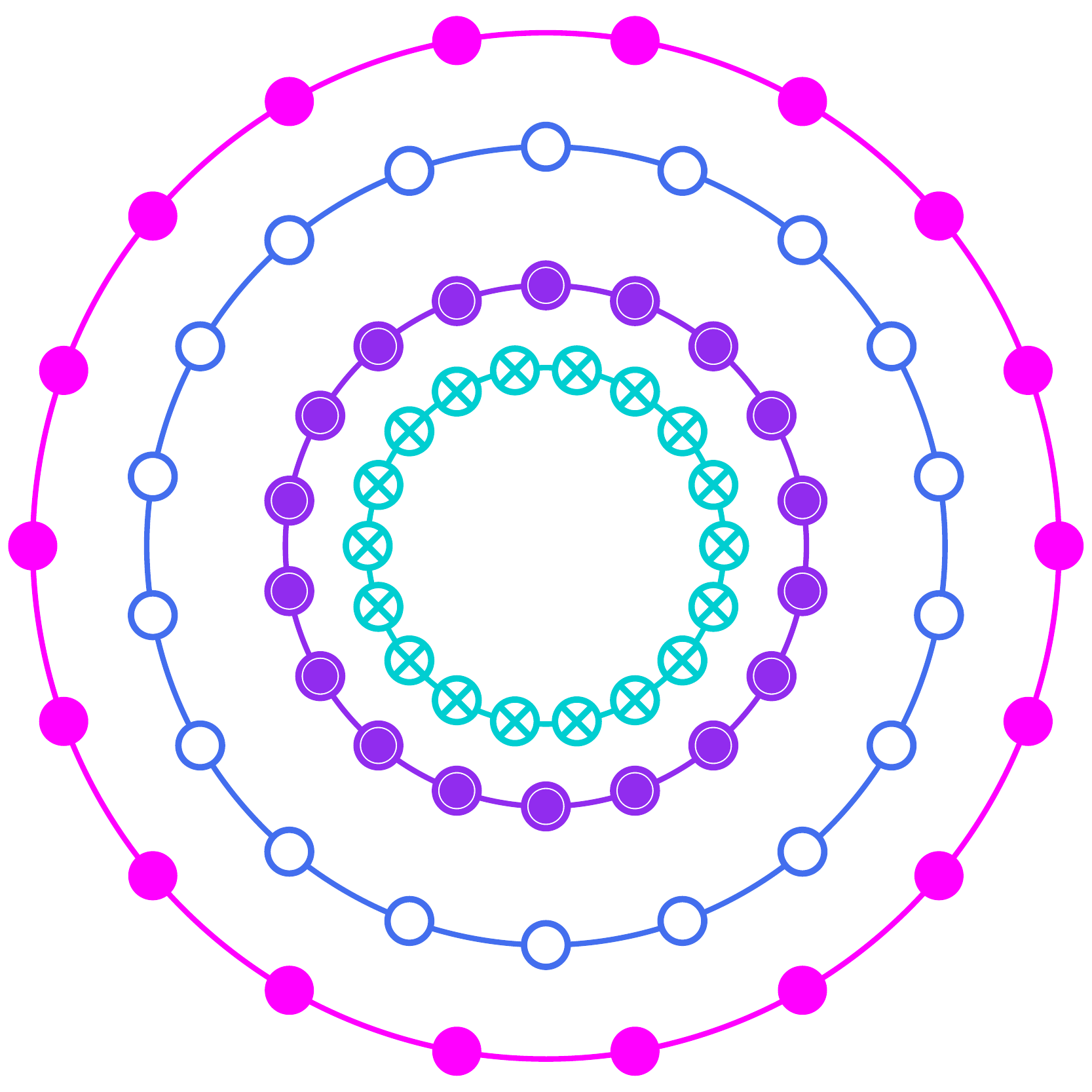}
	\centerline{$q: \cos\frac{\pi}{18}, \cos\frac{2\pi}{9}, 
	\frac{1}{2}, \cos\frac{7\pi}{18}$}
	\centerline{$\theta: 0, \pi/18, \pi/18, 0$}
	\includegraphics[scale=0.10]{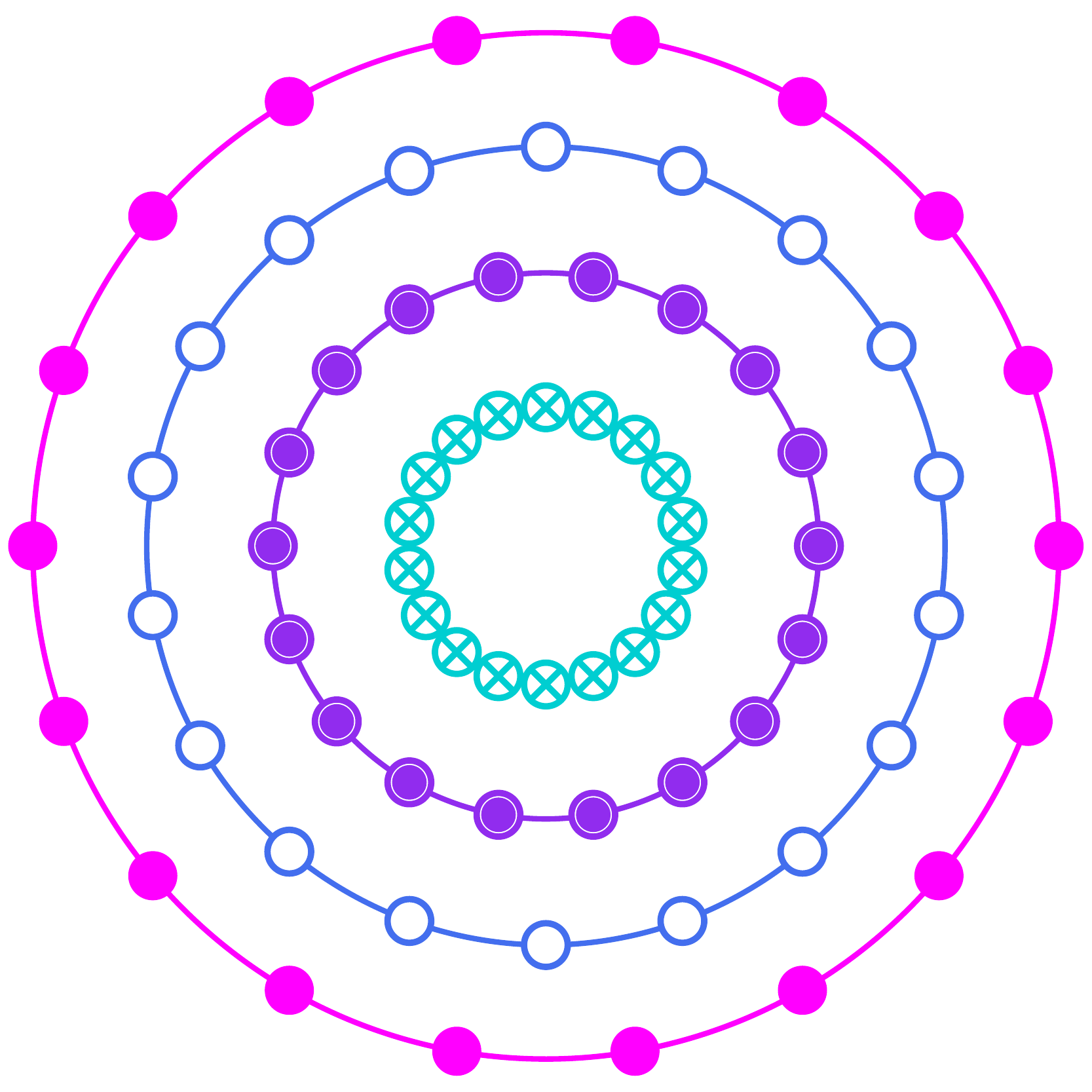}
	\centerline{$q: \cos\frac{5\pi}{18}, \frac{1}{2}, 
	\cos\frac{7\pi}{18}, \cos\frac{4\pi}{9}$}
	\centerline{$\theta: 0, \pi/18, 0, \pi/18$}
	\\
	\hline
	HC free energies \rule[85pt]{100pt}{0pt} &
	{
	\begin{equation*}
		\begin{aligned}
			&-9\epsilon^{*} \sum_{j=1}^{2}\hphi_{j}^{2}
			- 12 \sum_{j=1}^{2}\hphi_{j}^{3} 
			\\
			&- 36 \hphi_{1} \hphi_{2}^{2}
			+ \frac{459}{2} \sum_{j=1}^{2} \hphi_{j}^{4}
			\\
			&+ 108 \hphi_{1} \hphi_{2}^{3}
			+ 594 \hphi_{1}^{2} \hphi_{2}^{2}
		\end{aligned}
		\label{sm.eq:energy.18.2}
	\end{equation*}
	}
	\rule[55pt]{100pt}{0pt}
	&
	{
	\begin{equation*}
		\begin{aligned}
			&-9\epsilon^{*} \sum_{j=1}^{3}\hphi_{j}^{2}
			- 12 \sum_{j=1}^{3}\hphi_{j}^{3} 
			\\
			&- 36 \hphi_{1} \sum_{j=2}^{3}\hphi_{j}^{2}
			- 36 \hphi_{2}\hphi_{3} (2\hphi_{1}+\hphi_{3})
			\\
			&+ \frac{459}{2}\sum_{j=1}^{3}\hphi_{j}^{4} 
			+ 108 \hphi_2^3 (2\hphi_1 + \hphi_3)
			\\
			&+ 216 \hphi_3^3 (\hphi_1 + \hphi_2)
			+ 702 \hphi_2^2 \hphi_3^2
			\\
			&+ 702 \hphi_1^2 (\hphi_2^2 + \hphi_3^2)
			+ 756 \hphi_1^2 \hphi_2 \hphi_3
			\\
			&+ 432 \hphi_1 \hphi_2 \hphi_3 (\hphi_2 + \hphi_3)
		\end{aligned}
		\label{sm.eq:energy.18.3}
	\end{equation*}
	}
	\rule[25pt]{100pt}{0pt}
	&
	{
	\begin{equation*}
		\begin{aligned}
			&-9\epsilon^{*} \sum_{j=1}^{4}\hphi_{j}^{2}
			- 12 \sum_{j=1}^{4}\hphi_{j}^{3} 
			- 36 \hphi_1 \sum_{j=3}^{4} \hphi_j^2
			\\
			&- 36 \hphi_2^2 \hphi_3
			- 36 \hphi_4^2 (\hphi_2 + \hphi_3)
			\\
			&- 72 \hphi_2 \hphi_4 (\hphi_1 + \hphi_3)
			+ \frac{459}{2}\sum_{j=1}^{4}\hphi_{j}^{4} 
			\\
			&+ 108 \hphi_1 \hphi_3 (\hphi_1^2 + 2\hphi_3^2)
			+ 216 \hphi_4^3 \sum_{j=1}^{3} \hphi_j
			\\
			&+ 108 \hphi_2^3 (2\hphi_3 + \hphi_4)
			+ 324 \hphi_1 \hphi_2^2 \hphi_3
			\\
			&+ 432 \hphi_1 \hphi_2 \hphi_4 (\hphi_1 + \hphi_2 + \hphi_4)
			\\
			&+ 432 \hphi_2 \hphi_3 \hphi_4 (\hphi_2 + \hphi_4)
			+ 594 \hphi_1^2 \hphi_2^2
			\\
			&+ 702 (\hphi_1^2 + \hphi_2^2) (\hphi_3^2 + \hphi_4^2)
			\\
			&+ 702 \hphi_3^2 \hphi_4^2
			+ 756 \hphi_2 \hphi_3^2 \hphi_4
			\\
			&+ 540 \hphi_1 \hphi_3 \hphi_4^2
			+ 1296 \hphi_1 \hphi_2 \hphi_3 \hphi_4
		\end{aligned}
		\label{sm.eq:energy.18.4}
	\end{equation*}
	}
	\\
	\hline
	HC phase diagrams \rule[45pt]{100pt}{0pt} & 
	\includegraphics[scale=0.16]{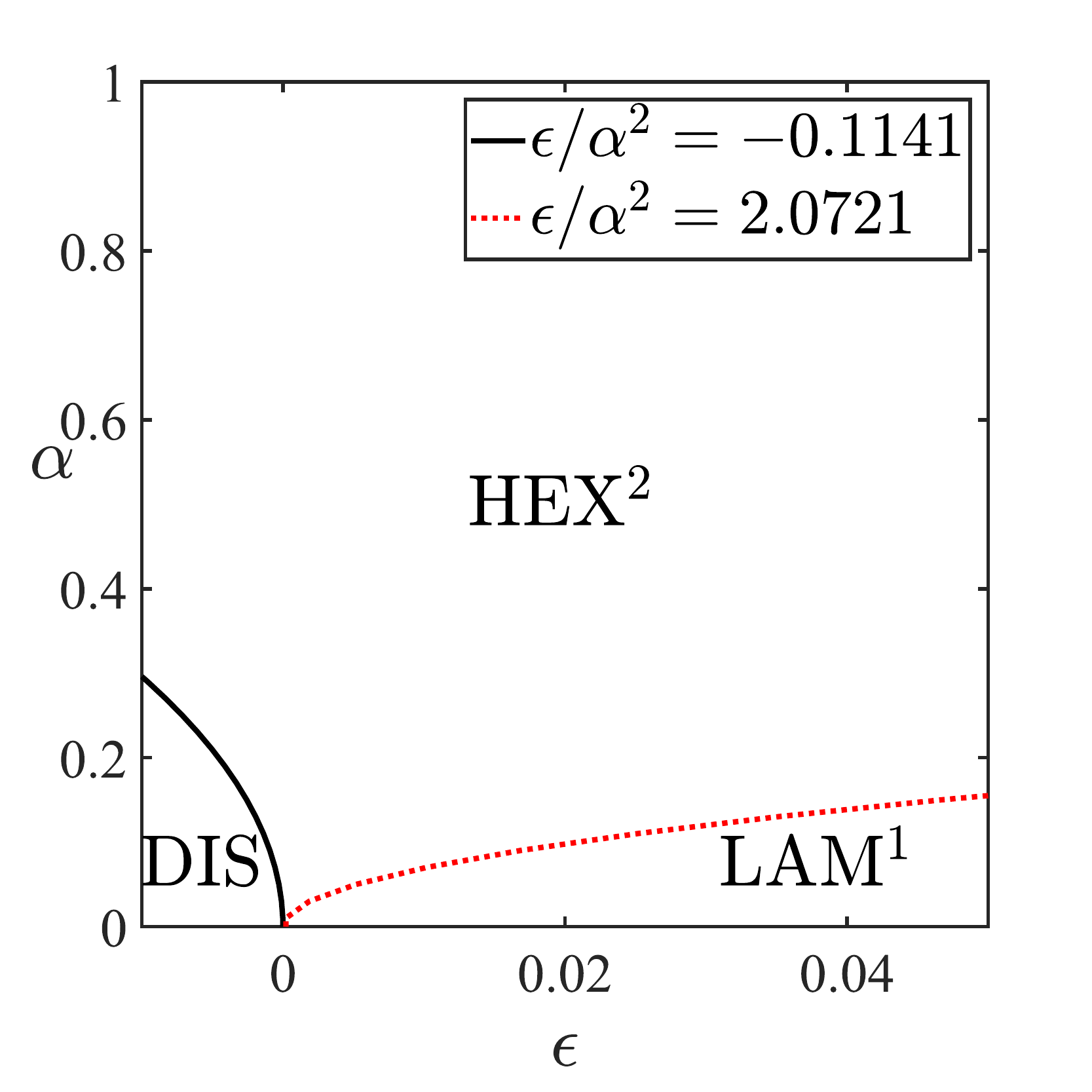} & 
	\includegraphics[scale=0.16]{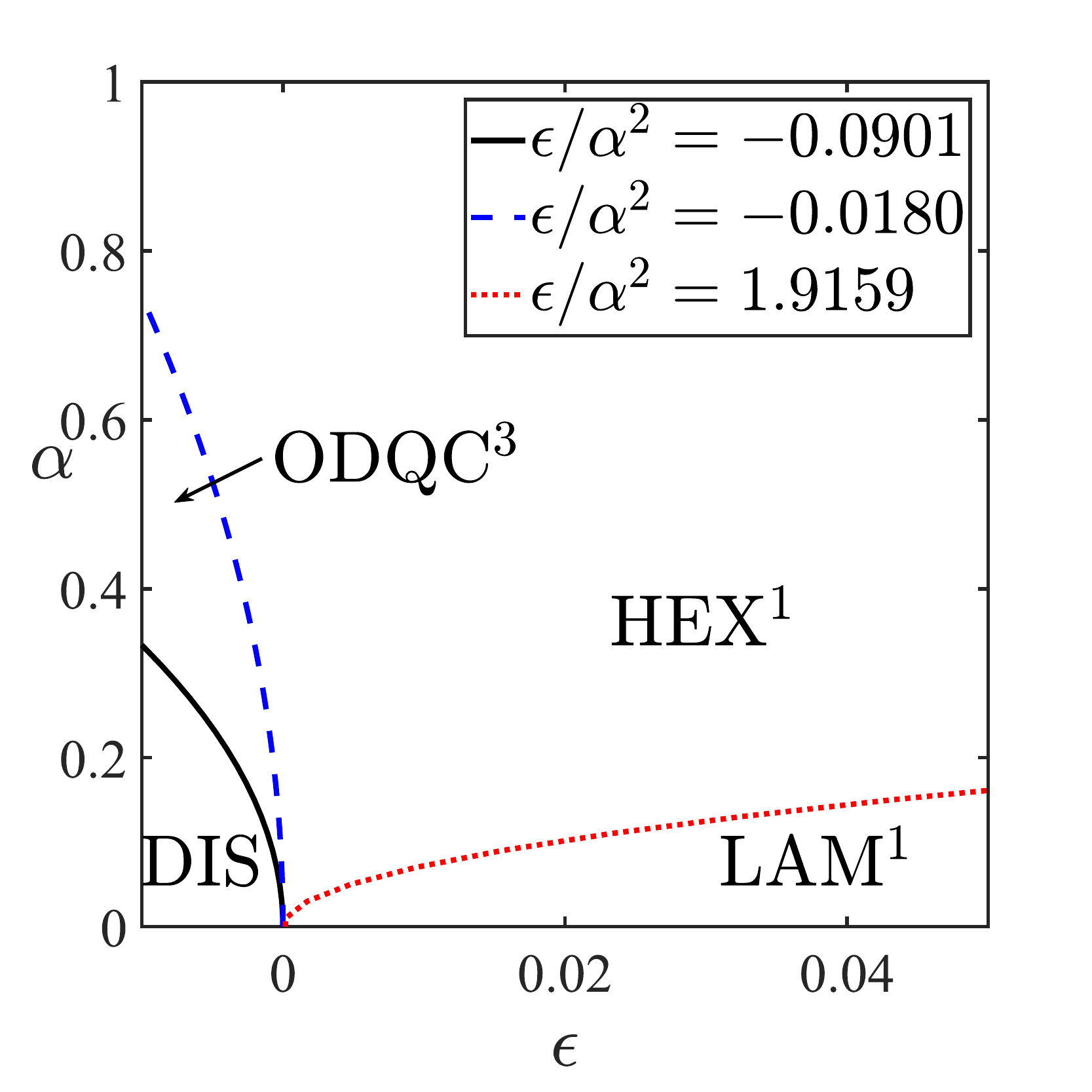} & 
	\includegraphics[scale=0.16]{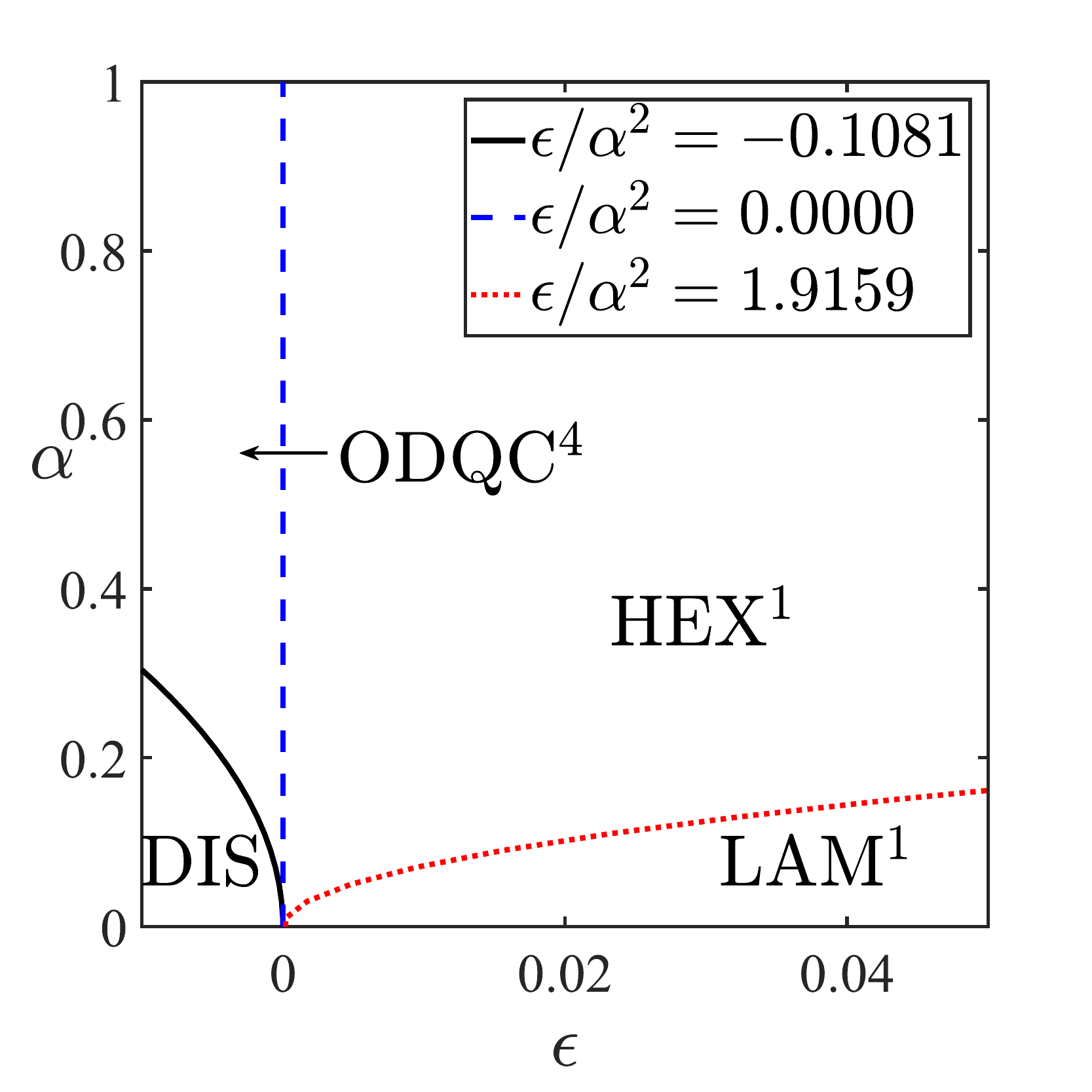} \\
	\hline
\end{longtable*}

\begin{figure*}[htbp]
	\centering
	\subfigure[OQC ($8$-fold)]{\label{sm.fig:high.8}
		\includegraphics[scale=0.25]{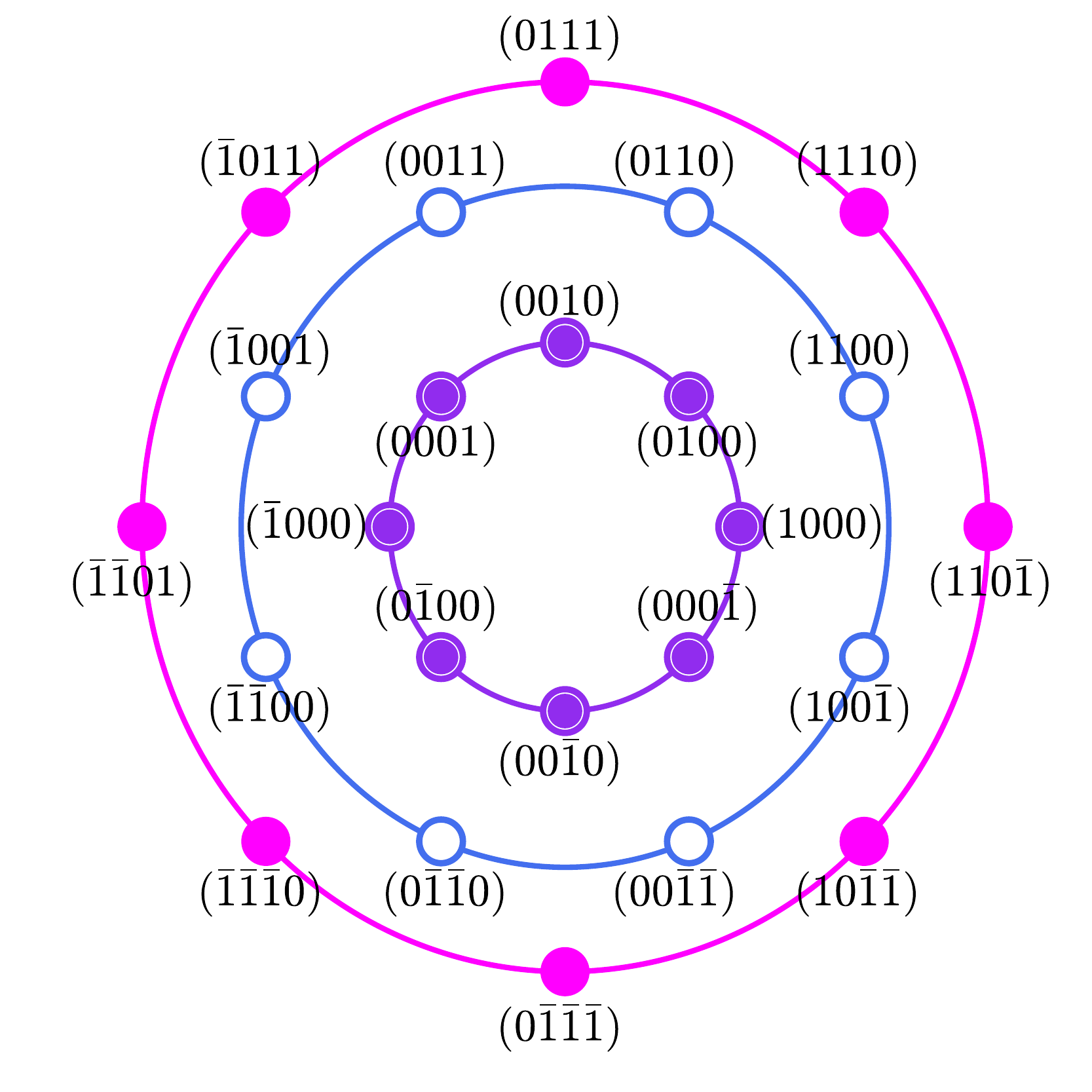}}
	\subfigure[DQC ($10$-fold)]{\label{sm.fig:high.10}
		\includegraphics[scale=0.25]{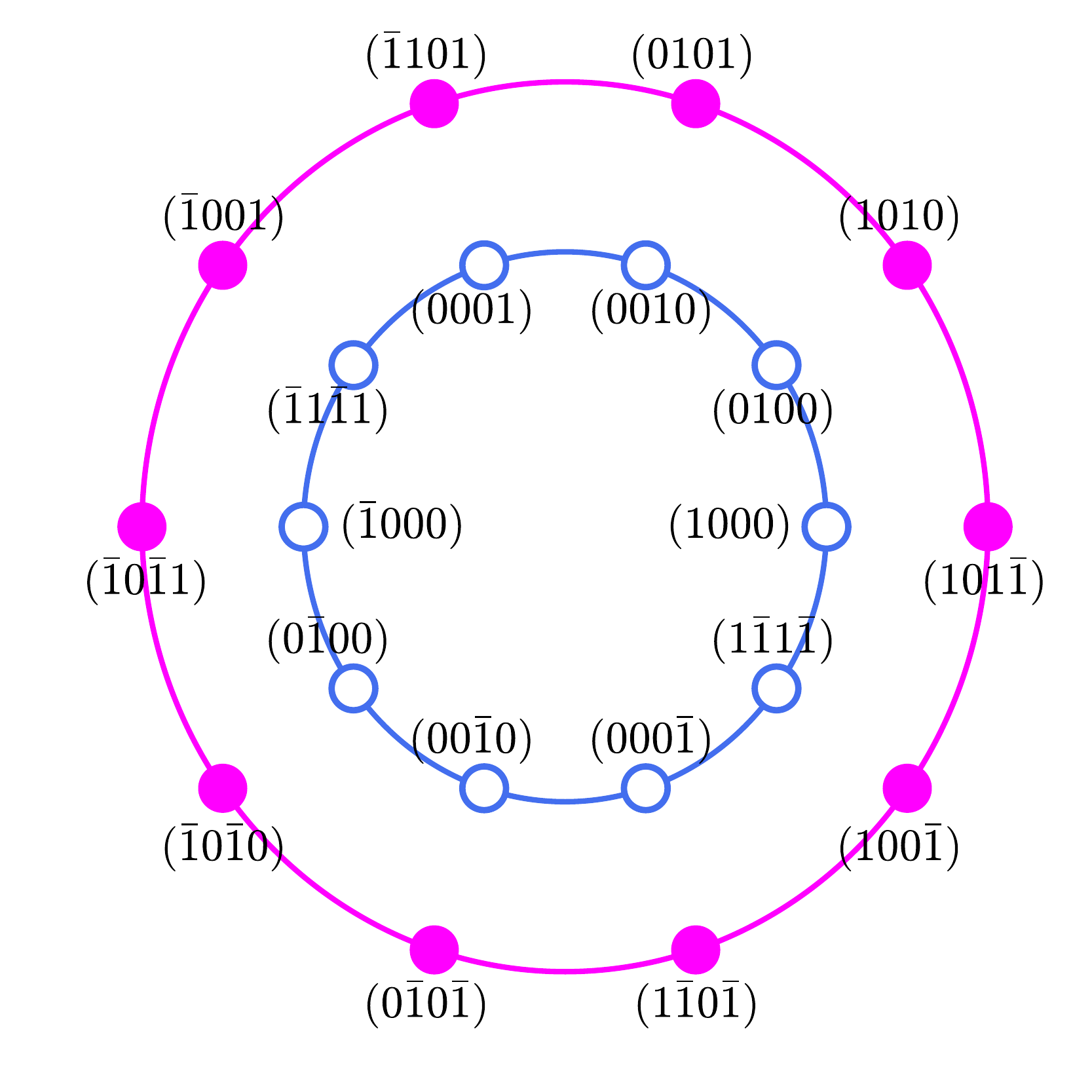}}
	\vfill
	\subfigure[DDQC ($12$-fold)]{\label{sm.fig:high.12}
		\includegraphics[scale=0.25]{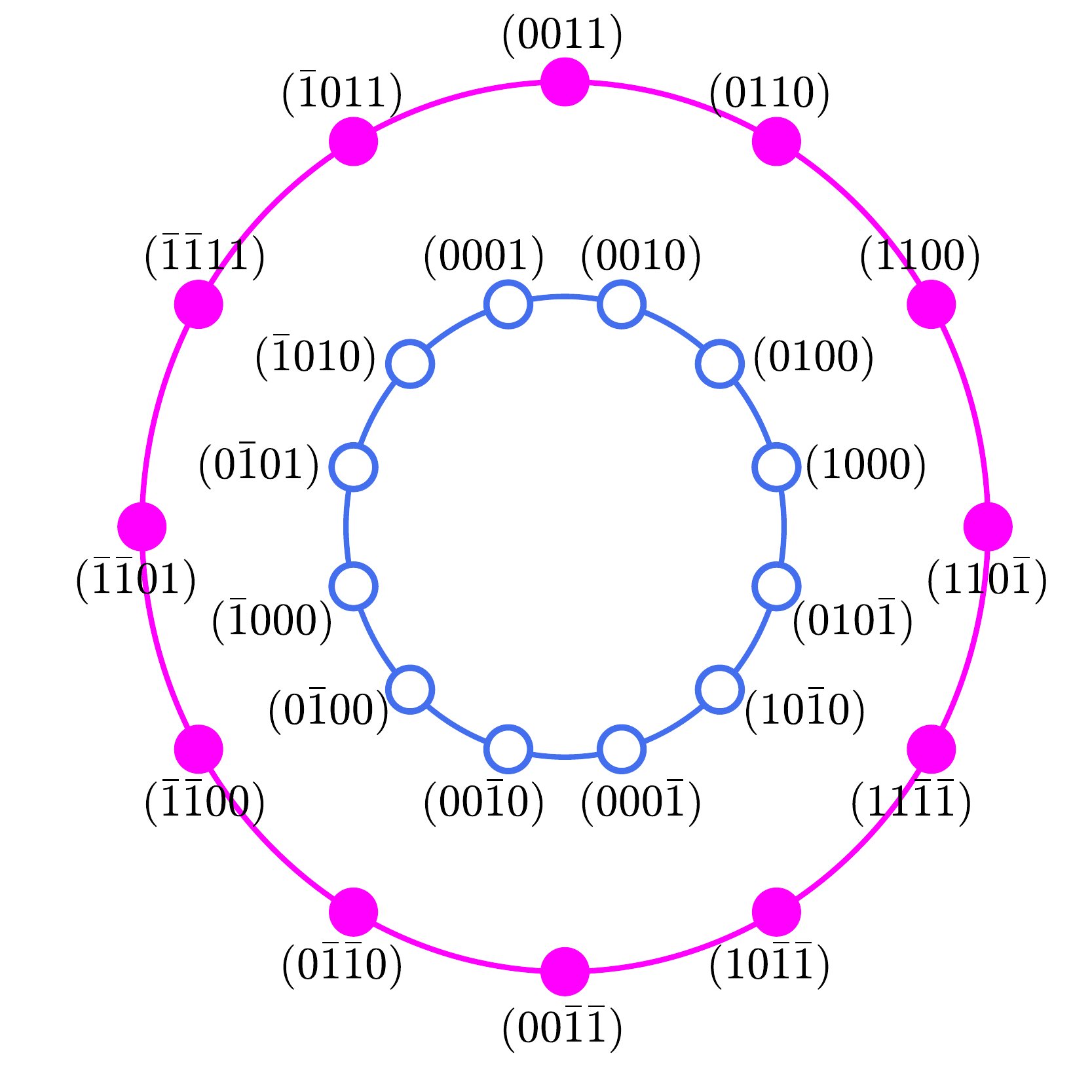}}
	\subfigure[TDQC ($14$-fold)]{\label{sm.fig:high.14}
		\includegraphics[scale=0.25]{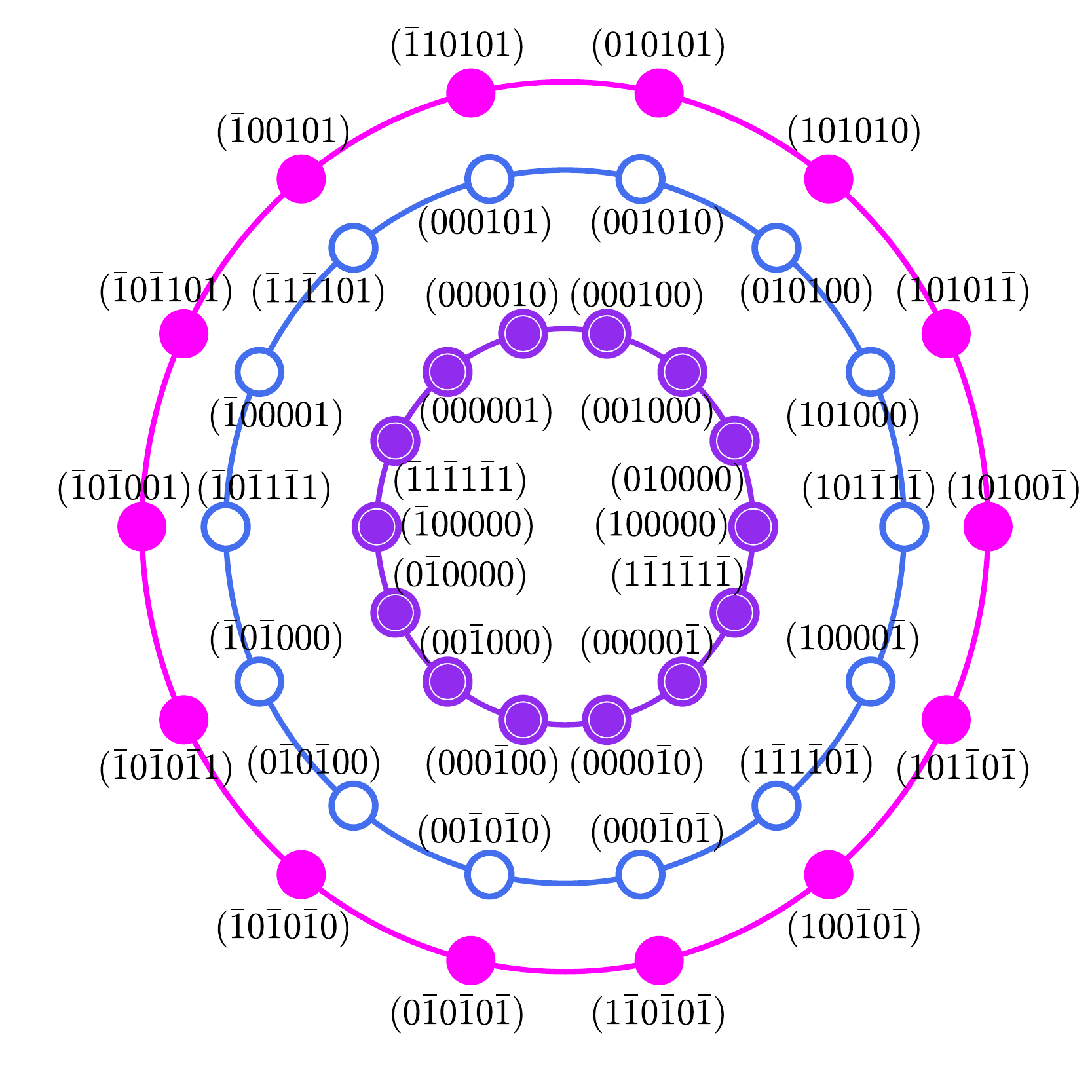}}
	\vfill
	\subfigure[HDQC ($16$-fold)]{\label{sm.fig:high.16}
		\includegraphics[scale=0.25]{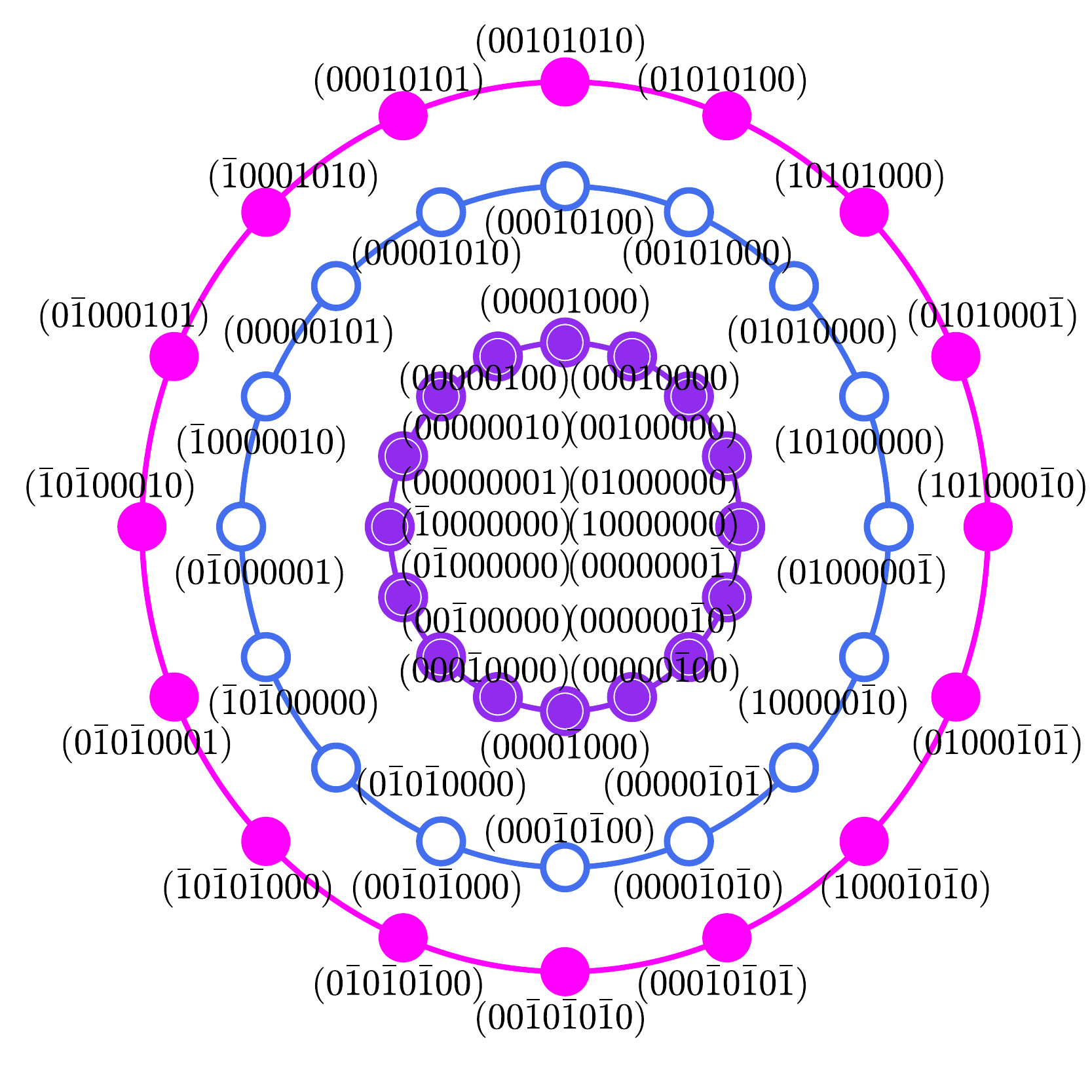}}
	\subfigure[ODQC ($18$-fold)]{\label{sm.fig:high.18}
		\includegraphics[scale=0.25]{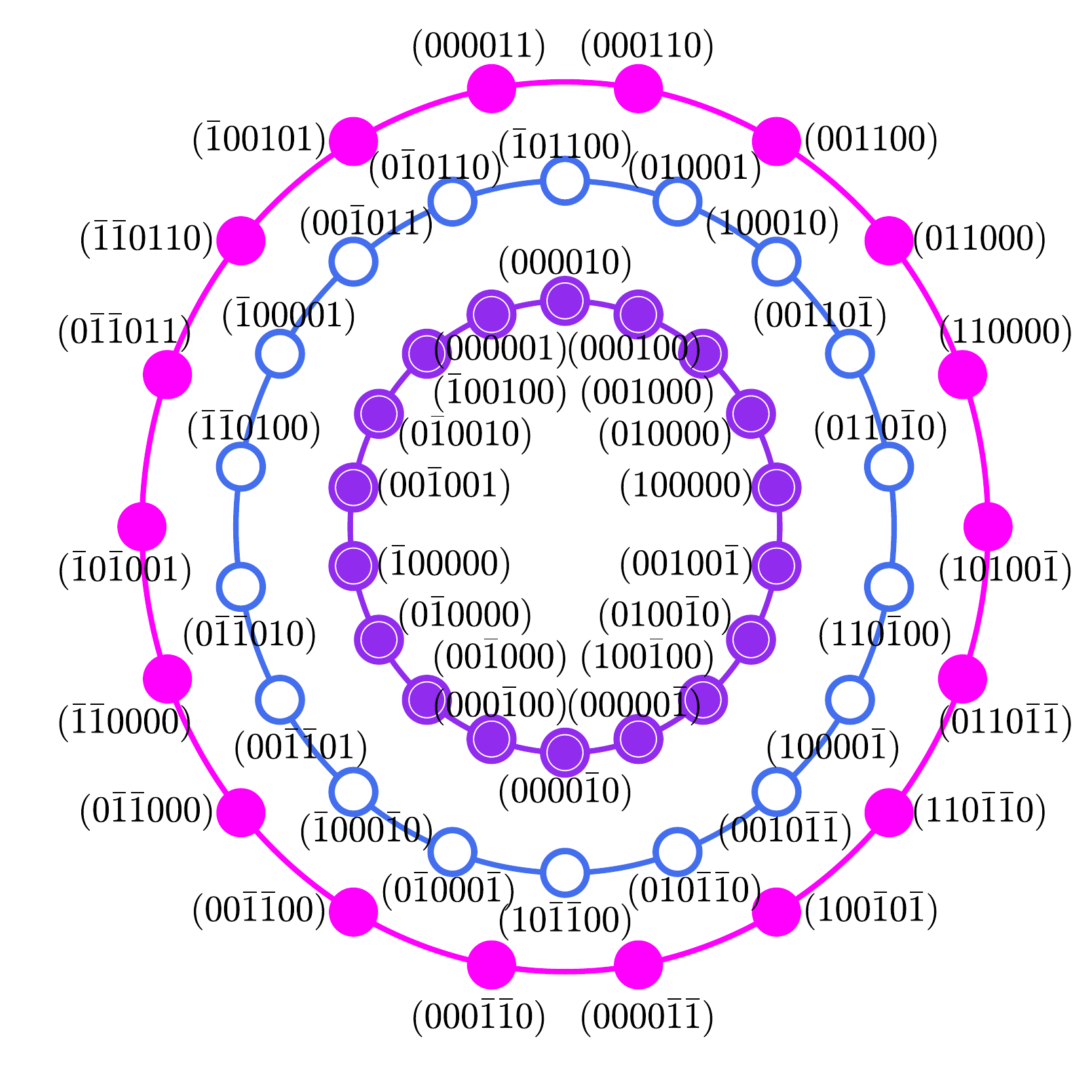}}
	\caption{\label{sm.fig:high}
		The high-dimensional coordinates of optimal primary RLVs of $2n$-fold
		QCs ($n=4,\ldots,9$).
	} 
\end{figure*}

\newpage
Under SC, we further evaluate $2n$-fold QCs and free energies accurately by the projection
method\,\cite{Jiang2014}.
To be consistent with the rescaling model
\begin{equation*}
	\begin{aligned}
		\mcF_{m}[\phi(\br)]
		& = \bbint\left( -\frac{\epsilon}{2}\phi^{2} 
			- \frac{\alpha}{3} \phi^{3} + \frac{1}{4}\phi^{4} \right)\,d\br
		\\ & ~~~~
		+ \frac{1}{2} \bbint \Big[ \prod_{j=1}^{m}(\nabla^{2}+q_{j}^{2}/q_*^2) 
			\phi(\br) \Big]^{2}\,d\br,
	\end{aligned}
\end{equation*}
we measure the length scales of $2n$-fold QCs in units of $q_*$.
$q_*$ takes the minimal value of $\{q_j\}_{j=1}^{m}$.
Concretely, $q_*$ equals to $\cos(3\pi/8)$ for OQC and HDQC, $\cos(3\pi/10)$ for DQC,
$1/2$ for DDQC and ODQC, $\cos(5\pi/14)$ for TDQC.
These projection matrices are
\begin{equation}
	\mcP_{OQC} = 
	\begin{pmatrix}
		1 & \cos\frac{\pi}{4} & 0 & \cos\frac{3\pi}{4} \\
		0 & \sin\frac{\pi}{4} & 1 & \sin\frac{3\pi}{4}
	\end{pmatrix},
	\label{sm.eq:pm.8}
\end{equation}
\begin{equation}
	\mcP_{DQC} =
	\begin{pmatrix}
		1 & \cos\frac{\pi}{5} & \cos\frac{2\pi}{5} & \cos\frac{3\pi}{5} \\
		0 & \sin\frac{\pi}{5} & \sin\frac{2\pi}{5} & \sin\frac{3\pi}{5}
	\end{pmatrix},
	\label{sm.eq:pm.10}
\end{equation}
\begin{equation}
	\mcP_{DDQC} =
	\begin{pmatrix}
		1 & \cos\frac{\pi}{6} & \cos\frac{\pi}{3} & 0 \\
		0 & \sin\frac{\pi}{6} & \sin\frac{\pi}{3} & 1
	\end{pmatrix},
	\label{sm.eq:pm.12}
\end{equation}
\begin{equation}
	\mcP_{TDQC} =
	\begin{pmatrix}
		1 & \cos\frac{\pi}{7} & \cos\frac{2\pi}{7} & \cos\frac{3\pi}{7} &
		\cos\frac{4\pi}{7} & \cos\frac{5\pi}{7} \\
		0 & \sin\frac{\pi}{7} & \sin\frac{2\pi}{7} & \sin\frac{3\pi}{7} &
		\sin\frac{4\pi}{7} & \sin\frac{5\pi}{7}
	\end{pmatrix},
	\label{sm.eq:pm.14}
\end{equation}
\begin{equation}
	\mcP_{HDQC} =
	\begin{pmatrix}
		1 & \cos\frac{\pi}{8} & \cos\frac{\pi}{4} & \cos\frac{3\pi}{8} & 0 
		& \cos\frac{5\pi}{8} & \cos\frac{3\pi}{4} & \cos\frac{7\pi}{8} \\
		0 & \sin\frac{\pi}{8} & \sin\frac{\pi}{4} & \sin\frac{3\pi}{8} & 1
		& \sin\frac{5\pi}{8} & \sin\frac{3\pi}{4} & \sin\frac{7\pi}{8}
	\end{pmatrix},
	\label{sm.eq:pm.16}
\end{equation}
\begin{equation}
	\mcP_{ODQC} = 
	\begin{pmatrix}
		1 & \cos\frac{\pi}{9} & \cos\frac{2\pi}{9} & 
		\cos\frac{\pi}{3} & \cos\frac{4\pi}{9} & 
		\cos\frac{5\pi}{9} \\
		0 & \sin\frac{\pi}{9} & \sin\frac{2\pi}{9} & 
		\sin\frac{\pi}{3} & \sin\frac{4\pi}{9} & 
		\sin\frac{5\pi}{9}
	\end{pmatrix}.
	\label{sm.eq:pm.18}
\end{equation}
In these projection matrices, the number of columns corresponds to the
dimensionality of the high-dimensional space.
\Cref{sm.fig:high} presents the corresponding high-dimensional coordinates of optimal primary RLVs. 
In \Cref{sm.tab:sc}, we show the stationary patterns and SC phase diagrams of $2n$-fold QCs.
The SC phase diagrams show that the HC minimal models can stabilize desired QCs under SC.

\newpage
\begin{longtable*}{|m{0.12\linewidth}<{\centering}|
	m{0.25\linewidth}<{\centering}|
	m{0.25\linewidth}<{\centering}|
	m{0.25\linewidth}<{\centering}|}
	\caption{
		\label{sm.tab:sc}
		The stationary patterns and SC phase diagrams of $2n$-fold QCs in the
		minimal Landau models designed by ISM ($n=4,\ldots,9$).
		We apply the projection method to compute these results.
		Set $\epsilon=-0.01$, $\alpha=0.4$ for $16$-fold QC and $\alpha=0.5$ for others.
		We show the diffraction patterns and morphologies of the stationary QCs.
		Diffraction patterns only plot the RLVs with intensities greater 
		than $10^{-6}$.
		In phase diagrams, we use superscripts to denote the number of 
		valid length scales.
	} \\*
	\endfirsthead
	\multicolumn{4}{c}{\Cref{sm.tab:sc} (continued): diffraction
	patterns and morphologies, SC phase diagrams}%
	\endhead
	\endfoot
	\endlastfoot
	\hline
	Desired QCs & SC diffraction patterns & 
	SC morphologies & SC phase diagrams \\
	\hline
	OQCs ($8$-fold) \rule[60pt]{100pt}{0pt} &
	\includegraphics[scale=0.14]{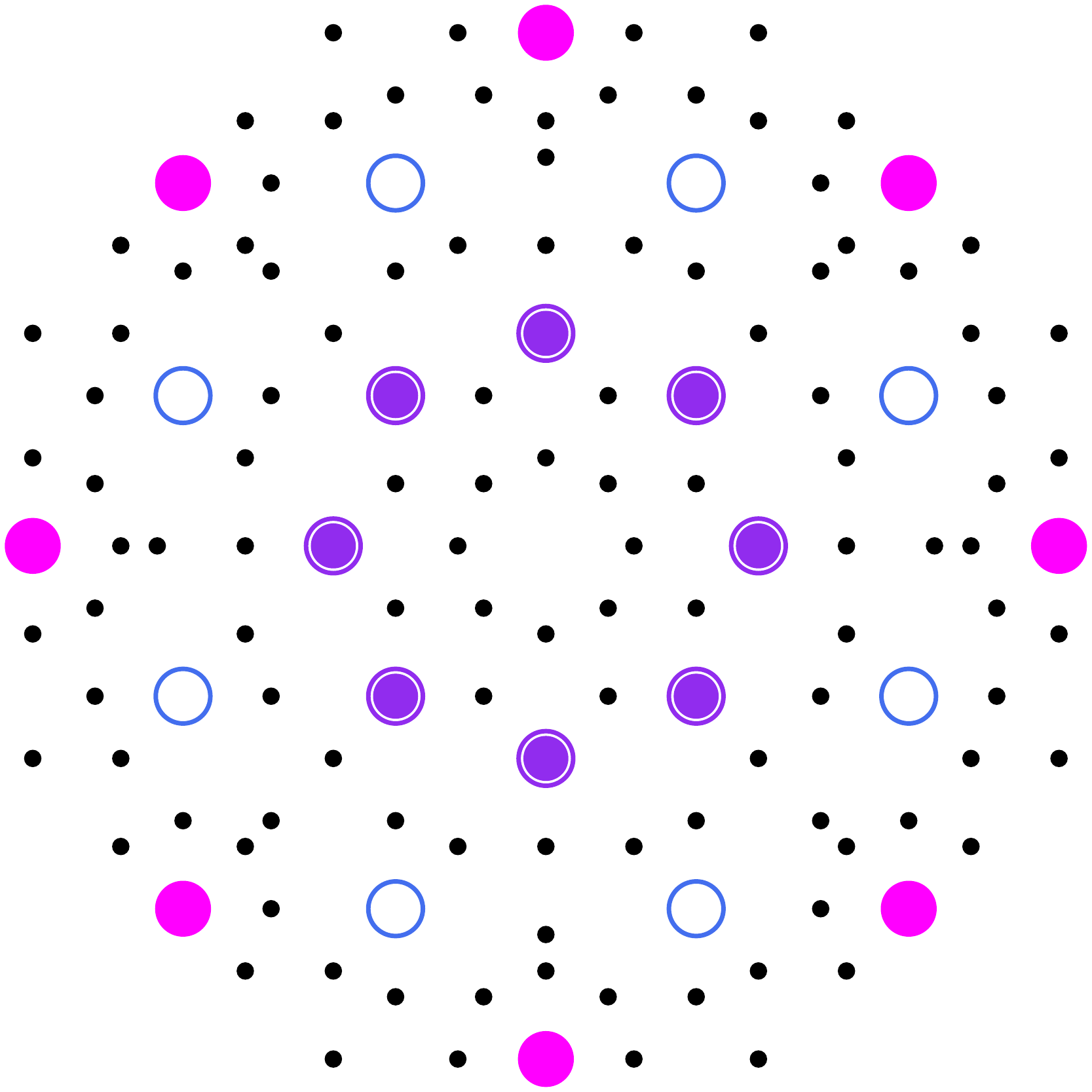}
	\centerline{$q: \cos\frac{\pi}{8}, \cos\frac{\pi}{4}, \cos\frac{3\pi}{8}$} &
	\includegraphics[scale=0.16]{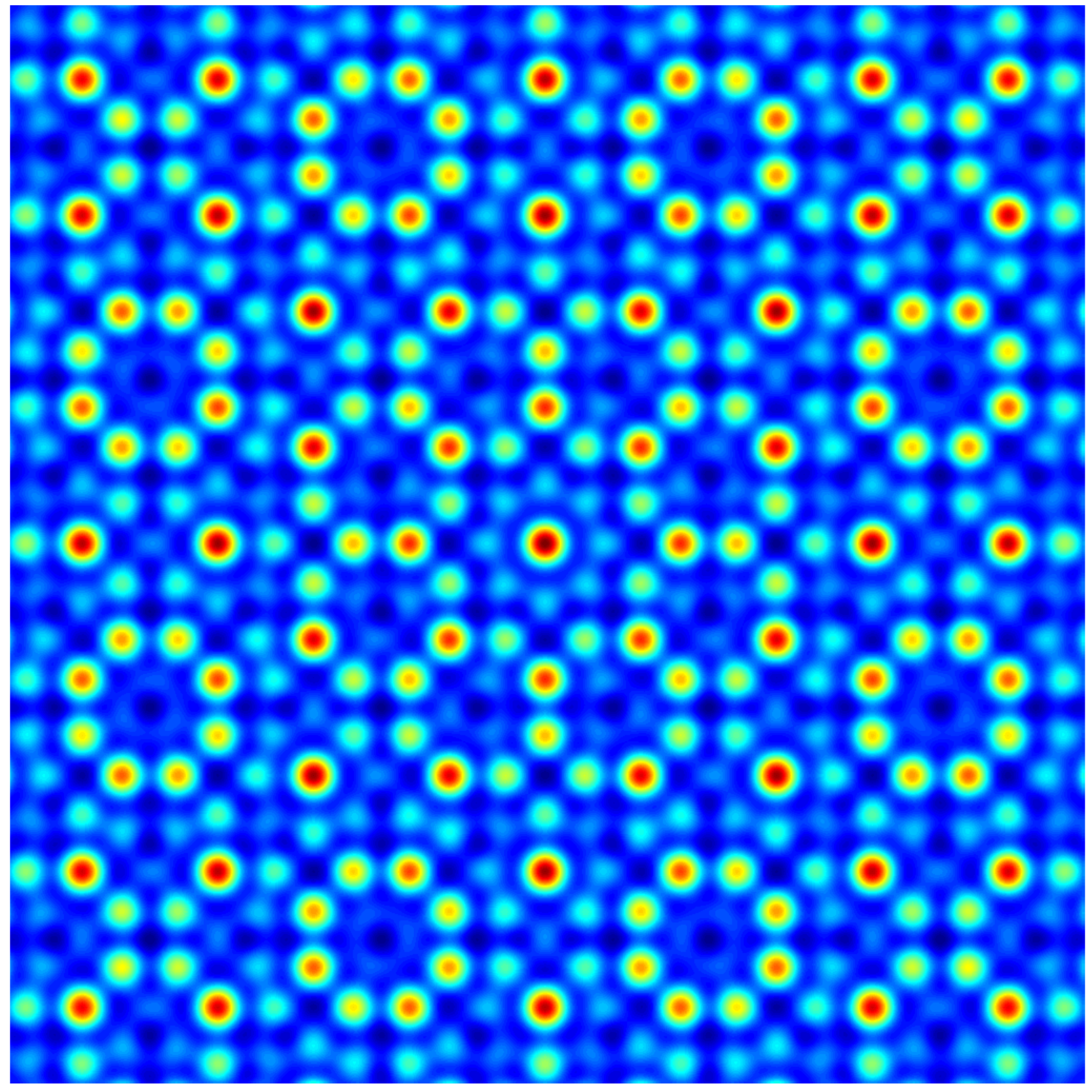} &
	\includegraphics[scale=0.16]{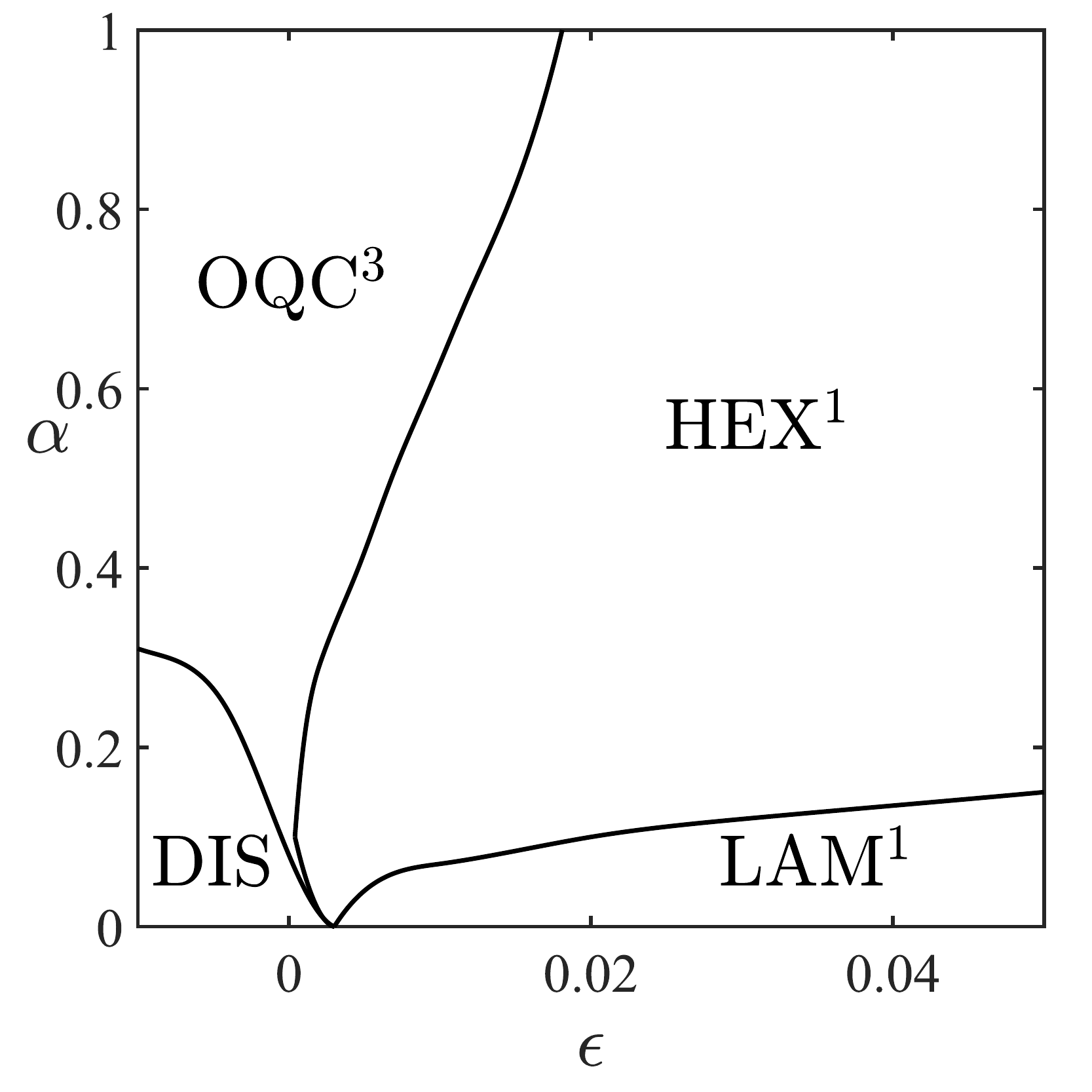} \\
	\hline
	DQCs ($10$-fold) \rule[60pt]{100pt}{0pt} & 
	\includegraphics[scale=0.14]{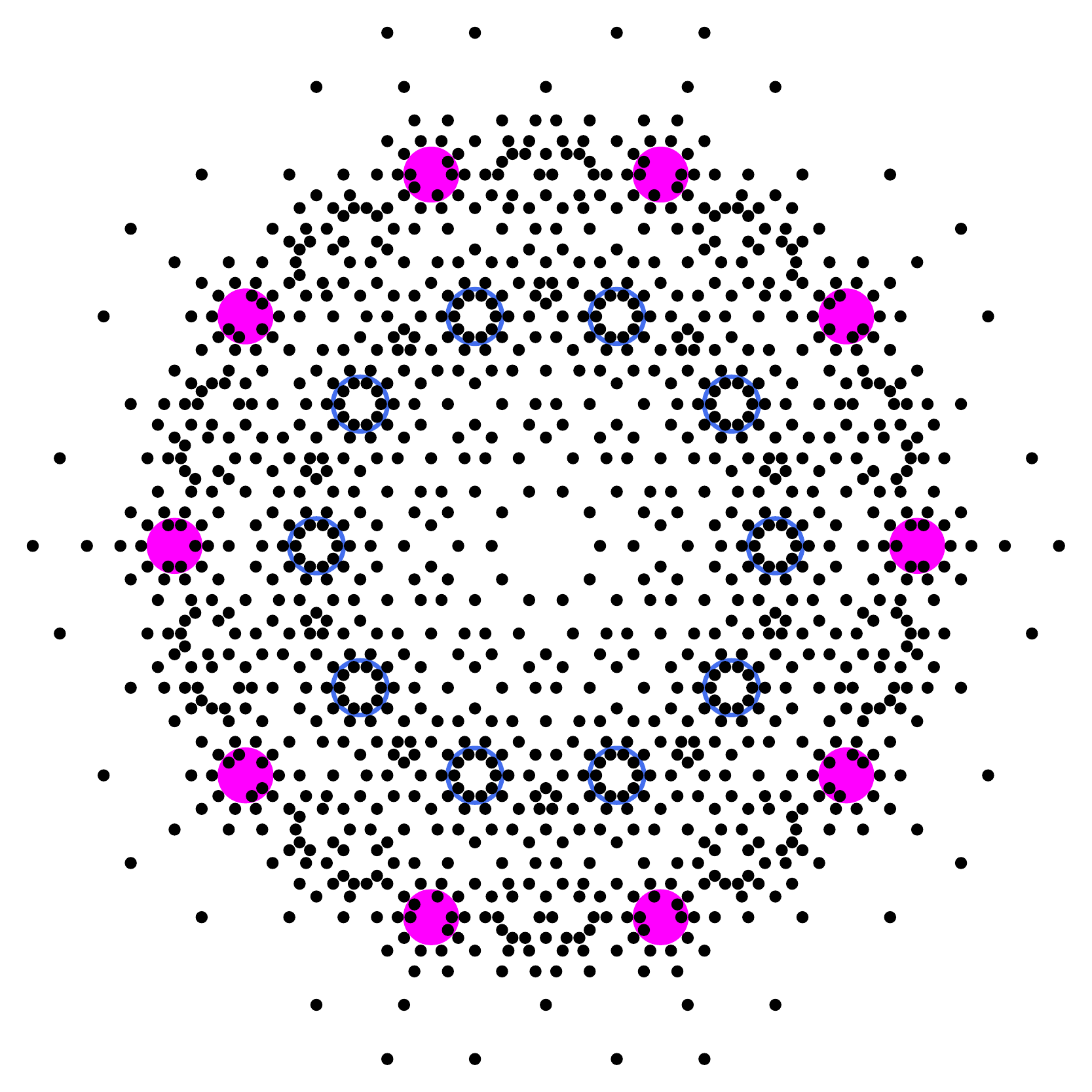}
	\centerline{$q: \cos\frac{\pi}{10}, \cos\frac{3\pi}{10}$} &
	\includegraphics[scale=0.16]{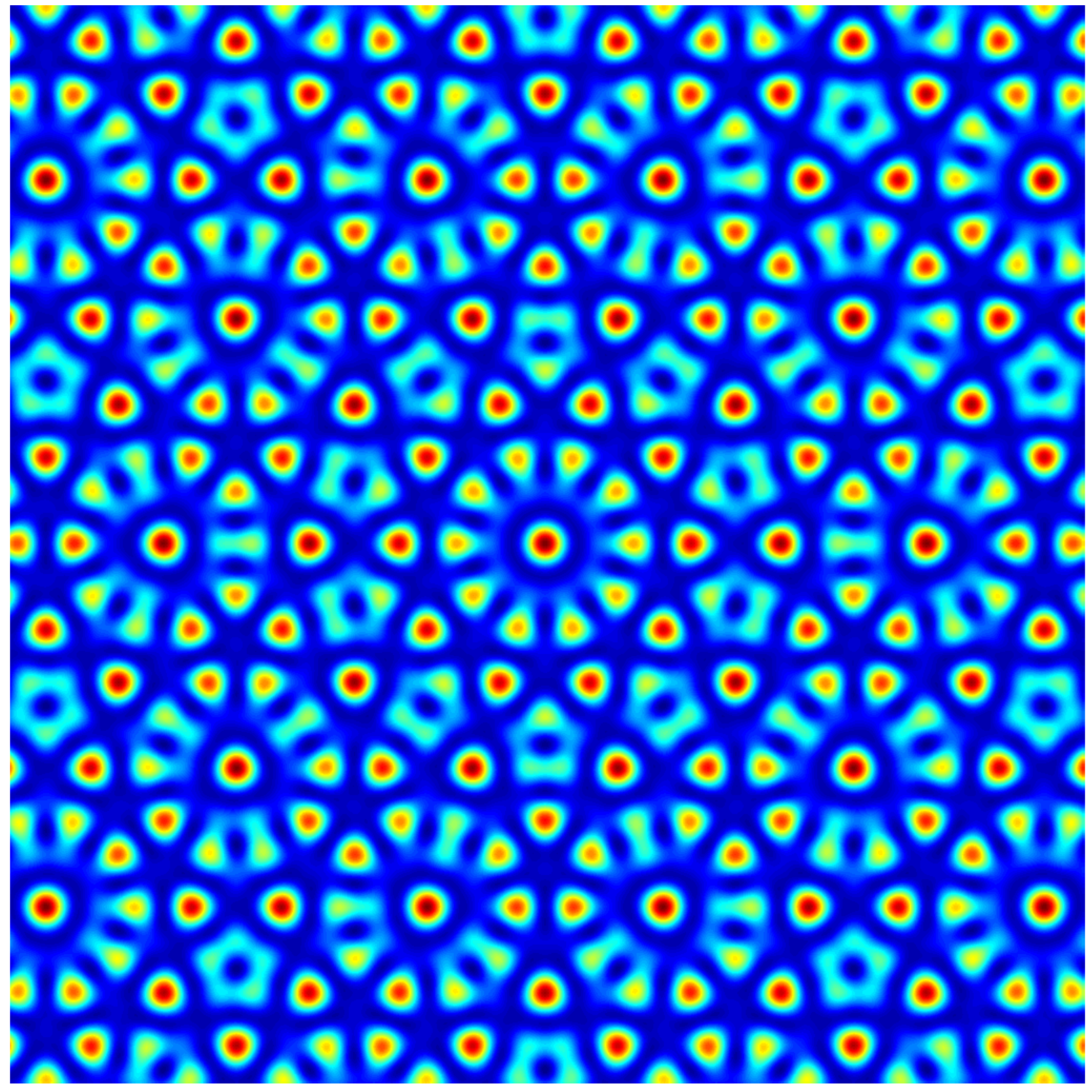} &
	\includegraphics[scale=0.16]{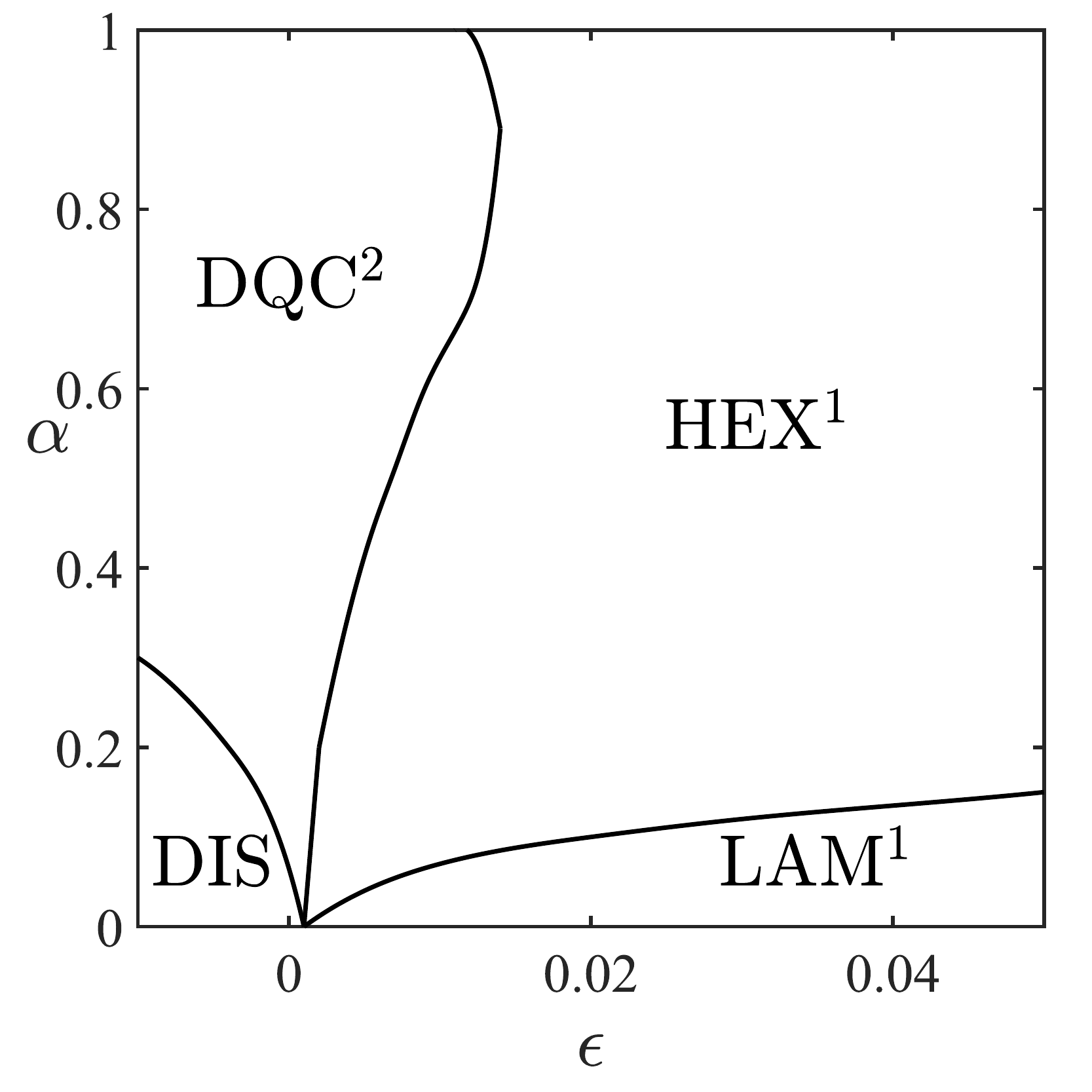} \\
	\hline
	DDQCs ($12$-fold) \rule[60pt]{100pt}{0pt} & 
	\includegraphics[scale=0.14]{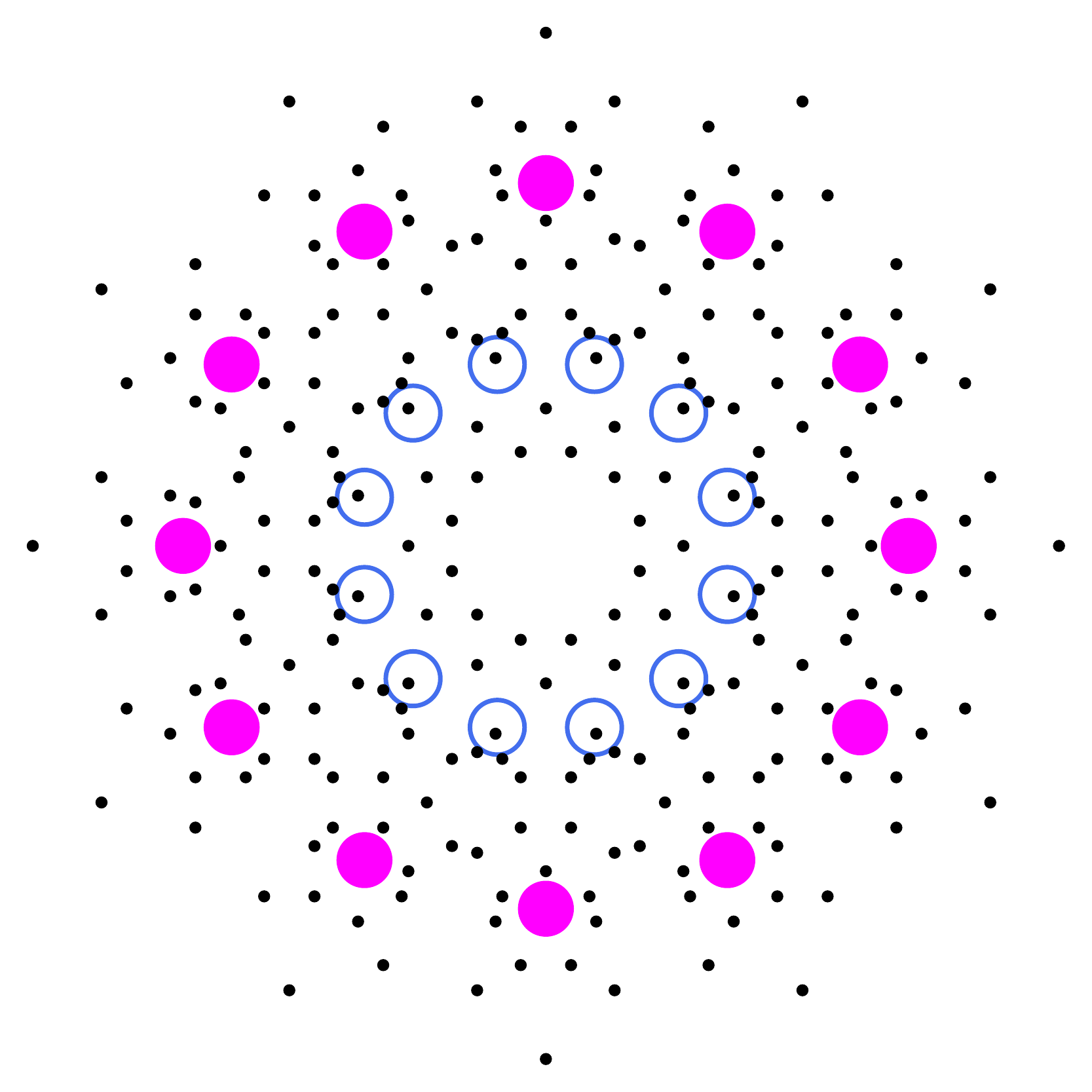}
	\centerline{$q: \cos\frac{\pi}{12}, \frac{1}{2}$} &
	\includegraphics[scale=0.16]{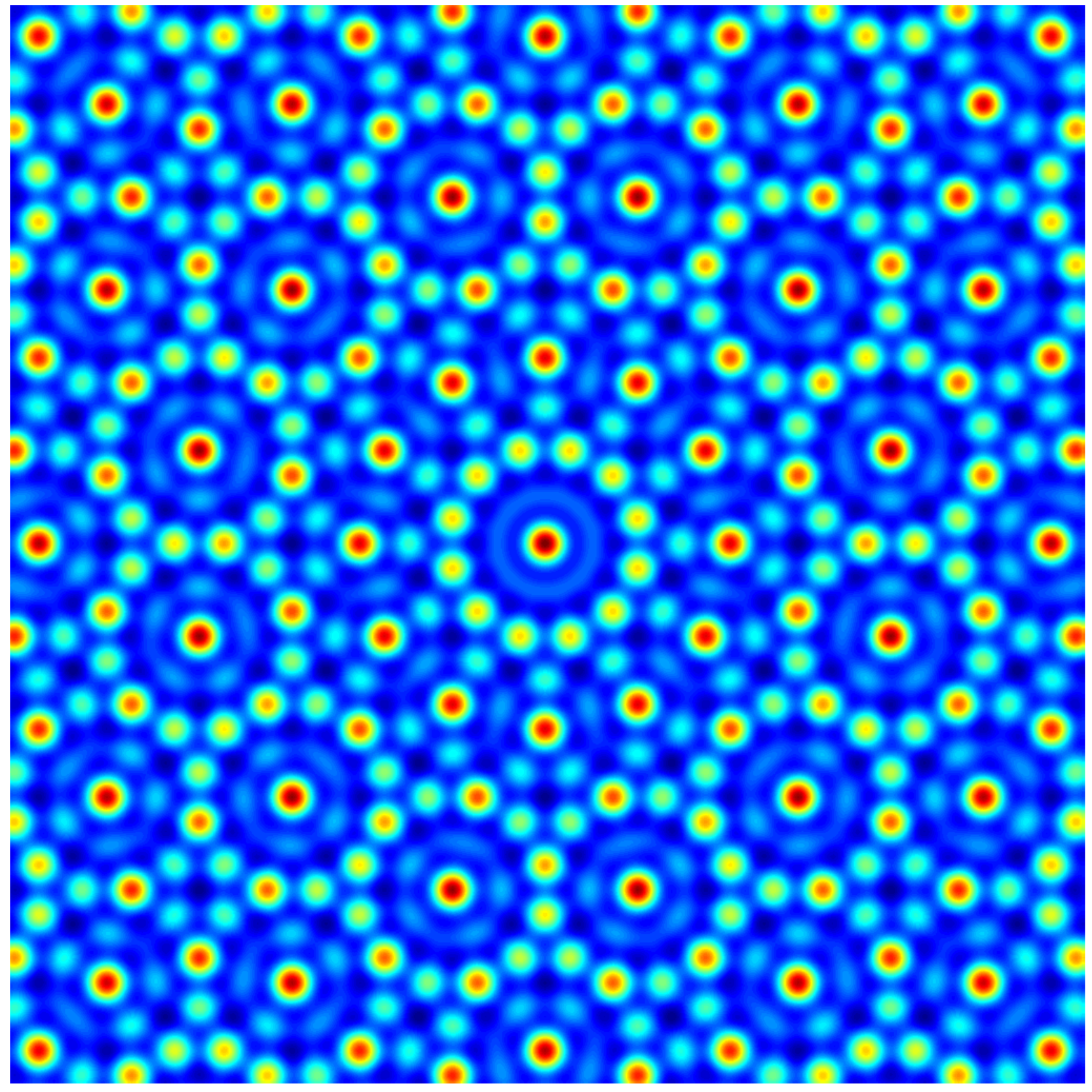} &
	\includegraphics[scale=0.16]{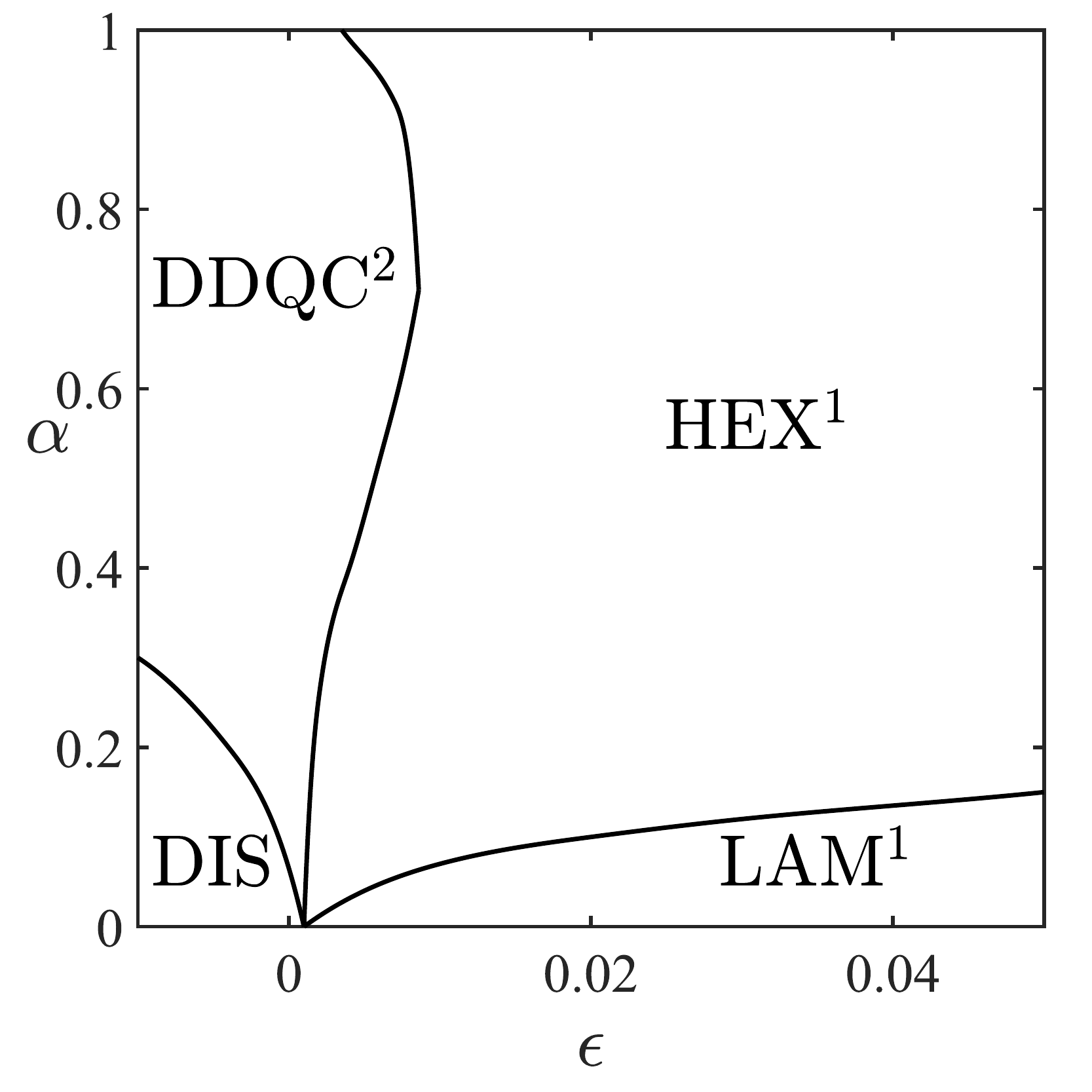} \\
	\hline
	TDQCs ($14$-fold) \rule[60pt]{100pt}{0pt} &
	\includegraphics[scale=0.14]{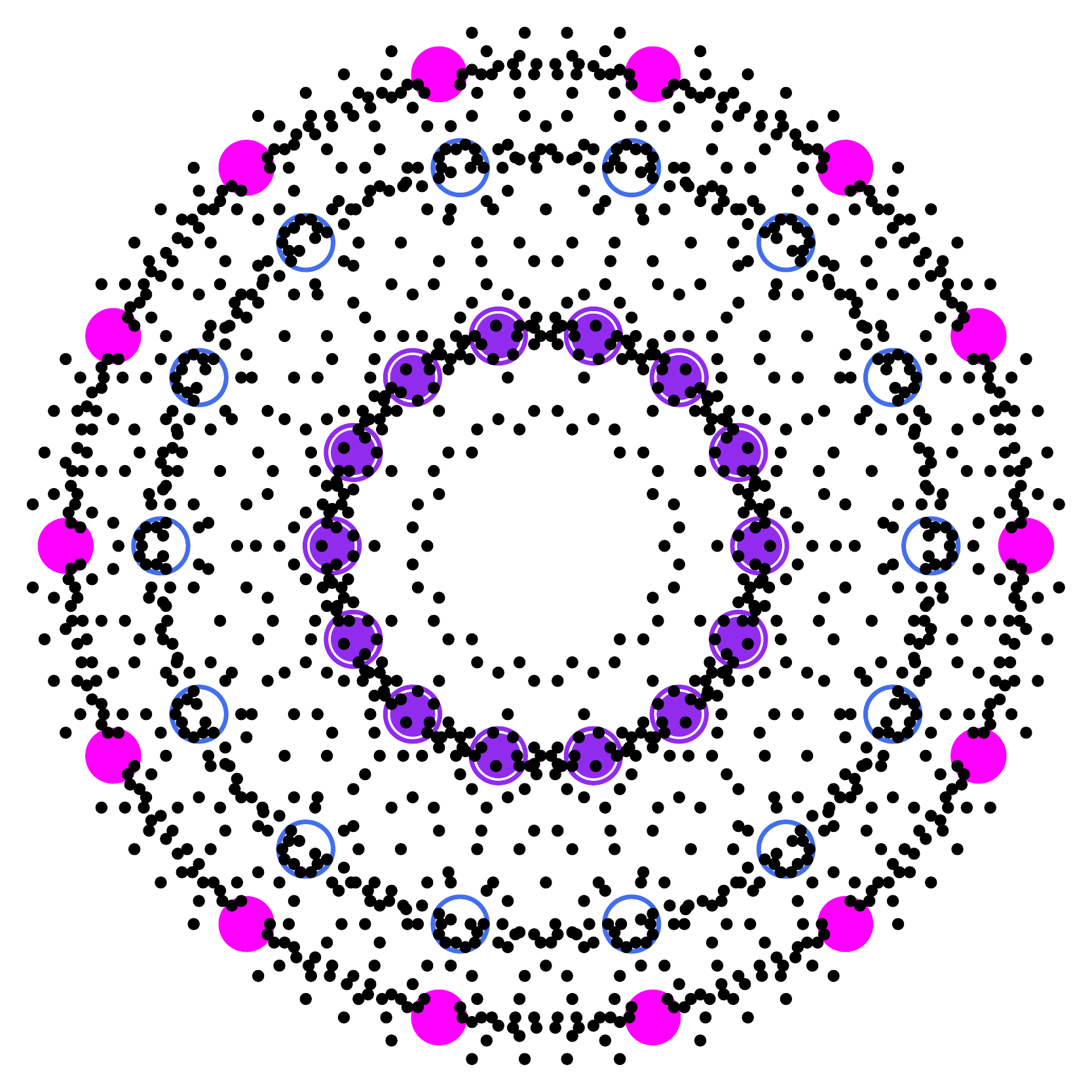}
	\centerline{$q: \cos\frac{\pi}{14}, \cos\frac{3\pi}{14}, \cos\frac{5\pi}{14}$} &
	\includegraphics[scale=0.16]{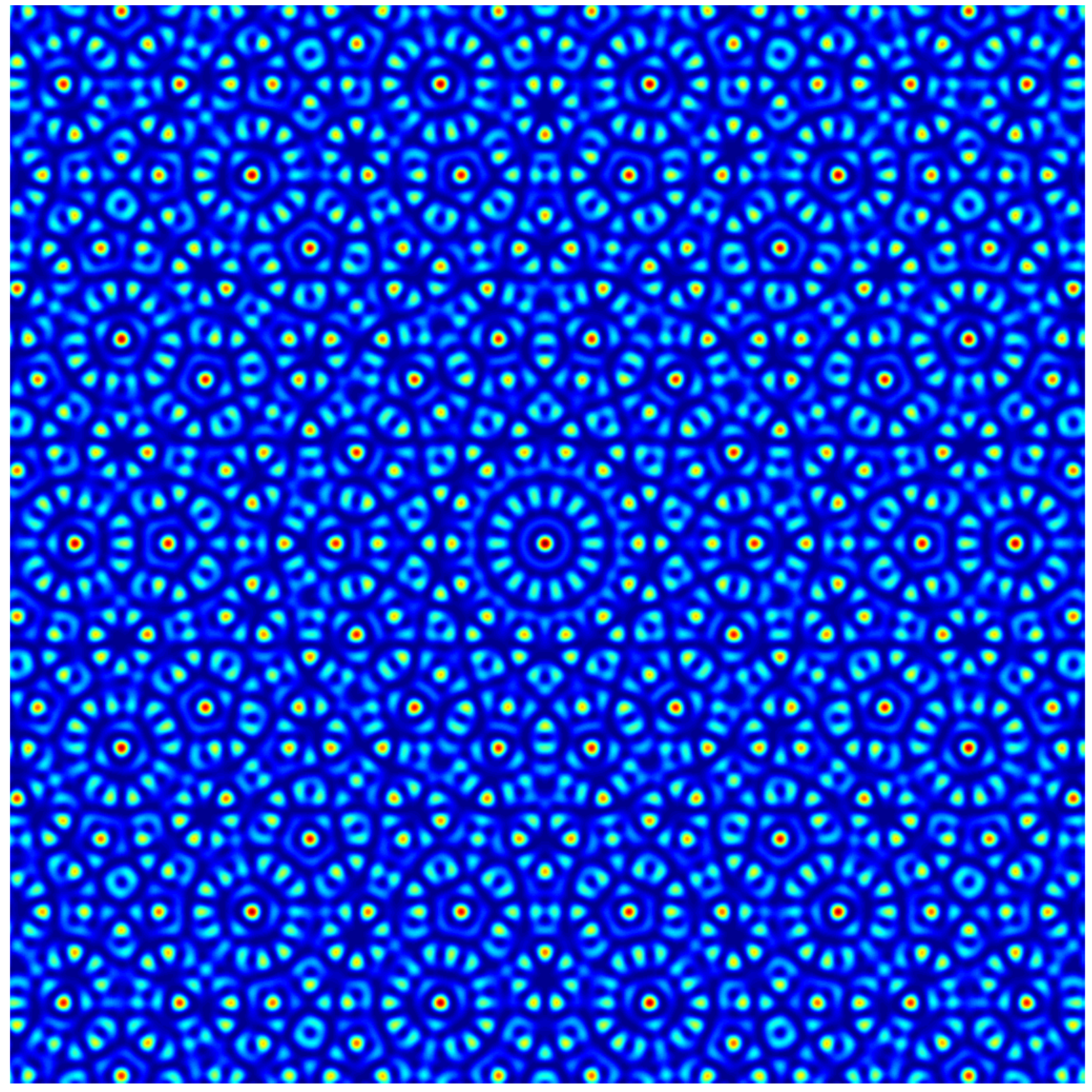} &
	\includegraphics[scale=0.16]{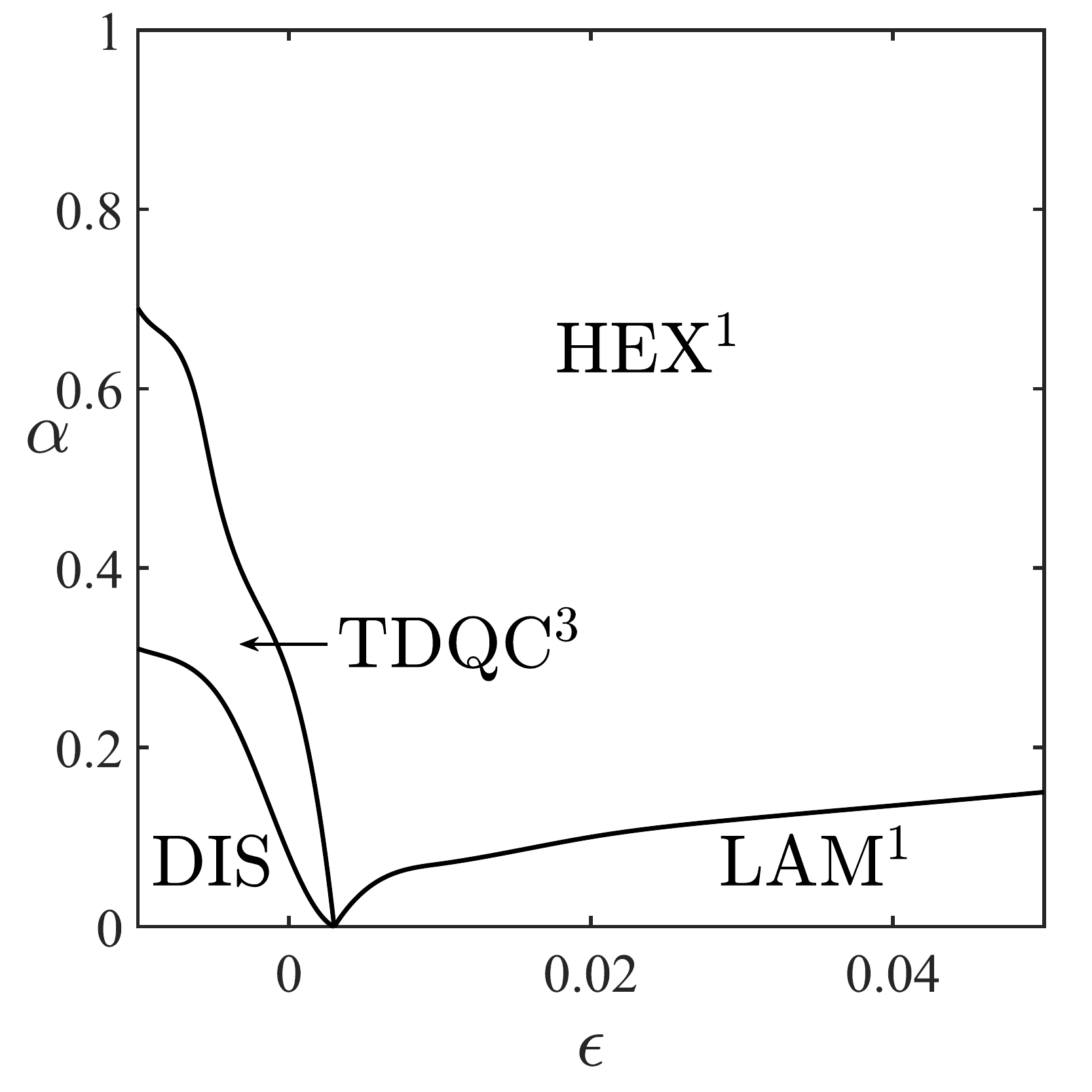} \\
	\hline
	HDQCs ($16$-fold) \rule[60pt]{100pt}{0pt} &
	\includegraphics[scale=0.14]{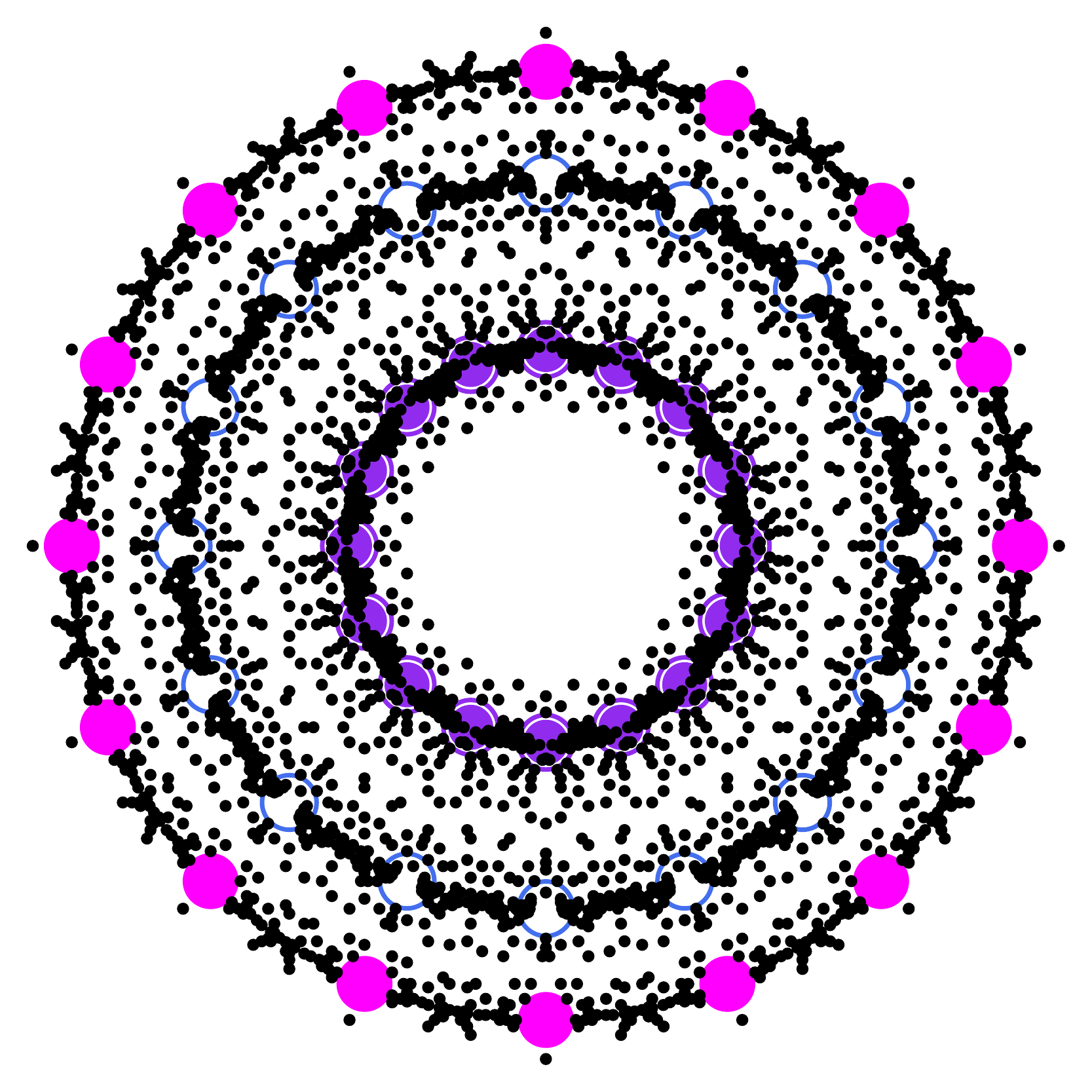}
	\centerline{$q: \cos\frac{\pi}{8}, \cos\frac{\pi}{4}, \cos\frac{3\pi}{8}$} &
	\includegraphics[scale=0.16]{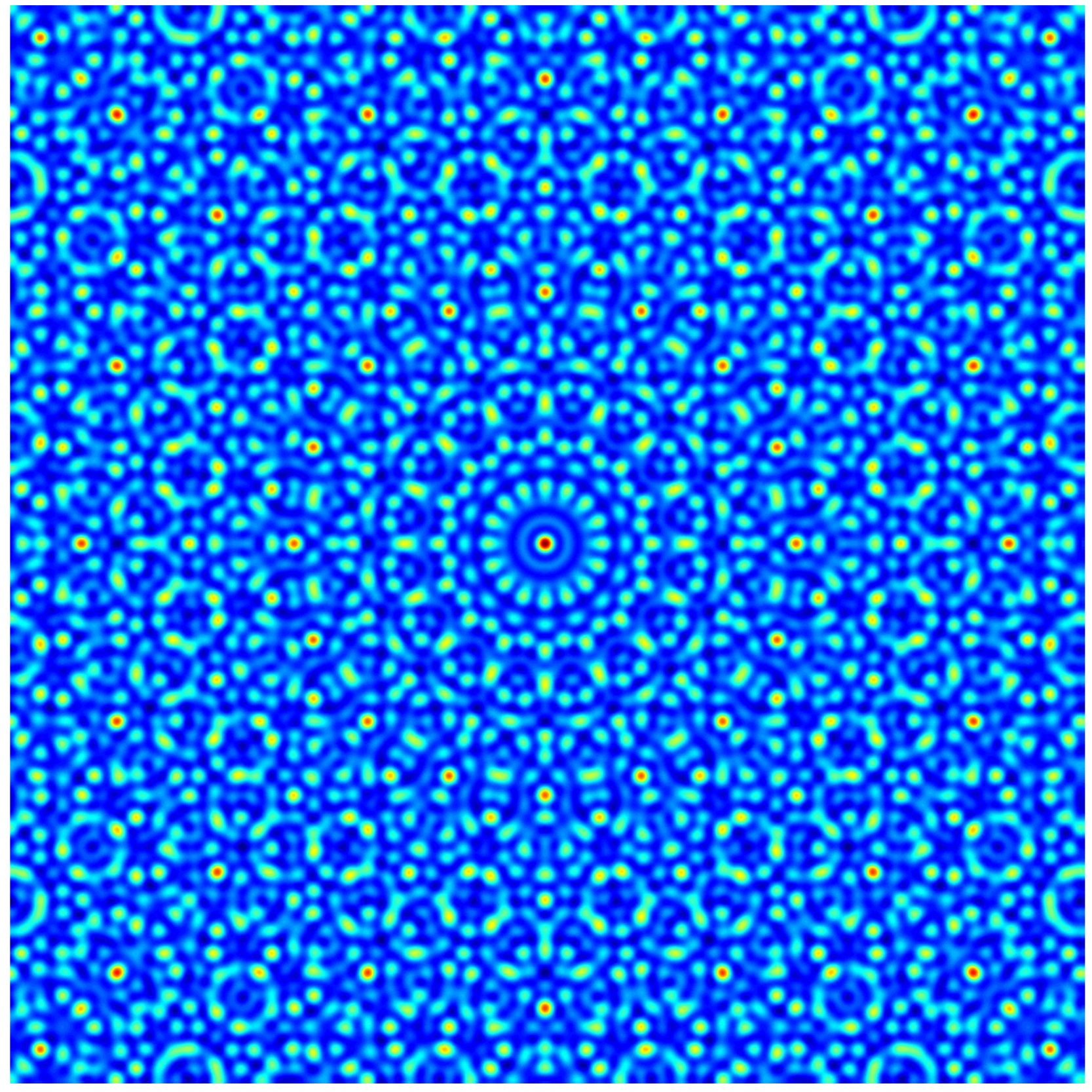} &
	\includegraphics[scale=0.16]{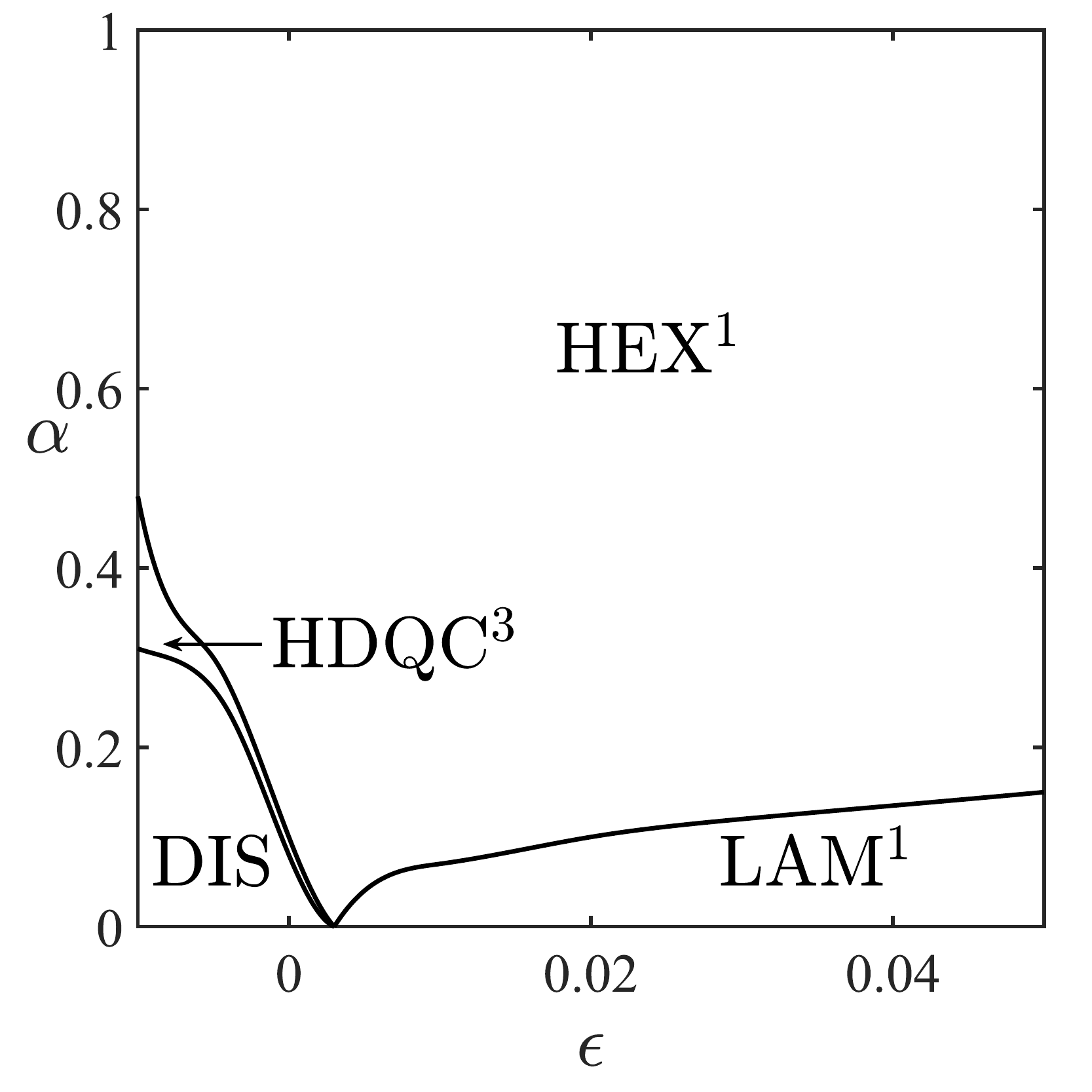} \\
	\hline
	ODQCs ($18$-fold) \rule[60pt]{100pt}{0pt} &
	\includegraphics[scale=0.14]{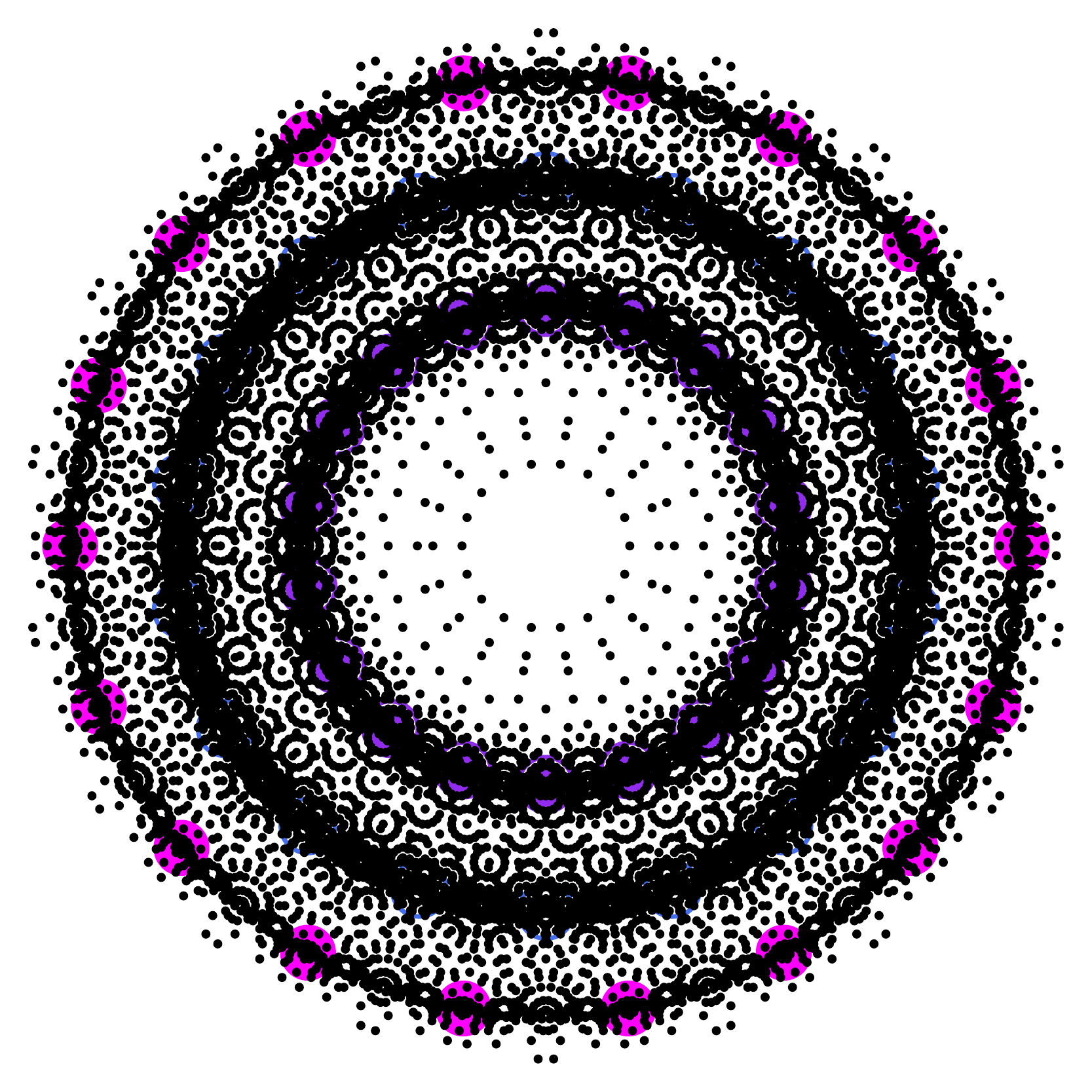}
	\centerline{$q: \cos\frac{\pi}{18}, \cos\frac{2\pi}{9}, \frac{1}{2}$} &
	\includegraphics[scale=0.16]{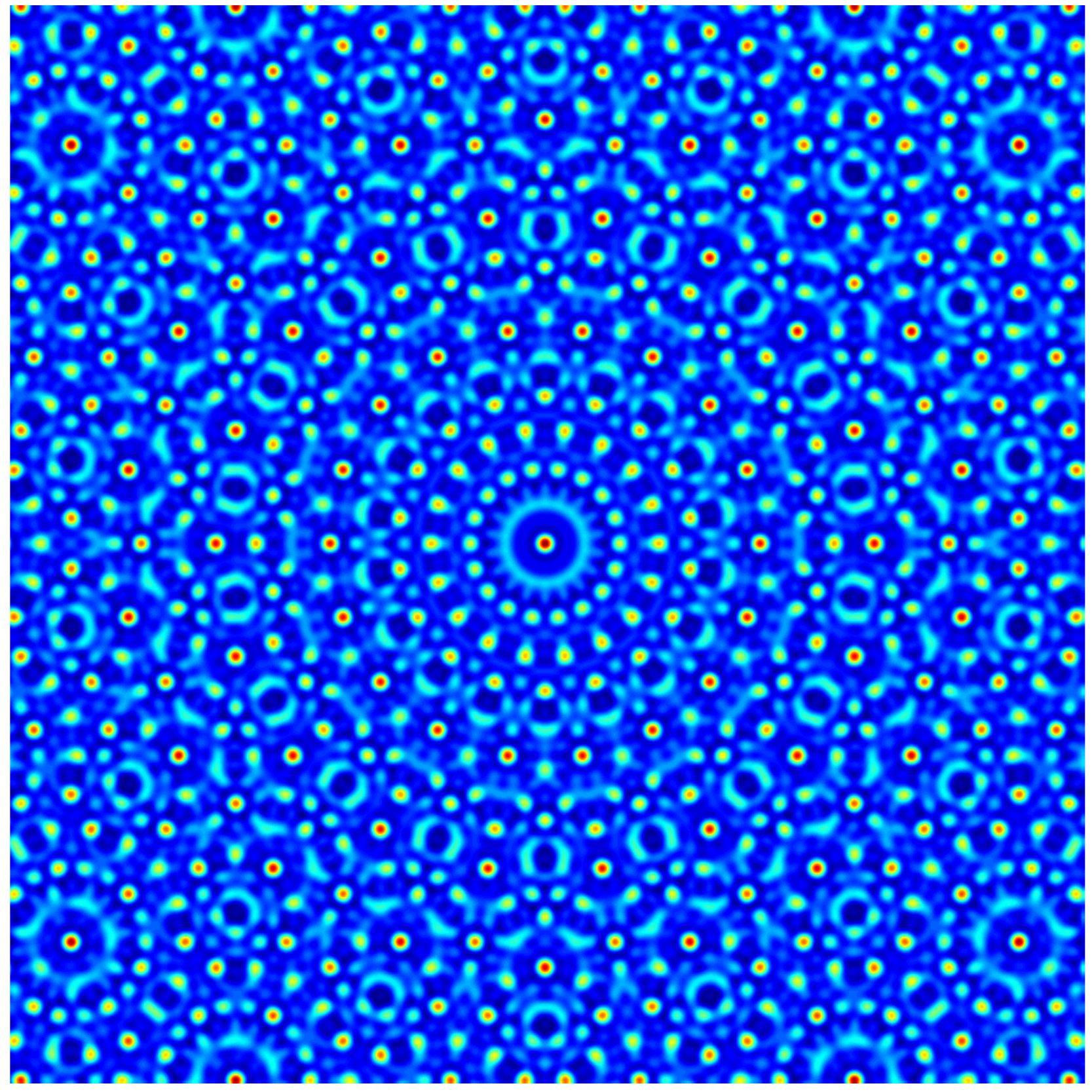} &
	\includegraphics[scale=0.16]{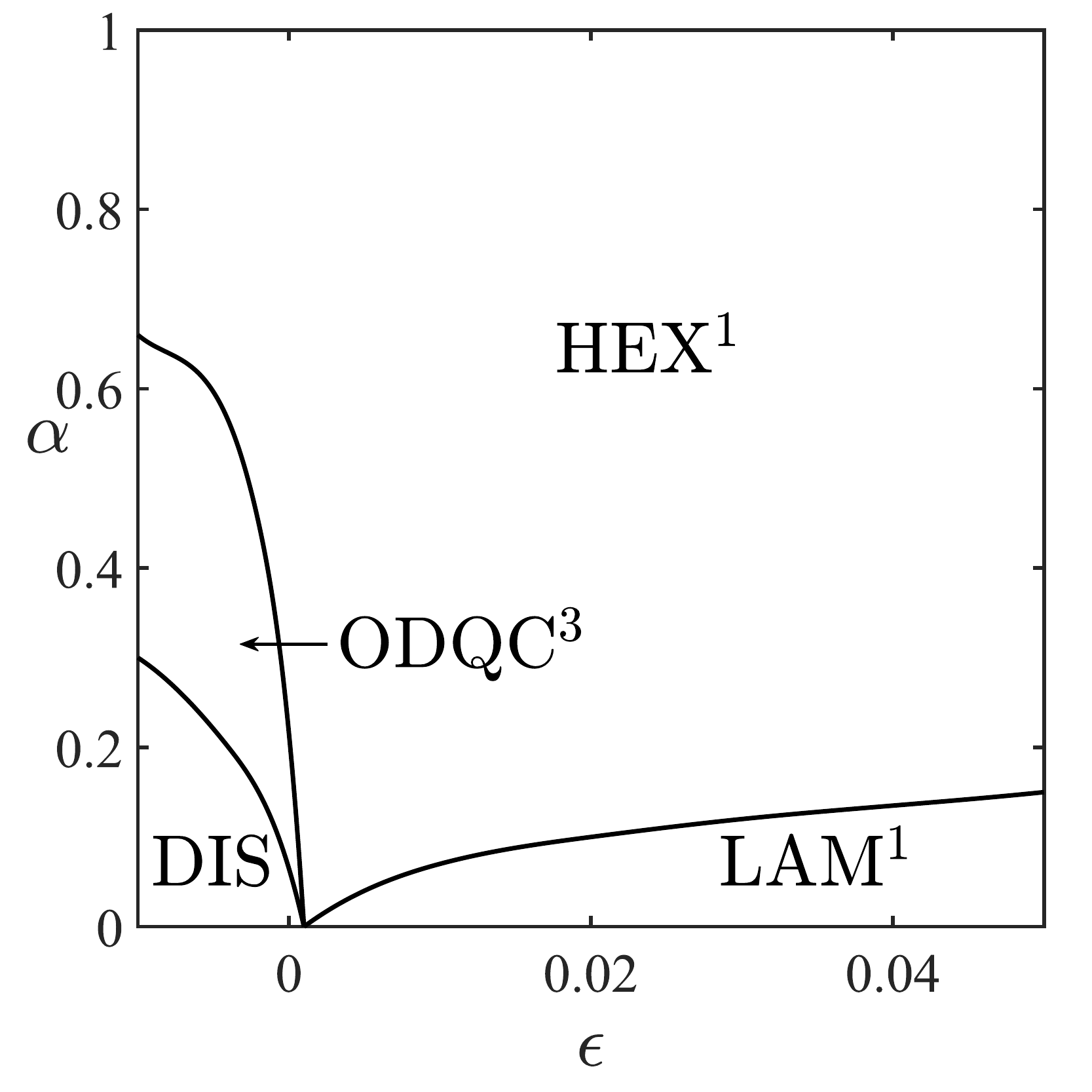} \\
	\hline
\end{longtable*}

\balance


\bibliography{ref}
\bibliographystyle{apsrev4-1}

\end{document}